\documentclass[12pt]{article}

\usepackage{amsmath}
\usepackage{graphicx}
\usepackage{latexsym}
\usepackage{amsfonts}
\usepackage{amssymb}
\usepackage{epsfig}
\usepackage{pstricks}

\makeatletter
\@addtoreset{equation}{section}
\makeatother

\newcommand{\beqn}{\begin{eqnarray}}
\newcommand{\eeqn}{\end{eqnarray}}
\newcommand{\be}{\begin{equation}}
\newcommand{\ee}{\end{equation}}
\newcommand{\non}{\nonumber \\}

\newcommand{\C}{{\sigma}}
\newcommand{\zt}{\zeta}
\newcommand{\A}{A}

\newcommand{\thba}[2]{\vartheta[\!\!\begin{array}{c}{\phantom{}\vspace{-.5mm}\scriptstyle#1}%
                        \\[-1.6mm]{\scriptstyle #2}\end{array}\!\!]}

\newcommand{\K}{K\"{a}hler}
\newcommand{\N}{{\cal N}}

\def\be{\begin{equation}}
\def\ee{\end{equation}}
\def\bea{\begin{eqnarray}}
\def\eea{\end{eqnarray}}

\newskip\humongous \humongous=0pt plus 1000pt minus 1000pt

\newif\ifdtup

\begin{document}


\begin{titlepage}
\begin{flushright}
arXiv:0804.3961\\
LMU-ASC 18/08, MPP-2008-26\\
SU-ITP-2008-08, YITP-2008-25\\
\end{flushright}
\vspace{.3cm}
\begin{center}
\baselineskip=16pt {\bf \LARGE  Update of  D3/D7-Brane Inflation  }
\vskip2mm  {\bf \LARGE on $K3\times T^{2}/\mathbb{Z}_{2}$}\\
\vspace*{0.6cm} \vfill
 {\Large
Michael Haack${}^a$, Renata Kallosh${}^{a,b,c}$, Axel Krause${}^{a}$, \\ \vskip2mm  Andrei Linde${}^{a,b,c}$, Dieter L\"{u}st${}^{a,d}$    and Marco Zagermann${}^d$
} \\
 \vspace*{5mm} \vfill
 {\small
 ${}^a$ Arnold Sommerfeld Center for Theoretical Physics \\
 Department f\"{u}r Physik, Ludwig-Maximilians-Universit\"{a}t M\"{u}nchen,  \\
 Theresienstra{\ss}e 37, 80333 Munich, Germany\\ \vskip2mm
 ${}^b$ Department of Physics, Stanford University, Stanford, CA 94305 \\
${}^c$ Yukawa Institute for Theoretical Physics, Kyoto, Japan \\ \vskip2mm
 ${}^d$  Max-Planck-Institut f\"{u}r Physik, F\"{o}hringer Ring 6,
 80805 Munich,  Germany \\ \vskip4mm  }
\end{center}
 \vfill
 \begin{center}
 {\bf Abstract}
 \end{center}

\noindent We update the D3/D7-brane inflation model on $K3\times T^{2}/\mathbb{Z}_{2}$
with branes and fluxes. For this purpose, we study the low energy theory
including $g_s$ corrections to the gaugino condensate superpotential that
stabilizes the $K3$ volume modulus. The gauge kinetic function is verified
to become holomorphic when the original ${\cal N}=2$ supersymmetry is
spontaneously broken to ${\cal N}=1$ by bulk fluxes. From the underlying
classical $\mathcal{N}=2$ supergravity, the theory inherits a shift symmetry
which provides the inflaton with a naturally flat potential. We analyze the
fate of this shift symmetry after the inclusion of quantum corrections. The
field range of the inflaton is found to depend significantly on the complex
structure of the torus but is independent of its volume. This allows for a
large kinematical field range for the inflaton. 
Furthermore, we show that the D3/D7  
model may lead to a realization of the recent CMB fit by Hindmarsh et al.\ 
with an 11\% contribution from cosmic strings and a spectral index close to 
$n_s=1$. On the other hand, by a slight change of the parameters of the 
model one can strongly suppress the cosmic string contribution and reduce 
the spectral index $n_{s}$ to fit the WMAP5 data in the absence of cosmic 
strings. We also demonstrate that the inclusion of quantum corrections 
allows for a regime of eternal D3/D7 inflation.

\vspace{2mm} \vfill \hrule width 3.cm
\end{titlepage}
\tableofcontents{}
\newpage

\section{Introduction}
During the past couple of years, the attempts to derive
 viable inflationary models from string theory have led
to a large number of very interesting scenarios, many of which
are, e.g., described  in the reviews \cite{Infreviews}. One of
the insights gained from this work was that a serious
phenomenological discussion of such models may require very explicit computations that also take into account
various types of stringy quantum corrections. What is more, the
results of these explicit computations may
 be quite unexpected, leading to surprising phenomenological
properties, as was nicely demonstrated in the recent updates
\cite{Baumann:2006th,Baumann:2007np,Baumann:2007ah} of the original KKLMMT model \cite{KKLMMT} of $D3/\overline{D3}$-brane
inflation in a warped throat geometry.\footnote{The 
idea of brane inflation with the inflaton as an inter-brane 
distance was proposed in \cite{Dvali:1998pa}.}

In this paper, we would like to revisit another inflationary model, the
D3/D7-brane inflation model \cite{Dasgupta:2002ew},
which also admits quite explicit calculations of the relevant quantum
corrections, and, moreover, has phenomenological properties that look
very interesting in view of some recent work on cosmic
strings and the CMB \cite{Bevis:2007gh,Pogosian:2008am}.  According
to \cite{Pogosian:2008am}, the recent puzzle of some
high $l$ excess power in CMB data from the ACBAR experiment, reported
in \cite{Reichardt:2008ay}, might possibly be considered as an evidence
for the existence of cosmic strings with tensions near the observational
bound.  In that case, the fit to the data in \cite{Bevis:2007gh}
requires $n_s\approx 1$ which is a prediction of D-term hybrid inflation
\cite{Binetruy:1996xj,Kallosh:2003ux,Dvali:2003zh,Binetruy:2004hh}
in the regime of very small couplings \cite{Kallosh:2003ux}.
The fact that a contribution of cosmic strings eases the tension
between D-term hybrid inflation and observational data and makes this
model consistent with $n_{s} \approx 1$ was noticed earlier in 
\cite{Battye:2006pk}.

To be concrete, the model we would like to study in this paper
is D3/D7-brane inflation on the background
$K3\times T^{2}/\mathbb{Z}_{2}$. This model has been introduced in
\cite{Dasgupta:2002ew} and further studied in
\cite{Hsu:2003cy}-\cite{Dasgupta:2008hw}.
One of the reasons  to study the D3/D7-model on
$K3\times T^{2}/\mathbb{Z}_{2}$ is its high computability. Type IIB string theory compactified on $K3\times T^{2}/\mathbb{Z}_{2}$ is related to M-theory compactified on $K3\times K3$  \cite{Tripathy:2002qw,Aspinwall:2005ad} and is associated with 4D, ${\cal N}=2$ supergravity \cite{N21} specifically described in  \cite{Angelantonj:2003zx,D'Auria:2004qv,D'Auria:2004td}.
Bulk moduli  stabilization in these models was studied in a series of  papers,
and it is one of the best understood string theory models with
stabilization of all bulk moduli \cite{Aspinwall:2005ad}. In its simplest
incarnations this model does not contain the
D-branes necessary to describe the Standard
Model of  particle physics at low energies
(see \cite{Angelantonj:2002ct,Kiritsis:2003mc,Blumenhagen:2006ci} for a review
on D-brane models with Standard Model like properties).
Therefore, the D3/D7-system studied in this paper
should be regarded as a brane/flux module, which is responsible
for inflation and
moduli stabilization, and which has to be complemented by
additional D-branes in order
to obtain realistic Standard Model phenomenology at lower energies.

In the D3/D7-brane inflationary model, an attraction between a D3- and a D7-brane is
triggered by a non-self-dual world volume flux on a D7-brane,  which
we will henceforth call the Fayet-Iliopoulos (FI)  D7-brane.
If both branes are spacetime-filling, and the D7-brane wraps
the $K3$-factor, the transverse interbrane distance on
$T^{2}/\mathbb{Z}_{2}$ plays the role of the inflaton.
A distinguishing feature of this model (as compared, e.g.,
with $D3/\overline{D3}$-brane inflation) is that the supersymmetry breaking
during the slow-roll de Sitter phase is spontaneous, and hence well-controlled.
More precisely, the supersymmetry breaking can be understood in terms
of a two-step process: Certain \emph{bulk} three-form fluxes on
$K3\times T^{2}/\mathbb{Z}_{2}$ may
spontaneously break the original $\mathcal{N}=2$ supersymmetry
preserved by the geometry to $\mathcal{N}=1$. In the resulting
effective $\mathcal{N}=1$ theory, the \emph{world volume}
fluxes on the D7-brane
then give rise to a D-term potential. Assuming the volume
modulus of the K3-factor to be fixed, this D-term potential
is non-zero for sufficiently large D3-D7-distance, breaking supersymmetry
spontaneously to $\mathcal{N}=0$.
This final spontaneous supersymmetry breaking induces a Coleman-Weinberg type
one-loop correction to the scalar potential that drives the D3-brane
towards the FI D7-brane. This motion corresponds to the phase of slow-roll inflation.

The resulting model is a stringy version of a hybrid D-term inflation
model \cite{Binetruy:1996xj,Kallosh:2003ux,Dvali:2003zh,Binetruy:2004hh}
with a waterfall stage at the end in which a charged scalar field
condenses.\footnote{This condensing field corresponds to a particular
state of the strings stretching between the D3- and D7-brane, which
becomes tachyonic at a certain  critical interbrane  distance due to
the world volume flux on the D7-brane. The D3-brane is then dissolved
on the D7-brane as an instanton, and $\mathcal{N}=1$ supersymmetry
becomes restored \cite{Dasgupta:2002ew}.}
As a D-term inflation model,  D3/D7-brane
inflation, a priori, does not suffer from the generic supergravity
eta-problem of F-term inflation models. The main problem of D-term
inflation  is instead the cosmic string production during the
waterfall stage,  when  the spontaneous breaking of the underlying
$U(1)$-symmetry takes place and the D-flatness condition is restored.

Depending on the value of the gauge coupling, $g$, of that $U(1)$,
two parameter regimes  have been studied for this model
in \cite{Kallosh:2003ux}:\\

\noindent\textbf{Regime A:}  If $g\geq 2\times 10^{-3}$, the last 60
e-foldings of inflation start far away from the bifurcation point
where the local de Sitter minimum turns into a de Sitter maximum.
In this regime, the cosmic string tension  is too large, and,
unless quantum corrections are taken into account and/or suitable modifications of the setup are made, this regime is ruled out
observationally. The spectral index
in this regime turns out to be $n_{s}\approx 0.98$.   \\

\noindent\textbf{Regime B:} For very small gauge coupling,
$g \ll  2 \times 10^{-3}$, the cosmic string tension can be
lowered to acceptable values, but the spectral index increases to
$n_{s}\approx 1$. In this parameter regime, inflation takes place
near the bifurcation point of the scalar potential.\\

For WMAP1, a spectral index $n_{s}\approx 0.98$  was
a very good fit to the data, and
a lot of work on D3/D7-brane inflation at that time
focused on curing the cosmic string
problem in regime A, e.g., by turning the cosmic strings
into so-called semi-local cosmic strings
\cite{Urrestilla:2004eh,Binetruy:2004hh,Dasgupta:2004dw,Urrestilla:2007sf}. In \cite{Urrestilla:2007sf} it was found that the upper bound on the semilocal cosmic string tension is three times higher than the one for the local Abelian strings which are produced at the end of standard D-term inflation. Another possibility to suppress cosmic strings in D-term inflation and to  lower the spectral index is by using higher order corrections to the \K \, potential \cite{Seto:2005qg}.
More recent data now tend to prefer slightly smaller values for the
spectral index, but the latest WMAP5 result \cite{Komatsu:2008hk} of $n_s =
0.96^{+0.014}_{-0.013}$  still
has large enough
error bars to accommodate $n_s\approx 0.98$ at the two sigma level.

Our goal in
this paper is to clarify some theoretical issues of D3/D7-brane
inflation and to
explore the possible parameter space of its predictions so as to be
prepared when more precise data become available in the future (for
instance from the
Planck satellite).  We give a particular emphasis also to the role
of the cosmic strings in this model \cite{Kallosh:2003ux,Dvali:2003zh,Binetruy:2004hh,Dasgupta:2004dw}. It has been realized recently that more  information on the evolution of cosmic string networks and updated numerical simulations will be required to compare the observational data with theory, see for example the recent work
\cite{Urrestilla:2007yw,Urrestilla:2008jv}. The possible general relevance of cosmic strings for
string theory was emphasized in \cite{Copeland:2003bj}.

Part of our original motivation for revisiting the  D3/D7-brane inflation model
in this context is the recent work \cite{Bevis:2007gh}, which
argues that a spectral index $n_{s}\approx 1$ might actually be compatible
with WMAP3 CMB data if an  11\% contribution to the CMB  due to cosmic strings
is allowed and properly taken into account.\footnote{This fit of the data is currently being revisited with account of WMAP5 data (M. Hindmarsh, private communication).} This would make the above-described regime B a case of phenomenological interest.

The second main motivation for our analysis derives from the insights gained in
\cite{Baumann:2006th,Baumann:2007np,Baumann:2007ah} that quantum corrections
may significantly alter the phenomenological properties of a model
in an unexpected way and that one may use tree-level supergravity
methods to compute some of these corrections in situations where
the use of conformal field theory methods is quite delicate.
In this paper, we perform similar computations for the
D3/D7-model, considering also the effects
of volume stabilization.

Volume stabilization in this model is achieved by a non-perturbative
F-term potential
due to either Euclidean D3-brane instantons or gaugino condensation
on stacks of D7-branes, which may arise after
spontaneous breaking of supersymmetry to ${\cal N}=1$. In this paper, we only focus on the volume of the $K3$-factor (the other K\"{a}hler moduli
could be stabilized by Euclidean D3-brane instantons \cite{Aspinwall:2005ad}\footnote{These Euclidean D3-instantons necessarily wrap the $T^{2}/\mathbb{Z}_{2}$-factor.
As the only open string dependence of the resulting superpotentials is via
the transverse distance between the spacetime filling
D3-brane and the corresponding D3-instanton, these
superpotentials are independent of the
D3-brane position along $T^{2}/\mathbb{Z}_{2}$
and, hence, the inflaton. \label{independent}}). Moreover, we restrict ourselves to
the mechanism of gaugino condensation. This implies a  constraint
on the charged matter spectrum of the brane setup, which has to
allow for the presence
of a non-perturbative superpotential from gaugino condensation
(for the case of Euclidean D3-branes analogous constraints were discussed in
\cite{Witten:1996bn,Gorlich:2004qm,Aspinwall:2005ad}). We will come back to this in
section \ref{gcsupo}. It should be possible to
obtain a charged matter content in the D3/D7 inflation model
which allows for gaugino condensation by considering appropriate
fluxes \cite{Gorlich:2004qm,Cascales:2004qp,LMRS,Jockers:2005zy}.
We will assume this in the following, without considering a concrete
model. The compatibility of D-term potentials from worldvolume fluxes
and gaugino condensation on D7-branes was subject of refs.
\cite{Burgess:2003ic,Binetruy:2004hh,Choi:2005ge}.

We consider non-perturbative
superpotentials of the form $W\sim\exp (-a f_{D7})$, where $a$ is a
constant and $f_{D7}$
denotes the gauge kinetic function of the D7-brane gauge fields in
the effective $\mathcal{N}=1$ supergravity description. This gauge kinetic
function depends \emph{holomorphically} on the moduli. At tree-level it is
just equal to the (complexified) $K3$-volume modulus, but, generically, it
also exhibits a dependence on the (complex) D3-brane positions, $y_{3}$,
on $T^{2}/\mathbb{Z}_{2}$ due to open string one-loop effects \cite{Berg:2004ek}. This
dependence on $y_{3}$ together with the dependence of the K\"{a}hler potential
on $y_{3}$ in general would lead to a $y_{3}$-dependence of the corresponding
non-perturbative F-term potential
that is supposed to fix the volume of $K3$. In analogy to the case described
in \cite{KKLMMT}, one might therefore fear that slow-roll inflation along
(a real slice of) $y_{3}$ might suffer a severe interference with this
volume stabilizing F-term potential, leading to an eta-problem.

In ref.\ \cite{Hsu:2003cy}, however, it was observed that, to a certain
approximation, the 4D theory descending from the
$K3\times T^{2}/\mathbb{Z}_{2}$-compactification features a
shift-symmetry along the real part of $y_{3}$, which may
protect that direction from getting
a large mass from the non-perturbative F-term potential. More precisely,
if one follows \cite{Angelantonj:2003zx,D'Auria:2004qv,D'Auria:2004td}
and describes the ``lowest order'' theory at the $\mathcal{N}=2$ level in terms
of a \emph{cubic} holomorphic prepotential $\mathcal{F}(t)=c_{ijk} t^i t^j t^k$,
where $t^i$ denote the moduli living in vector multiplets, the resulting
K\"{a}hler potential, $K$, only depends on the imaginary parts of the moduli:
$K=K(t^{i}-\bar{t}^{i})$.\footnote{``Lowest order'' here refers to an expansion
of the prepotential for large values of the dilaton $u$ and the
$K3$ volume modulus $s$, cf.\ eqs.~\eqref{sdef1} and \eqref{udef}. It is
not an expansion in the string coupling. We will come back to this point, for instance, below eq.~\eqref{bm} in section \ref{qc}.} Moreover, for the particular cubic prepotential suggested in \cite{Ferrara:1996wv,Antoniadis:1996vw}, it was observed in \cite{Hsu:2003cy} that the relevant D7-gauge couplings (still at the $\mathcal{N}=2$ level) likewise do not depend on the real part of $y_{3}$. This inflaton shift symmetry is not expected to survive all quantum corrections
(e.g.\ the generic threshold corrections to the D7-gauge couplings that we
mentioned in the previous paragraph), but it raises the hope that the violations of the shift symmetry can be kept small, at least in certain parameter regimes.

It is the purpose of this paper to address this and other important
features of the low energy effective theory of the D3/D7 inflationary model and to apply the resulting insights
to the  discussion of brane configurations with promising cosmology. This
includes, in particular, configurations with cosmology along the lines
of \cite{Bevis:2007gh}, i.e.\ with
$n_s\approx 1$, which relies on a future detection of cosmic strings. However, the quantum corrections could also lead to a smaller spectral index, closer to the WMAP5 value without the account of cosmic strings.

On our way towards this goal, we focus on the following points:
\begin{enumerate}
\item As we mentioned above, an important ingredient in the D3/D7-brane
inflation scenario with volume stabilization via gaugino condensation
is the partial spontaneous supersymmetry breaking from $\mathcal{N}=2$ to $\mathcal{N}=1$.
As for the gauge couplings, there is a profound difference
between $\mathcal{N}=2$ and $\mathcal{N}=1$ supergravity: In $\mathcal{N}=1$
supergravity, the gauge kinetic function must be holomorphic, whereas in
$\mathcal{N}=2$ supergravity it is in general not. Thus, in a first step
we verify that the gauge kinetic function indeed becomes holomorphic after
integrating out the fields that become massive in the $\mathcal{N}=1$
minimum.\footnote{This holomorphicity problem is different from the
``rho-problem'' discussed in \cite{Berg:2004ek,Giddings:2005ff,Baumann:2006th},
which may be present even in $\mathcal{N}=1$ theories that do not
arise from spontaneously breaking $\mathcal{N}=2$. We will come back to the ``rho-problem'' in section \ref{qc}.} This is an important
consistency check which also allows us to trace the shift symmetry through the
process of partial supersymmetry breaking $\mathcal{N}=2 \rightarrow \mathcal{N}=1$.
This will be the main content of section \ref{secholomorphic}.

\item The purely cubic prepotential of ref.\ \cite{Ferrara:1996wv,Antoniadis:1996vw}
by itself certainly gives rise to a consistent $\mathcal{N}=2$ supergravity theory in 4D.
It was stressed in \cite{Berg:2004ek}, however, that the cubic prepotential
already contains part of the open string threshold corrections. It is thus
important to know precisely  which quantum corrections it really  captures
in order to understand possible additional corrections that might
break the inflaton shift symmetry. To this end, we have to uncover how the
theory induced by the cubic prepotential is related to the full 10D theory
including all the relevant corrections. This will be the main content of
section \ref{qc}. As it is not clear how to generalize the
world sheet calculation of \cite{Berg:2004ek} to the case with RR fluxes, we
use the closed string dual Green's function method
\cite{Giddings:2005ff, Baumann:2006th, Baumann:2007ah}. The results are, however,
consistent with \cite{Berg:2004ek}. This was already noticed in
\cite{Giddings:2005ff, Baumann:2006th}, but here we
fill in some missing steps to make the application of the Green's function method
to the $K3\times T^{2}/\mathbb{Z}_{2}$ model more concrete
(see also appendix \ref{10to4}).
We end section \ref{qc} with drawing a parallel between the form of the
gaugino condensate superpotential in the case at hand and the one found
in the warped throat case \cite{Baumann:2006th}.

\item As we discuss further in the main text, the D7-brane stack on
  which gaugino condensation takes place should be at a different
  position on $T^{2}/\mathbb{Z}_{2}$ than the D7-brane on which
  world-volume flux is supposed to attract the D3-brane (see, e.g.,
  Fig. \ref{pillowfig} in section \ref{braneconfigs} for a possible realization). Otherwise, the
  $K3$-volume is destabilized after inflation. As an analogy consider the  result of the recent computation for the volume
  stabilizing superpotential  in $D3/\overline{D3}$ inflation \cite{Baumann:2007ah}, where
$
W=  A \left[ 1- \left(\frac{\phi_{D3}}{\phi_{D7}}\right)^{\frac{3}{2}}\right]^{1/N} e^{-a\rho}
$ with $N$ being the rank of the gauge group and $\rho$ denoting the volume modulus.
When the position of the mobile D3-brane, $\phi_{D3}$, coincides with the position, $\phi_{D7}$, of the volume stabilizing D7-branes, i.e.,  $\phi_{D3}=  \phi_{D7}$, the superpotential vanishes.
This means that the D3-brane  has to move towards the anti-D3
  brane in the direction opposite to the stack of D7-branes that is responsible for the volume stabilization, so that at the exit from inflation at $\phi_{D3}= 0$ the superpotential acquires  a simple KKLT-type  form \cite{Kachru:2003aw}
$W_{exit}=  A e^{-a\rho}$ with constant $A$.
In our model, the role of the attracting anti-D3-brane is played by the
  D7-brane with the world-volume flux on it, i.e., by the  FI  D7
  brane. It should thus likewise be placed away from the stack of the volume
 stabilizing D7's so as to avoid the destabilization of the volume at the exit from inflation.

\item The quantum corrections to the D7-brane gauge coupling, and thus to the
non-perturba\-tive superpotential, indeed break the shift symmetry of the
real part of $y_3$, and in general the real part of $y_3$ is no longer a distinguished direction.
It becomes then a matter of fine-tuning to obtain a direction in field space that is flat enough to support inflation, cf.\ also
\cite{McAllister:2005mq}. For example, one could imagine
that there are values for the complex structure of the torus for which the quantum corrections are small. A more basic question, which one can pose even without a flat direction (but which has an important bearing on describing inflation in this model in case there is a flat direction)
is the question of the kinematical field range
of the canonically normalized D3-brane coordinate. We will see in section \ref{inflaton} that this field range can be much larger than
usually assumed when the torus is very asymmetrical, i.e., when the imaginary part of  its complex structure
is either very large or very small. If the corresponding direction happens
to be flat enough for slow roll inflation this would allow for a large
field variation during inflation. Although this is promising, since
it relaxes the Lyth bound on the tensor-to-scalar ratio \cite{Lyth},
we will find in section \ref{cosmology} that the actual tensor
modes in the D3/D7 model are tiny (at least for the conventional situation in which the vacuum energy during inflation is dominated by the D-term).

\item Given the improved understanding of the low energy effective theory
gained from studying the previous points, one can then look for the most
interesting and controllable brane configurations and study their
cosmology. This is sketched in section \ref{braneconfigs}. We leave
a more detailed study to future work \cite{pheno}.

\item In section \ref{cosmology}, we give a first discussion
  of the cosmological
  properties of the updated D3/D7 model. In particular, we discuss the
  possibility to reproduce the fit to the data in \cite{Bevis:2007gh}
  by using the simplest brane configuration
  discussed in section \ref{braneconfigs}. We also study a more
  general situation in which the string theory corrections derived
  in the preceding sections of the paper play an interesting
  role for cosmology.
  In particular, we find that these corrections can lead to a
  deviation from the standard D-term inflation
  scenario that allows for more flexibility to adjust the values
  of the cosmic string
  tension and the spectral index. The corrections can also lead to a
  maximum in the potential, which allows for a regime of
  eternal inflation.
\end{enumerate}
Our results are summarized in section \ref{concl}. To begin with, we introduce a few general facts of the
$K3\times T^{2}/\mathbb{Z}_{2}$ compactification
in section \ref{d3d7inflation}.
Some technical details are collected in the Appendices.


\section{Effective action for D3/D7-brane inflation on $K3\times T^{2}/\mathbb{Z}_{2}$}
\label{d3d7inflation}

In this section, we collect some background material on the low energy
effective action that describes the
D3/D7 inflationary model on $K3\times T^{2}/\mathbb{Z}_{2}$. This will
flesh out some of the statements in the introduction and  serves as our
starting point for a more refined analysis of this effective action in
subsequent sections.

We begin with a few general remarks on type IIB string theory on
$K3\times T^{2}/\mathbb{Z}_{2}$. The $\mathbb{Z}_{2}$ orientifold
operation involves the product
\begin{equation} \label{z2}
 \mathbb{Z}_{2}=\Omega \cdot (-1)^{F_{L}} \cdot \mathcal{I}\ ,
\end{equation}
where $\Omega$ denotes the worldsheet orientation reversal, $F_{L}$ is the
spacetime fermion number in the left-moving sector, and $\mathcal{I}$ reflects
the coordinates on $T^{2}$. The operation $\mathcal{I}$ on the torus has four
different fixed points, and the geometry of the space $T^{2}/\mathcal{I}$ has
the shape of a ``pillow'' with each of the four corners being the location of
an O7-plane that fills out $K3$ and the non-compact part of spacetime. Global
cancellation of the 7-brane tadpole requires the presence of 16 D7-branes that
likewise wrap $K3$ and fill the 4D spacetime. In order to obtain a configuration
with constant dilaton, four D7-branes have to sit on top of each of the four
O7-planes so as to cancel the 7-brane charges locally. The geometry of this
configuration is locally flat, but the deficit angles of $\pi$ at each
fixed point lead to a globally spherical topology. Distributing the D7-branes
differently deforms the configuration away from the orientifold limit and
generically requires a description in terms of
F-theory \cite{Vafa:1996xn}.\footnote{For a nice discussion of the geometry of $T^{2}/\mathcal{I}$
see, for instance, section 6.3 of \cite{Dabholkar:1997zd} and for general
introductions to orientifolds, see \cite{Angelantonj:2002ct,Blumenhagen:2006ci}.}

As the O-planes and the D7-branes both wrap $K3$, they induce altogether minus 24 units
of D3-brane charge \cite{Tripathy:2002qw}, which has to be canceled by D3-branes
and/or background flux such that
\begin{equation} \label{tadpol}
 \frac{1}{2}N_{\rm flux}+N_{D3}=24\ ,
\end{equation}
where
\begin{equation}
 N_{\rm flux}=\frac{1}{(2\pi)^4(\alpha^{\prime})^{2}}\int_{K3\times T^{2}}
H_{3}\wedge F_{3}
\end{equation}
with the integral being evaluated on the covering torus (which
explains the factor of $1/2$ in front of $N_{\rm flux}$ in \eqref{tadpol},
cf.\ also \cite{Kachru:2002he}).
Here, $H_{3}$ and
$F_{3}$ denote the NSNS and RR three-form field strengths, respectively.

The geometry of $K3\times T^{2}/\mathbb{Z}_{2}$ preserves
$\mathcal{N}=2$ supersymmetry in 4D, and the light fields can be
grouped into the $\mathcal{N}=2$ supergravity multiplet as
well as several $\mathcal{N}=2$ vector and hypermultiplets.
A very clear correspondence between 4D and 10D fields can be
given at or near the orientifold limit. Let us first consider
this correspondence in the case without D3-branes.


\subsection{Classical effective action without D3-branes}
\label{withoutd3}

The NSNS- and RR-two-forms with one leg along the non-compact
directions and one
along the torus give rise to four vector fields in 4D. One linear
combination of these four vectors corresponds to the 4D graviphoton,
whereas the other three enter three vector multiplets. The three
complex scalars of these vector multiplets
are\footnote{Note that our definition of the complex
structure modulus $t$ differs slightly
from \cite{D'Auria:2004qv}. Following \cite{D'Auria:2004qv}, we use
the mostly minus signature for the metric. Thus, the
imaginary part of $t$ is negative in our definition (in contrast to the
conventions used in Polchinski's book \cite{Polchinski:1998rq},
for instance). This is, however,
consistent with the value $t = -i$ found in \cite{D'Auria:2004qv}
in ${\cal N}=1$ supersymmetric minima. Note, moreover, that
the definition of $u$ refers to the orientifold point at which
the D7-brane charge is canceled locally. In general, its definition
includes the D7-brane scalars, in analogy to the T-dual situation
with D9/D5-branes discussed in \cite{Antoniadis:1996vw}; see also
\cite{D'Auria:2004qv}.   Finally, we would like to mention that,
throughout the paper, we use the notation of Ferrara and collaborators
for the vector multiplets
\cite{Angelantonj:2003zx,D'Auria:2004qv,D'Auria:2004td}.
A different notation is more common in
a large part of the literature on the heterotic and type I string,
denoting the dilaton by $S$, K\"ahler moduli by $T$ and complex structure
moduli by $U$. To have a quick reference guide, we here give the relevant
permutation to relate the two notations:
\begin{eqnarray}
s &\rightarrow& T \nonumber \\
t &\rightarrow& U \nonumber \\
u &\rightarrow& S \nonumber \ .
\end{eqnarray}
 \label{fnconventions}}
\begin{eqnarray}
s&=&C_{(4)}-i \textrm{Vol}(K3)\ ,\label{sdef1}\\
t&=&\frac{g_{12}}{g_{11}}+i\frac{\sqrt{\det g}}{g_{11}}\ , \label{t} \\
u&=& C_{(0)}-ie^{\varphi} \label{udef}
\end{eqnarray}
which denote, respectively, the $K3$-volume modulus with its
axionic RR-partner $C_{(4)}$, the $T^2$ complex structure
modulus and the axion-dilaton.

The position moduli of the 16 D7-branes on the torus are denoted by $y_7^{k}$ $(k=1,\ldots,16)$. Depending on where one chooses the origin of these coordinates, they could obviously be defined in various ways. A very convenient way to define them for brane configurations close to the orientifold limit is to use $y_{7}^{1,2,3,4}$ for the complex positions of branes number 1-4 with respect to fixed point number 1, and similarly, to use $y_{7}^{5,6,7,8}$ to  denote the positions of the branes number
5-8 with respect to fixed point number 2,  and so forth. In this notation,
$y_{7}^{k}=0$ for all $k=1,\ldots,16$ thus would mean that there are
four D7-branes sitting on top of each O7-plane, and we are at the
orientifold limit with constant dilaton. However, other parametrizations
are also possible (see e.g.\ \cite{LMRS} for a more detailed account).
The complex scalars $y_{7}^{k}$ live in 16 additional vector multiplets,
with the corresponding vector fields given by the respective
D7-brane gauge fields.

Classically, the moduli space of the vector multiplet sector
is described by the special K\"{a}hler manifold
\begin{equation}
\mathcal{M}_{V}\cong \left( \frac{SU(1,1)}{U(1)}\right)_{s}\times
\frac{SO(2,18)}{SO(2)\times SO(18)}\ , \label{SO(18)coset}
\end{equation}
where the first factor is parametrized by $s$, and the remaining scalars
$(t,u,y_{7}^{k})$ span the second factor.
This geometry can be obtained from the following cubic prepotential:
\begin{equation}
\mathcal{F}(s,t,u,y_7^k)=stu-\frac{1}{2}sy_7^k y_7^k\ .   \label{prepnoD3}
\end{equation}
In F-theory language, the dilaton, $u$, corresponds to the complex structure of the
elliptic fiber of a second $K3$ factor, which we
will denote by $\widetilde{K3}$. In this picture, the $SO(2,18)/(SO(2)\times SO(18))$
factor of $\mathcal{M}_{V}$ describes the complex structure moduli space of
$\widetilde{K3}$. It should be noted that, far away from the orientifold limit,
the convenient separation of the scalars into closed and open string moduli is in general no longer possible, and the 10D meaning of e.g.\ $y_{7}^{k}$ as  brane positions is less clear \cite{LMRS}.

The isometry group of $\mathcal{M}_{V}$ has an obvious subgroup
$SU(1,1)_{s}\times SU(1,1)_{t} \times SU(1,1)_{u}$ (cf.\ Appendix \ref{dualsym}),
which contains the discrete subgroup
$SL(2,\mathbb{Z})_{s}\times SL(2,\mathbb{Z})_{t}\times SL(2,\mathbb{Z})_{u}$.
The group $SL(2,\mathbb{Z})_u$ is just the usual IIB S-duality group relating
strong and weak string coupling, whereas $SL(2,\mathbb{Z})_{s}$ corresponds
to a T-duality group associated with the size of $K3$.
The group $SL(2,\mathbb{Z})_t$, finally,
describes modular transformations of the two-torus, i.e., conformal
transformations that preserve its complex structure.
This symmetry will be relevant in some of the following discussions. 
We therefore stress that it 
is present also in the orientifold theory where the internal 
space is $K3 \times T^{2}/\mathbb{Z}_{2}$ (i.e.\ there is a $T^2$-factor only 
in the covering space). This can be understood by noticing that the 
$\mathbb{Z}_{2}$ symmetry, which inverts the torus coordinates, commutes with the 
$SL(2,\mathbb{Z})_t$ transformation, cf.\ \eqref{sl2zony3}. Alternatively, it 
follows from the fact that the T-dual theory (with two T-dualities along the 
$T^2$-directions)
would have an actual torus factor in the compactification space. 

The remaining moduli of the original $K3$-factor, as well as the torus volume and the remaining
axions from the  RR-four-form with two legs along K3 and two legs along
$T^{2}/\mathbb{Z}_{2}$ live in altogether 20 hypermultiplets and
parametrize, at tree-level, the quaternionic K\"{a}hler manifold
\begin{equation}
\mathcal{M}_{H}=SO(4,20)/(SO(4)\times SO(20))\ .
\end{equation}
This manifold has 22 translational isometries
along  the 22 real axionic directions, $C^{I}$ $(I=1,\ldots,22)$,
which descend  in the above-mentioned way from the RR-four-form.
These 22 axions transform in the vector representation of
$SO(3,19)\subset SO(4,20)$, and hence decompose into an $SO(3)$
triplet $C^{m}$ $(m=1,2,3)$ and an $SO(19)$-vector $C^{a}$ $(a=1,\ldots,19)$.
We will sometimes refer to the $C^{m}$ and $C^{a}$ as, respectively,
positive and negative norm axions. As we will further discuss in section \ref{secholomorphic}, bulk three-form fluxes will lead to gaugings of
some of the shift symmetries, $C^{I}\rightarrow C^{I}+\alpha^{I}$, of these axions.


\subsection{Classical effective action with D3-branes}

The inclusion of D3-branes introduces additional open string moduli\footnote{The presence of the D3-branes also
modifies the definition of the scalar ${\rm Im}(s)$ of eq.\ \eqref{sdef1}; it is
not given by the $K3$ volume anymore, cf.\ the discussion in section
\ref{qc} and a related discussion in \cite{Baumann:2006th}).}: the D3 positions on $K3$, which live in additional hypermultiplets, and the D3
positions on $T^{2}/\mathbb{Z}_{2}$, which are part of additional vector
multiplets (which also include the D3-brane gauge fields). As the most relevant
fields during D3/D7-brane inflation all live in vector multiplets, we focus on
that sector in the following.\footnote{ The waterfall fields are in
hypermultiplets, but they vanish during inflation. }

In the case without D3-branes, we have encountered a natural (lowest order)
description of the special K\"{a}hler geometry of the vector multiplet moduli space in terms
of the complex structure moduli of the elliptically fibered $\widetilde{K3}$,
which leads  to the symmetric space (\ref{SO(18)coset}) based on the cubic
prepotential (\ref{prepnoD3}). The D3-brane positions do not have such a natural
geometric description
in F-theory, and it is a priori not clear whether they can be included,
at least in some approximation, in an equally elegant way. In refs.\
\cite{Ferrara:1996wv,Antoniadis:1996vw},
however, a simple, ``lowest order'', description in terms
of another, extended, cubic prepotential was proposed:
\begin{equation}
\mathcal{F}(s,t,u,y_7^k,y_3^r)=stu-\frac{1}{2}sy_7^k y_7^k
-\frac{1}{2}uy_3^r y_3^r\ ,  \label{prep}
\end{equation}
where $y_{3}^{r}$ $(r=1,\ldots, N_{D3})$ denote the complex D3-brane
positions on $T^{2}/\mathbb{Z}_2$. As already mentioned in the introduction,
\eqref{prep} can be viewed as the leading term in an expansion for large values of $s$ and $u$, but does not
capture all the (open string) 1-loop corrections (although it contains already some of them). We will come back to
this point in section \ref{qc}.

As an important remark, we note that the special K\"{a}hler manifold
following from the prepotential (\ref{prep}) is no longer a symmetric space
(although it is still homogeneous) and that the discrete symmetries
$SL(2,\mathbb{Z})_{s}\times SL(2,\mathbb{Z})_{t}\times SL(2,\mathbb{Z})_{u}$
are partially broken \cite{D'Auria:2004qv}. This is further elaborated on
in Appendix \ref{dualsym}, where it is also shown that the
$SL(2,\mathbb{Z})_{t}$ symmetry is restored by including the full 1-loop effects
(more concretely, we show this for the K\"ahler potential and the gauge
couplings in the case without fluxes, but one may expect it to hold more
generally).


\subsection{Towards hybrid D-term inflation}

In order to recover D3/D7-brane inflation on $K3\times T^{2}/\mathbb{Z}_{2}$
as a hybrid D-term inflation model in $\mathcal{N}=1$ supergravity, one needs to
take into account a  few additional ingredients:
\begin{enumerate}
\item \emph{Three-form fluxes}\\
Three-form fluxes on $K3\times T^{2}/\mathbb{Z}_{2}$ generically stabilize
the moduli $(t,u,y_{7}^{r})$ and may lead to spontaneous partial
supersymmetry breaking $\mathcal{N}=2\rightarrow\mathcal{N}=1$. In terms
of the 4D, $\mathcal{N}=2$ supergravity description, the fluxes induce
charges for some of the 4D fields, and one has a gauged supergravity theory
with a nontrivial scalar potential associated with the gauging. The critical
points of this potential may preserve $\mathcal{N}=2,1,0$ supersymmetry. In
an $\mathcal{N}=1$ vacuum,
one of the two $\mathcal{N}=2$ gravitini (together with some of the other fields)
gains a mass. Integrating out  these massive fields gives an effective theory with
$\mathcal{N}=1$ supersymmetry and the remaining moduli $(s,y_{3}^{r})$.
This partial supersymmetry breaking is studied in section \ref{secholomorphic}.

\item \emph{Gaugino condensation on wrapped D7-branes}\\
The volume modulus, $s$, of the $K3$ can be stabilized by non-perturbative
superpotentials, either due to Euclidean D3 instantons or gaugino condensation
on a stack of D7-branes wrapping the $K3$. For simplicity, we will consider
only gaugino condensation in this paper and assume the existence of a suitable
stack of D7-branes. The resulting non-perturbative superpotential is then of
the schematic form
\begin{equation}
W_{np}=A(y_{3},y_{7},u,t)e^{-ias}\ ,  \label{ias}
\end{equation}
where $a$ is some positive
constant and $A$ denotes a function of the other moduli
(or their stabilized values after those moduli are fixed) and
possibly of charged matter fields.\footnote{The $u$ dependence
of $A$ might arise, for instance, via a correction to the
D7-brane gauge coupling from world-volume fluxes.}
The resulting supergravity F-term potential
(which also contains a contribution, $W_{0}$, from
the flux superpotential) can then stabilize the
modulus $s$, just as in \cite{Kachru:2003aw}.
We will discuss this in more detail in section \ref{qc}.

\item \emph{World volume flux on another D7-brane}\\
As mentioned in the introduction, the inflaton potential in the original
D3/D7 inflationary model is generated by  spontaneous supersymmetry breaking
due to a non-selfdual world volume flux on another D7-brane, which then triggers
an attraction of a nearby D3-brane towards that D7-brane. In 4D, the supersymmetry
breaking due to the worldvolume fluxes can be attributed to a non-vanishing
D-term potential.
As mentioned in the introduction, it is important that  the D7-brane with the world volume
flux is different from the D7-branes on which gaugino condensation
takes place and that both types of D7-branes are at different locations on
$T^{2}/\mathbb{Z}_{2}$. The reason for this is that the function $A(y_{3},\ldots)$
entering the non-perturbative superpotential  (\ref{ias})
vanishes if the D3-brane sits on top of the D7-branes responsible for the
gaugino condensation \cite{Ganor:1996pe}. If the gaugino condensation D7-branes and those with
worldvolume flux were the same, this would lead to volume destabilization at the end
of inflation, when the D3-brane dissolves as an instanton on the D7-branes.
The situation is thus similar to the setup described in
\cite{Baumann:2007np,Baumann:2007ah}, where the mobile D3-brane
also moves away from the volume stabilizing D7-branes and approaches the
anti-D3-brane at the tip of the throat. The analogue of the anti-D3-brane
would then be the D7-brane with world volume flux in our
setup.
\end{enumerate}


\subsection{Inflaton shift symmetry}

As described in \cite{KKLMMT} for warped D3-brane inflation, the volume
stabilization with non-per\-tur\-ba\-tive F-term potentials can easily ruin
an otherwise successful inflationary model. In the scenario described in
\cite{KKLMMT}, this is due to the dependence of the non-perturbative F-term
potential on the D3-brane position, which is in general induced by the
K\"{a}hler potential and the $y_3$-dependence
of the analogue of our function
$A$ in the superpotential. Using fine-tuning,
one might hope to balance these effects in some cases so as to yield
valuable inflationary potentials, but the works
\cite{Baumann:2007np,Baumann:2007ah}
showed that this might be more difficult than naively expected.

Despite some superficial similarity with the situation in
\cite{KKLMMT}, the D3/D7-brane model on $K3\times T^{2}/\mathbb{Z}_{2}$,
based on the cubic prepotential (\ref{prep}),
appears to behave differently in this respect. Namely, as was argued in \cite{Hsu:2003cy},
the $\mathcal{N}=2$ theory with prepotential (\ref{prep})
features a shift symmetry for the K\"{a}hler potential and the D7-brane gauge kinetic function along the real parts of the D3-brane position
moduli, $y_{3}^{r}$.\footnote{The K\"{a}hler potential is actually independent of \emph{all}
real parts of \emph{all} scalars whenever the holomorphic prepotential is purely cubic.}
This in turn would imply a shift symmetry in
the non-perturbative F-term potential.

If we assume that the D7-brane with the non-self-dual
worldvolume flux sits at $y_7=0$ (we are from now on suppressing the indices
$k$ and $r$ of the D7- and D3-brane coordinates wherever it
does not cause confusion), the attractive force it exerts on
a mobile D3-brane only depends on the
absolute value, $|y_{3}|$, of that D3-brane's position \cite{Dasgupta:2002ew}.
Hence, if we assume that the initial position of the D3-brane
has $\textrm{Im}(y_{3})=0$, the D3-brane is attracted towards
the flux D7-brane along the $\textrm{Re}(y_{3})$ direction,
which is unaffected by the non-perturbative F-term
potential.\footnote{In fact, the F-term potential
inherits a strong dependence on $\textrm{Im}(y_{3})$
from the K\"{a}hler potential
that would drive $\textrm{Im}(y_{3})\rightarrow 0$.
$\textrm{Im}(y_{3})= 0$
thus seems to be a natural initial condition (at least
if threshold corrections are negligible). }
We would thus get
a valid D-term inflation scenario with the $s,t,u,y_7$ moduli stabilized.

It should be noted that if the K\"{a}hler potential and the relevant
gauge couplings had been functions of $|y_{3}|$ instead of
$\textrm{Im}(y_{3})$ (as would be the case, e.g., for a ``canonical''
K\"{a}hler potential $K=|y_{3}|^{2}$),
one would also have had a shift symmetry along the phase of
$y_{3}$.\footnote{This phase is
a compact direction in field space, but so is $\textrm{Re}(y_{3})$
due to the compactness of the torus.}
However, in that case, also the attractive potential between the
D7-brane with world volume flux and the D3-brane  would be independent
of the phase of $y_{3}$, and one would have a completely flat direction
and no inflation. It is thus important that the shift symmetry is along
a direction in field space along which the inflationary potential is not flat.

There are a few possible caveats in the above considerations. For
one thing, the treatment of \cite{Hsu:2003cy} was entirely in the
framework of $\mathcal{N}=2$
 supergravity, and the spontaneous partial supersymmetry breaking
to $\mathcal{N}=1$ induced by bulk fluxes was not yet taken into account.
This transition to $\mathcal{N}=1$ supergravity, however, is an important step.
First, it is a prerequisite for gaugino condensation, which is impossible in $\mathcal{N}=2$ supersymmetry. Second, it is necessary to verify the holomorphicity of the resulting $\mathcal{N}=1$ gauge kinetic function. Third, one needs to make explicit how the shift symmetry is inherited by the effective $\mathcal{N}=1$ theory that descends from the cubic prepotential (\ref{prep}).
We will consider the effects of partial supersymmetry breaking in section \ref{secholomorphic}.

Another possible caveat in our above arguments in favor of the inflaton
shift symmetry is that the shift symmetry is a consequence of the special
cubic form of the prepotential (\ref{prep}). This is in particular true for the
shift symmetries of the K\"{a}hler potential, which  are generic consequences
of cubic prepotentials, but otherwise non-generic. In fact, we already mentioned
the threshold corrections of the D7-brane gauge couplings due to stretched D3-D7
strings, which are not completely captured by a purely cubic
prepotential. The generic breaking of the shift symmetry by these corrections
can be quite easily seen for the D7-brane gauge coupling and the resulting gaugino condensate superpotential. Concretely, the threshold corrections sensitively depend on the masses
of the D3-D7 strings, which in turn depend on the distance between these branes.
Therefore, they generically induce a
$y_{3}$-dependence of the D7-brane gauge couplings, which will then be
visible as a non-trivial function $A(y_{3},\ldots)$ in the non-perturbative
superpotential (\ref{ias}). As this function (like the loop corrected
$\mathcal{N}=1$ gauge kinetic function it descends from) has to be
\emph{holomorphic}, it must also depend on the real part of $y_{3}$,
if it is to depend on $y_{3}$ at all. This would then violate
the shift symmetry.
However, the dependence of $A$ on the other (already fixed) moduli
such as $t$ or $u$ opens up the possibility that one might be able to tune
violations of the inflaton shift symmetry to a small violation that
does not change much the original desired D-term inflationary scenario.

In order to trace the above-mentioned caveats, it is evidently important to first
take a closer look at the step of partial supersymmetry breaking, which
has been neglected in the literature so far. Afterwards, we will come back
to the question of additional quantum corrections which might
lead to a breaking of the shift symmetry.


\section{Holomorphicity and partial $\mathcal{N}=2\rightarrow \mathcal{N}=1$ SUSY breaking}
\label{secholomorphic}

In 4D, $\mathcal{N}=1$ supergravity,  vector fields can have non-minimal kinetic terms of the form
\begin{equation} \label{n1notation}
{g}^{-1/2} \mathcal{L}_{\textrm{kin}} = -\frac{1}{4} \textrm{Re}(f_{\Lambda\Sigma}) F^{\Lambda}_{\mu\nu} F^{\mu\nu \Sigma} + \frac{1}{8} \textrm{Im}(f_{\Lambda\Sigma})\epsilon^{\mu\nu\rho\sigma} F^\Lambda_{\mu\nu}F^\Sigma_{\rho\sigma}\ ,
\end{equation}
where $g$ is the metric determinant,  $f(z)_{\Lambda\Sigma}$ denotes the gauge kinetic function (or, more generally, the gauge kinetic matrix), which can depend at most \emph{holomorphically} on the scalar fields $z^i$ $(i=1,\ldots,n_{C})$ of $n_{C}$ chiral multiplets.
Here, $\Lambda,\Sigma,\ldots=1,\ldots,n_{V}$, where $n_{V}$ is the number of vector multiplets.
In $\mathcal{N}=1$ supergravity, the gauge kinetic function is completely independent of the K\"{a}hler geometry of the scalar manifold.

In $\mathcal{N}=2$ supergravity, the kinetic terms of vector fields can also
be expressed in terms of the real and imaginary part\footnote{In order to  conform to large parts of the supergravity literature, we are following here the standard convention that the
  real and imaginary parts of the kinetic matrices appear in an opposite way for $\mathcal{N}=1$ and $\mathcal{N}=2$ supergravity.  The conventions are related by a simple redefinition of the form $\mathcal{N}_{\Lambda\Sigma} \rightarrow i{\mathcal{N}}_{\Lambda\Sigma}$. We are using the conventions of \cite{Ceresole:1995ca} for the $\mathcal{N}=2$ theory. \label{defi}  }
of a kinetic matrix $\mathcal{N}_{\Lambda\Sigma} (z,\bar{z})$:
\begin{equation} \label{n2notation}
{g}^{-1/2}\mathcal{L}_{\textrm{kin}} = \frac{1}{4}  \textrm{Im}(\mathcal{N}_{\Lambda\Sigma}) F^{\Lambda}_{\mu\nu} F^{\mu\nu \Sigma} + \frac{1}{8} \textrm{Re}(\mathcal{N}_{\Lambda\Sigma})\epsilon^{\mu\nu\rho\sigma} F^\Lambda_{\mu\nu}F^\Sigma_{\rho\sigma}\ .
\end{equation}
Here, the kinetic matrix $\mathcal{N}_{\Lambda\Sigma}(z,\bar{z})$ $(\Lambda,\Sigma,\ldots=0,1,\ldots,n_{V})$  is a function of the scalar fields, $z^{i}$ $(i=1,\ldots,n_{V})$, of $n_{V}$ $\mathcal{N}=2$ vector multiplets.
In contrast to the $\mathcal{N}=1$ case, however,
$\mathcal{N}_{\Lambda\Sigma}$ is in general \emph{not} a holomorphic function of the scalars $z^{i}$ (neither is it anti-holomorphic).\footnote{This is also different from \emph{rigid} $\mathcal{N}=2$ supersymmetry, where the kinetic matrix is still holomorphic.}

The non-holomorphicity of $\mathcal{N}_{\Lambda\Sigma}$ is manifest when
$\mathcal{N}_{\Lambda\Sigma}$ is derived from  a holomorphic prepotential $F(X(z))$, via the standard expression
\begin{equation}
{\cal N}_{\Lambda \Sigma}(z, \bar z) =  \bar F_{\Lambda\Sigma} + 2i \frac{\textrm{Im}(F_{\Lambda\Delta})\textrm{Im}(F_{\Sigma\Pi}) X^{\Delta}X^{\Pi}}{\textrm{Im}(F_{\Delta\Pi})X^\Delta X^\Pi}\ ,
\label{gaugecoupl}
\end{equation}
where $X^{\Lambda}(z)$ are homogeneous special coordinates on $\mathcal{M}_{V}$, and $F_{\Lambda\Sigma} \equiv \partial_{X^\Lambda}\partial_{X^\Sigma} F$.
In terms of the natural symplectic section
\begin{equation}
\Omega=(X^{\Lambda},F_{\Lambda}=\frac{\partial F}{\partial X^{\Lambda}})\ , \label{natsection}
\end{equation}
$F(X(z))$ also determines the K\"{a}hler potential on the vector multiplet moduli space by
\begin{equation}
K(z,\bar z)= -\ln[i({\bar{X}}^{\Lambda}F_{\Lambda} -{\bar{F}}_{\Lambda} X^{\Lambda})]\ .
\label{K}
\end{equation}
More generic symplectic sections, $\Omega^{\prime}=(X^{\Lambda \prime},F_{\Lambda}^{\prime})$,
where  $ F_{\Lambda}^{\prime}$ is not necessarily the derivative of a prepotential, can be obtained by
symplectic rotations of (\ref{natsection}). Whereas
the K\"{a}hler potential (\ref{K}) is manifestly symplectically invariant, and hence also
valid in the new basis $\Omega^{\prime}$, the gauge kinetic matrix does transform nontrivially.
A general expression for $\mathcal{N}_{\Lambda\Sigma}$ that is valid for any section is
\begin{equation}
 \bar{\mathcal{N}}_{\Lambda\Sigma}=h_{\Lambda|I}  (f^{-1})^{I}{}_{\Sigma}, \textrm{ where } f_{I}^{\Lambda}= \left( \begin{array}{c}\mathcal{D}_{i}X^{\Lambda} \\ \bar{X}^{\Lambda} \end{array} \right) \quad ; \quad  h_{\Lambda|I} = \left(\begin{array}{c} \mathcal{D}_{i}F_{\Lambda} \\ \bar{F}_{\Lambda} \end{array} \right) , \label{gaugecoupl2}
\end{equation}
where $\mathcal{D}_{i}$  denotes the K\"{a}hler covariant derivative (for more details on special K\"{a}hler geometry see \cite{N21,Ceresole:1995ca}). The non-holomorphicity of the matrix ${\mathcal{N}}_{\Lambda\Sigma}$ is clear from the expressions for the ``double size sections'' $f_{I}^{\Lambda}$ and $h_{\Lambda|I} $ which depend on holomorphic as well as non-holomorphic functions of the scalars.


\subsection{Special K\"{a}hler geometry of our model}

Using the cubic prepotential  \eqref{prep} as well as  $s=X^{s}/X^{0}$ etc. and $F(X^{\Lambda})=(X^{0})^{2}\mathcal{F}(X^{\Lambda}/X^{0})$ with $\Lambda=0, \ldots , (3+16+N_{D3})$,
one defines the conventional symplectic section (\ref{natsection}).

In order to have a 4D,  ${\cal N}=2$ supergravity description that makes the  duality symmetry of string theory manifest, however, it is necessary, even in absence of fluxes, to change the symplectic basis  to another  one in which the magnetic components $F_{\Lambda}$ are not derivatives of a prepotential. This basis corresponds to the Calabi-Visentini coordinates \cite{Calabi}, and its main feature is that the  new electric components $X^{'\Lambda}$ do not depend on $s$, the $K3$-volume. The dependence on $s$ enters only via the  magnetic components $F'_{\Lambda}$. In what follows, we will use only this new symplectic basis, and we will drop the primes on $X$ and on $F$ from now on. The theory will thus be defined by  the following new symplectic section \cite{Angelantonj:2003zx}:\footnote{The choice of this section is also the natural one for the gaugings due to the fluxes (see below).}
\begin{eqnarray}
X^0 &=& \frac{1}{{\sqrt{2}}}\,(1 - t\,u +
\frac{(y_7^k)^2}{2})\,\,\,\,,\,\,\,\,\, X^1 = -\frac{t +
u}{{\sqrt{2}}}\,,
\nonumber \\
X^2 &=&  -\frac{1}{{\sqrt{2}}}\,({1 + t\,u -
\frac{(y_7^k)^2}{2}})\,\,\,\,,\,\,\,\,\,X^3 = \frac{t -
u}{{\sqrt{2}}}\,,
\nonumber \\
X^k &=& y_7^k\,\,\,\,,\,\,\,\,\,X^r = y_3^r\,,
\nonumber\\
F_0 &=& \frac{s\,\left( 2 - 2\,t\,u + (y_7^k)^2 \right) +
u\,(y_3^r)^2}{2\,{\sqrt{2}}}\,\,\,\,,\,\,\,\,\,
F_1 = \frac{-2\,s\,\left( t + u \right)  +
  (y_3^r)^2}{2\,{\sqrt{2}}}\ , \nonumber\\
F_2&=&
  \frac{s\,\left( 2 + 2\,t\,u - (y_7^k)^2 \right)  -
  u\, (y_3^r)^2}{2\,{\sqrt{2}}}\,\,\,\,,\,\,\,\,\,
 F_3 = \frac{2\,s\,\left( -t + u \right)  +
 (y_3^r)^2}{{2\,\sqrt{2}}}\ , \nonumber\\
F_i &=& -
  s\,y_7^k
 \,\,\,\,,\,\,\,\,\,
F_r = -u\,y_3^r\, . \label{newbasis}
\end{eqnarray}
The corresponding gauge kinetic matrix, $\mathcal{N}_{\Lambda\Sigma}$, in this basis is a very complicated expression that covers four pages (see Appendix B in ref.\ \cite{D'Auria:2004qv}).

As the K\"{a}hler potential does not depend on the choice of the symplectic basis, it can be computed using formula (\ref{K}) and the original section (\ref{natsection}) derived from the prepotential (\ref{prep}).
The result is
\begin{equation}
K = -\ln \Big[-8\,({\rm Im}(s)\,{\rm Im}(t){\rm
Im}(u)-\frac{1}{2}\,{\rm Im}(s)\,({\rm
Im}(y_7^k))^2-\frac{1}{2}\,{\rm Im}(u)\,({\rm
Im}(y_3^r))^2)\Big] \ . \label{KP}
\end{equation}
In the following we will be interested in partial supersymmetry breaking to ${\cal N}=1$ by three-form fluxes.
As we will discuss, this typically fixes $u$ and $t$, but leaves the $K3$-volume $s$ as a light field.
The astute reader might notice that the form of the K\"ahler potential given in \eqref{KP}, 
even without D-branes, would not lead to a no-scale potential in the ${\cal N}=1$ effective 
theory of the flux compactification. 
The reason is that the full K\"ahler potential contains an additional part from the fields
which used to be in hypermultiplets before the supersymmetry breaking. They parametrize a 
K\"ahler-Hodge manifold, whose K\"ahler potential would render the full flux-induced potential 
of the no-scale type, cf.\ section 5.2 of \cite{ADFL}.


\subsection{Gauged supergravity and partial SUSY breaking}

Thus far, we have only discussed the K\"{a}hler potential and the
gauge kinetic matrix in the effective $\mathcal{N}=2$ supergravity
theory that describes type IIB string theory on
$K3\times T^{2}/\mathbb{Z}_{2}$ with D3- and D7-branes.
Without background fluxes, this theory would have no gauge
interactions (apart from the
gauge interactions on (stacks of) D-branes), and,
as a consequence of extended supersymmetry, no scalar potential for
the moduli.

We therefore now reconsider the above compactification in the presence of the NSNS- and RR-three-form fluxes. They lead to non-trivial 4D gauge interactions
and hence a scalar potential that can stabilize some of the moduli and may lead to spontaneous partial supersymmetry breaking. Due to the orientifold projection, the only non-trivial three-form flux components have two legs along the $K3$  and one leg along the torus. As such, they can induce gauged shift symmetries  of the form
 \begin{equation}
 \mathcal{D}_{\mu}C^{I}=(\partial_{\mu}C^{I} - q_{\Lambda}^{I}A_{\mu}^{\Lambda}) \label{der2}
\end{equation}
 for the 22 axions $C^{I}$
 in the hypermultiplet sector, as is easily  seen by, e.g.,  reducing the 10D kinetic term
 \begin{equation}
 \tilde{F}_{5}\wedge \ast \tilde{F}_{5},
\end{equation}
with
\begin{equation}
\tilde{F}_{5}=dC_{4}-\frac{1}{2}C_{2} \wedge H_{3} +\frac{1}{2}B_{2}\wedge F_{3} \label{F5}
\end{equation}
 to four dimensions.
 Because of extended supersymmetry, the above gauging entails  a nontrivial scalar potential, as mentioned earlier.
In the language of gauged supergravity, the constants $q_{\Lambda}^{I}$ in the above covariant derivatives (\ref{der2}) are the components of constant Killing vectors on the scalar manifold.\footnote{The $q_{\Lambda}^{I}$ might be called ``gauge couplings'', as they parametrize the minimal coupling of the vectors to the scalars. On the other hand, the entries of the kinetic
 matrix $\mathcal{N}_{\Lambda\Sigma}$ are sometimes also referred to as ``gauge couplings'', as they can be understood as moduli dependent generalizations of
 the standard expressions $g^{-2}F_{\mu\nu}F^{\mu\nu}$. To avoid confusion, we will refer to the $q_{\Lambda}^{I}$ as ``charges'' and reserve the term ``gauge couplings'' for the moduli dependent entries of the kinetic matrix of the vector fields.} They parametrize the three-form fluxes as we now describe.

The three-form fluxes  can be expanded as follows
\begin{eqnarray}
F_3&=&\alpha_1 \wedge dx^1 + \alpha_2\wedge dx^2\ , \nonumber\\
H_3&=&\beta_{1}\wedge dx^1 + \beta_{2}\wedge dx^2\ ,
\end{eqnarray}
where $\alpha_{1,2},  \beta_{1,2} \in H^{2}(K3,\mathbb{Z})$, and $x^1$ and $x^2$
are real coordinates on the two-torus with complex structure $t$: $x=x^1+tx^2$. The second
cohomology group $H^{2}(K3,\mathbb{Z})$ is isomorphic to a lattice $\Gamma^{3,19}$  with the
inner product of two harmonic two-forms given by $\alpha\cdot \beta:=\int_{K3}\alpha\wedge\beta$.
Using an orthogonal basis $\eta_{I}=(\eta_{m},\eta_{a})$ of $H^{2}(K3,\mathbb{Z})$ with three
positive norm forms, $\eta_{m}$,  and 19 negative norm forms, $\eta_{a}$, the gauge charges
$q_{\Lambda}^{I}$ introduced in (\ref{der2}) can be read off from the expansions \cite{LMRS}:
\begin{eqnarray}
q_{0}^{I}\eta_{I}&=& \frac{1}{\sqrt{2}}(\beta_{1}-\alpha_{2})\ ,\nonumber\\
q_{1}^{I}\eta_{I}&=& \frac{1}{\sqrt{2}}(-\beta_{2}-\alpha_{1})\ ,\nonumber\\
q_{2}^{I}\eta_{I}&=& \frac{1}{\sqrt{2}}(\beta_{1}+\alpha_{2})\ , \nonumber\\
q_{3}^{I}\eta_{I}&=& \frac{1}{\sqrt{2}}(-\beta_{2}+\alpha_{1})\ ,
\end{eqnarray}
with $q_{\Lambda}^{I}=0$ for $\Lambda>3$.
These coefficients $q_{\Lambda}^{I}$ refer to the new symplectic basis
(\ref{newbasis}), and they show that each of the four bulk vector fields
$A_{\mu}^{0}$, $A_{\mu}^{1}$, $A_{\mu}^{2}$, $A_{\mu}^{3}$ is a non-trivial linear
combination of NSNS- and RR-fields.

The bulk fluxes do not change the form of the cubic prepotential
or the corresponding
symplectic section. The K\"{a}hler potential and the gauge kinetic matrix
will therefore not be directly affected by the bulk fluxes either (at least
at the leading order we are discussing in this section),
and are thus still given by the moduli-dependent (but flux-independent)
expressions mentioned earlier.\footnote{To be more precise, ``on-shell''
there is always  an indirect influence of the fluxes on the gauge kinetic
matrix, because the fluxes dynamically fix some of the moduli  at  values
that depend themselves on the particular values of the bulk fluxes.
Inserting the vevs of these moduli back  into the gauge kinetic matrix
introduces then an implicit dependence on the flux parameters. In addition,
tadpole cancellation might require also a different number of branes
in the presence and absence of fluxes, making the number of open string
moduli  and hence the dimension of the moduli space flux-dependent.
As was shown in \cite{LMRS}, the ``backreaction'' of fluxes on the 4D
effective theory is even a bit more severe for  worldvolume fluxes on
the D7-branes, as these can change already the leading order prepotential,
and hence the ``off-shell'' theory.
We will ignore these effects of worldvolume fluxes in this
paper. \label{fluxprepot}}

The scalar potential that is induced by the gauging (\ref{der2}),
is quadratic in the $q_{\Lambda}^{I}$ and depends on the scalar fields,
$\Phi$, of 4D supergravity,
\begin{equation}
V= q_{\Lambda}^{I}\, V{}_{IJ}^{\Lambda \Sigma}(\Phi) \,
q_{\Sigma}^{J}\ \label{Vequation}
\end{equation}
with some field-dependent matrix
$V{}_{IJ}^{\Lambda \Sigma}(\Phi)$.\footnote{$\Phi$ is meant to
include both the vector multiplet scalars as well as the
hypermultiplet scalars here.}

Depending on the choice of the flux parameters $q_{\Lambda}^{I}$,
the vacua of this potential
can preserve $\mathcal{N}=2,1,0$ supersymmetry. For bulk fluxes,
the supersymmetry conditions were first derived from the 10D
perspective in \cite{Tripathy:2002qw}, where it was found that an $\mathcal{N}=1$
vacuum requires exactly two  flux parameters
with positive norm (positive norm here refers to the $(3,19)$-signature metric
on the lattice $\Gamma^{(3,19)}$ of possible flux parameters, $q_{\Lambda}^{I}$, as
described by the index $I=(m,a)$) and none with lightlike norm.
$\mathcal{N}=2$ vacua, by contrast, allow for at most two different
flux vectors $q_{\Lambda}^{I}$, and they both have to be of negative norm. In
\cite{ADFL,Angelantonj:2003zx,D'Auria:2004qv,D'Auria:2004td},
these conditions were recovered in a 4D,
$\mathcal{N}=2$ supergravity approach. A further extension that also
includes the effects of worldvolume fluxes was given in \cite{LMRS}
in the language of F-theory.

In the following, we are interested in vacua with a partial
breakdown of supersymmetry, $\mathcal{N}=2 \rightarrow \mathcal{N}=1$.
Such partial supersymmetry breaking requires the use of a symplectic
section with $F_{\Lambda} \neq \partial_{\Lambda}F$ \cite{CGP}, which
is in line with our use of the symplectic section (\ref{newbasis}).
As already mentioned, partial supersymmetry breaking gives mass to
some of the fields, including one of the two gravitini, and below
this mass scale we expect an effective $\mathcal{N}=1$  description.
This effective $\mathcal{N}=1$ theory should now have a \emph{holomorphic}
gauge kinetic matrix, even though the corresponding $\mathcal{N}=2$
expression it descends from is highly
\emph{non}-holomorphic.\footnote{This restoration of
holomorphicity should be a general
feature of all $\mathcal{N}=2$ supergravity theories that are
spontaneously broken to $\mathcal{N}=1$ (for various aspects of
partial supersymmetry breaking see, e.g., \cite{ADF}).} In the
following subsection and in Appendix \ref{Apphol}, we
verify this restoration of holomorphicity for
 some representative cases that are relevant for our later discussions and discuss the manifestation of the shift symmetry in the $\mathcal{N}=1$ theory.

As mentioned earlier, this holomorphicity problem is
something different from  the ``rho-problem'' mentioned in refs.
\cite{Berg:2004ek,Giddings:2005ff,Baumann:2006th}. The ``rho-problem''
has to do with the proper relation
between 10D and 4D variables in the dimensional reduction process,
whereas the holomorphicity problem raised here is a purely 4D issue.


\subsection{Emergence of holomorphicity after partial SUSY breaking}
\label{secholomorphic3}
We follow refs. \cite{D'Auria:2004qv,D'Auria:2004td} and for simplicity consider in detail only the case when
each vector field $A_{\mu}^{0,1,2,3}$ gauges at most one of the 22 axionic directions
 $C^{I}$. Furthermore, we will, for the sake of simplicity, only consider the
 following set of potentially non-vanishing gauge charges:
 \begin{eqnarray}
 g_{0}&:=&q_{0}^{m=1}\ , \nonumber\\
   g_{1}&:=&q_{1}^{m=2}\ , \nonumber\\
   g_{2}&:=&q_{2}^{a=1}\ , \nonumber\\
   g_{3}&:=&q_{3}^{a=2}\ ,
 \end{eqnarray}
i.e., $A_{\mu}^{0}$ and $A_{\mu}^{1}$ may gauge spacelike $C^{I}$-directions,
whereas timelike $C^{I}$-directions may be gauged by $A_{\mu}^{2}$ and $A_{\mu}^{3}$.
As we are only considering bulk fluxes at the moment,
all $q_{\Lambda}^{I}$ for $\Lambda>3$ will be zero. For the sake of simplicity, we only
consider one D7- and one D3-brane coordinate, which we call $y_7$ and $y_3$, respectively;
the  other brane coordinates enter completely analogously.
The vector fields corresponding to $y_7$ and $y_{3}$ will be denoted by
$A_{\mu}^{4}$ and $A_{\mu}^{5}$, respectively.
 We  consider the following three special cases with regard to the holomorphicity
of the gauge couplings in more detail:\\
Case 1: $g_0,g_1\neq 0\qquad\qquad\quad $ $(\mathcal{N}=2  \rightarrow  \mathcal{N}=1)$\ , \\
Case 2: $g_0,g_1,g_2,g_3 \neq 0\qquad$  $\; (\mathcal{N}=2  \rightarrow  \mathcal{N}=1)$\ , \\
Case 3: $g_2,g_3\neq 0\qquad\qquad\quad$ $(\mathcal{N}=2  \rightarrow  \mathcal{N}=2)$\ , \\
where the charges that are not listed are always assumed to be zero. As case 1 is the one of most direct interest to us, we will content ourselves with discussing that case in the main text and refer to Appendix \ref{Apphol} for cases 2 and 3, where we will also discuss some more general gaugings.

\subsubsection{Case 1: $\mathcal{N}=2  \rightarrow  \mathcal{N}=1$ for
$g_0,g_1\neq 0$}
In this case, $A_{\mu}^{0}$ and $A_{\mu}^{1}$ gauge the spacelike directions
$C^{m=1}$ and $C^{m=2}$, respectively, whereas $A_{\mu}^{2,3,4,5}$ do not
participate in the gauging. According to the classification mentioned below
eq.\ (\ref{Vequation}),  one thus expects an $\mathcal{N}=1$ supersymmetric
vacuum in which some of the moduli $(s,t,u,y_7,y_3)$ are fixed.
To derive the vevs of these stabilized moduli, one sets half of the fermionic
supersymmetry variations to zero. To have an unbroken $\mathcal{N}=1$
supersymmetry in Minkowski vacua, one has to require that $g_{0}-g_{1}=0$
\cite{D'Auria:2004qv}. This corresponds to a constant Killing spinor in
spacetime.
One can also have AdS vacua with broken supersymmetry in this model and
in such a case the value of
$g_{0}\neq g_{1}$ is not restricted and defines the curvature of the
AdS space \cite{D'Auria:2004td}. In the $\mathcal{N}=1$ supersymmetric  case,
for the particular quaternionic shift symmetries we have chosen to gauge,
vanishing of half of the fermionic variations results in the following
conditions on the moduli \cite{D'Auria:2004qv}:
\begin{eqnarray}
u=t&=&-i \ , \nonumber\\
y_7&=&0  \ , \label{conditions}
\end{eqnarray}
i.e., the complex structure modulus, $t$, the axion-dilaton, $u$, and the D7-brane position,
$y_7$,  are frozen by the fluxes (smaller values for the dilaton $u$ are also
possible, cf.\ appendix \ref{Apphol3}). We are thus left with the $K3$-volume modulus, $s$, and
the position modulus, $y_{3}$, of the D3-brane.

The $\mathcal{N}=1$ K\"{a}hler potential after this partial supersymmetry breaking takes the form
\begin{equation} \label{Kfix}
K= -\ln \left( 4i (s-\bar s) + (y_3-\bar y_3)^2 \right)\ .
\end{equation}
It  simply inherits the shift symmetry in the real part of $y_{3}$ of the corresponding $\mathcal{N}=2$ K\"{a}hler potential.

In order to discuss the manifestation of the shift symmetry $y_3\rightarrow y_3 +\alpha$ $(\alpha \in \mathbb{R})$ also in the $\mathcal{N}=1$ D7-brane gauge coupling, we have to recover the proper \emph{holomorphic} $\mathcal{N}=1$ gauge couplings
first. Inserting (\ref{conditions}) into the 4-pages-long non-holomorphic gauge coupling matrix $\mathcal{N}_{\Lambda\Sigma}$ of  Appendix B of \cite{D'Auria:2004qv}, one notices that these expressions simplify drastically, and that  some of the $\mathcal{N}_{\Lambda\Sigma}$ are indeed holomorphic functions of the surviving scalars $s$ and $y_3$.
However, some other components are  also antiholomorphic and some are neither. The resolution is that not all of these couplings are present in the $\mathcal{N}=1$ theory, as some of the vector fields are not.
Indeed, $A_{\mu}^{0}$ and $A_{\mu}^{1}$ are massive due to their Stueckelberg coupling to the axionic scalars  and have to be integrated out.
The only remaining matrix entries  then involve the indices $\Lambda= 2,3, 4,5$. The matrix is in general non-diagonal: the two bulk vector field directions $\Lambda= 2,3$ are heavily mixed with themselves and  with the D3 vector field direction $\Lambda= 5$. The only direction which is completely disentangled from the rest is the vector field on the D7-brane, which corresponds to the $\Lambda=  4$ direction. The submatrix $\mathcal{N}_{\Lambda\Sigma}$ with $\Lambda,\Sigma\geq 2$ reads
\begin{equation}
\mathcal{N}_{\Lambda\Sigma}=  \left( \begin{array}{cccc}
-\bar{s} -\frac{i}{2}{\bar{y}_3}^{2} & \frac{1}{4}{\bar{y}_3}^{2} & 0 & -\frac{i}{\sqrt{2}}\bar{y}_3\\
\frac{1}{4}{\bar{y}_3}^{2}& -\bar{s} &0 &\frac{1}{\sqrt{2}}\bar{y}_3\\
0 & 0 &-\bar{s}&0\\
-\frac{i}{\sqrt{2}}\bar{y}_3 & \frac{1}{\sqrt{2}}\bar{y}_3 & 0 &-i
\end{array} \right), \label{coupling1}
\end{equation}
which is purely \emph{anti}holomorphic. Thus, if one defines the
$\mathcal{N}=1$ gauge kinetic function to be proportional to the
complex conjugate of the surviving $\mathcal{N}_{\Lambda\Sigma}$
components, one obtains purely holomorphic $\mathcal{N}=1$
gauge couplings, as desired.\footnote{It should be possible
to understand the structure of the off-diagonal terms in \eqref{coupling1}
from a higher dimensional point of view along the lines of
\cite{Berg:2003ri}.}

Let us now take a look at the shift symmetry for the D3-brane coordinate $y_3$,
\begin{equation}
y_3 \rightarrow y_3 + \alpha \ ,\qquad \alpha=\bar \alpha\ .
\label{shiftsymmetry}
\end{equation}
Obviously, there is a $y_3$-dependent mixing between the bulk vector
fields $A_\mu^2, A_\mu^3$ and the vector field $A_\mu^5$ of the D3-brane
as well as a $y_3$-dependent mixing among the bulk vector fields themselves.
On the other hand, there is no such mixing for the D7-brane gauge coupling,
and it is just given by the diagonal entry  $\mathcal{N}_{44}=-\bar{s}$,
which is $y_3$-independent. Thus, the shift symmetry (\ref{shiftsymmetry})
along the real part of the D3-brane coordinate is preserved for the
D7-brane gauge coupling also after the partial supersymmetry breaking
from $\mathcal{N}=2$ to $\mathcal{N}=1$ (and it is this gauge coupling
which appears in the non-perturbative superpotential via gaugino
condensation).
We next analyze the fate of this shift symmetry, which plays an
important role for the flatness of the inflaton potential, after
the inclusion of quantum corrections.


\section{Quantum corrections}
\label{qc}

In the previous section, we saw that, within the
framework of 4D, $\mathcal{N}=2$ gauged supergravity, holomorphicity is properly established when the theory is spontaneously broken to $\mathcal{N}=1$ supersymmetry. This was to be expected and confirms the internal consistency of 4D supergravity.

In this section, we infer the proper definition of the
variable ${\rm Im}(s)$ in terms of 10D quantities by performing a dimensional reduction of
the D7-brane DBI action (leading to the D7-brane gauge kinetic term) and the
D3-brane DBI action (leading to the D3-brane scalar kinetic
term).\footnote{The technical details of the calculation are
deferred to appendix \ref{10to4}.}
The place where an incomplete
analysis would have led to the analogue of the ``rho-problem'', can be
easily identified. Our analysis follows closely
the general discussion of \cite{Giddings:2005ff, Baumann:2006th, Baumann:2007ah},
which we adapt to the case at hand, but at some points we
can be a bit more concrete, as the torus metric is explicitly known (see also the 
discussion in section 4.1 of \cite{Burgess:2006cb}).
As in \cite{Berg:2004ek, Giddings:2005ff, Baumann:2006th}, it will become
clear that the ``rho-problem'' is really an artificial problem,
due to an incomplete consideration of $g_s$-corrections in the low
energy effective action.

Identifying the proper definition of the variable ${\rm Im}(s)$
also sheds some light on the issue of the shift symmetry
that we discussed in the last section. As the cubic prepotential
\eqref{prep} leads to a shift-symmetric potential in the low energy
effective action, it is important to know which effects (i.e.\ which
$g_s$-corrections) it already takes into account and which
are not contained. This is a prerequisite for an analysis about
how $g_s$-corrections might break the shift-symmetry. Related
discussions can be found in
\cite{Berg:2004ek, McAllister:2005mq, Giddings:2005ff, Baumann:2006th}.

In order to compactify the D3- and D7-brane DBI actions,
we split the 10D coordinates into
external coordinates $x^{\mu}$ $(\mu=0,1,2,3)$ and coordinates
$x^{m}$ $(m=1,2,3,4)$ and $x^{i}$ $(i=1,2)$
that parametrize the $K3$-space and the torus, respectively.
The 10D metric can then be written as
\begin{eqnarray}
 ds^2 &=& G_{\mu\nu}^{4D} dx^\mu dx^\nu + G_{mn}^{K3} dx^m dx^n
+ G_{ij}^{T^2} dx^i dx^j\nonumber \\
 &=& h^{-1/2} g_{\mu\nu} dx^\mu dx^\nu + h^{1/2} \big[ g_{mn} dx^m dx^n
+ g_{ij} dx^i dx^j  \big]
\nonumber \\
 &=&h^{-1/2} e^{-4U_1-2U_2} \tilde{g}_{\mu\nu} dx^\mu dx^\nu + h^{1/2}
\big[ e^{2U_1}\tilde{g}_{mn}
 dx^m dx^n + e^{2U_2} \tilde{g}_{ij} dx^i dx^j  \big]\ . \label{metric1}
\end{eqnarray}
In this expression, $h(x^m,x^i)$ is a warp factor, and $e^{2U_1}$ and
$e^{2U_2}$ are the breathing modes of
the two internal spaces, i.e., $\tilde{g}_{mn}$ and $\tilde{g}_{ij}$
denote metrics that correspond to a
fixed reference volume, $\tilde{V}^{K3}$ and $\tilde{V}^{T^2}/2$ (the
factor of $1/2$ comes from the orientifolding; locally the
metric on $T^2/\mathbb{Z}_2$ is the same as on $T^2$, though, so
that we did not distinguish them in \eqref{metric1}).
The Weyl-rescaling
$g_{\mu\nu}\rightarrow e^{-4U_1-2U_2} \tilde{g}_{\mu\nu}$ decouples
the two breathing modes from the 4D Einstein-Hilbert term.
This is analogous to the case of a
Calabi-Yau discussed in \cite{Baumann:2007ah}, with the
only difference that there are two independent breathing modes
in the case at hand.

Following \cite{Baumann:2006th}, we make the split
\begin{equation}
h=h_0+\delta h\ , \label{hexp}
\end{equation}
where $h_{0}$ is the constant zero mode of $h$ and
$\delta h$ is a function of the internal coordinates $(x^m,x^i)$
which integrates to zero over the whole internal space.
It also depends on the positions of the branes.

The DBI-action of a spacetime filling D7-brane that wraps
$K3$ contributes the term (we use the usual convention
${\rm Tr} (T^a T^b) = \tfrac{1}{2} \delta^{a b}$
for generators $T^a$ of $SU(N)$)
\begin{equation} \label{DBI7}
  S=-\frac{1}{8} T_7 (2\pi \alpha^{\prime})^2 \int_{K3} d^4 x
  \sqrt{G^{K3}} \int_{\mathbb{R}^{3,1}} d^4 x
  \sqrt{G^{4D}} F_{\mu\nu}^a F_{\rho\sigma}^a
G^{4D, \mu\rho} G^{4D, \nu\sigma},
\end{equation}
with the tension of a $Dp$-brane
 \begin{equation}
 T_{p}=\frac{1}{g_{s}(2\pi)^p(\alpha^{\prime})^{(p+1)/2}}\ .
 \end{equation}
The 4D part of this action is conformally invariant, and one can easily read
off the effective 4D gauge coupling $g$ as
 \begin{eqnarray}
 g^{-2} &=& \frac12 T_{7}(2\pi \alpha^{\prime})^2 \int_{K3}d^4 x \sqrt{G^{K3}}
= \frac12 T_{7}(2\pi \alpha^{\prime})^2
\int_{K3}d^4 x\, \sqrt{\tilde{g}^{K3}} e^{4U_{1}}h \\
 &=& \frac12 \frac{T_{3}}{(2\pi)^2} \int_{K3}d^4 x\, \sqrt{\tilde{g}^{K3}} e^{4U_{1}}h\ , \label{couplingg}
 \end{eqnarray}
where we have expressed $T_7=T_3(2\pi)^{-4} (\alpha^{\prime})^{-2}$
in terms of the D3-brane tension in the last line.

Thus, in order to determine the 4D gauge coupling, we need to know the
warp factor \eqref{hexp}, or rather its integral over $K3$.
The Einstein equation implies a Poisson equation for
$\delta h$ (which is why this method
is called the Green's function method),
whose integral over $K3$ we solve in appendix \ref{d7gc}. In this
way one can determine the dependence of the D7-brane gauge
coupling on the D3-brane scalars \cite{Giddings:2005ff,Baumann:2006th,Burgess:2006cb}.
For simplicity we just consider a single D3-brane coordinate and
D7-branes at the origin. In this case, the resulting
gauge coupling turns out to be
\begin{equation} \label{gaugecoupling}
g^{-2}= \frac12 \left[ \frac{T_{3}h_{0} \tilde{V}^{K3}}{(2\pi)^2} \right]
e^{4U_{1}} - \frac12
\frac{[\textrm{Im}(y_{3})]^{2} }{\textrm{Im}(t)}
- \frac{1}{(2\pi)^2} \ln |\vartheta_1(\sqrt{2 \pi} y_3,t)| + \ldots\ ,
\end{equation}
where the ellipsis stands for terms depending on the
complex structure $t$ but not depending on $y_3$, and the theta function $\vartheta_{1}$ is defined in Appendix \ref{thetaf}. The omitted terms are
the real part of a holomorphic function in $t$
but can not be determined by the Green's function method. Using
CFT methods they can be determined in the case without flux and
are proportional to $\ln |\eta(t)|$, cf.\
\cite{Antoniadis:1999ge,Berg:2004ek}.

Comparing the kinetic term of $y_3$ arising from the K\"ahler potential
\eqref{KP} with the one obtained from the D3-brane DBI action shows that
${\rm Im}(s)$ only contains the first two terms
of \eqref{gaugecoupling} but not the third. This we show explicitly in appendix \ref{seckaehler}.
Thus, the appropriate definition of ${\rm Im}(s)$ in terms of
10D quantities is
\begin{equation}
-\textrm{Im}(s) := \left[ \frac{T_{3}h_{0} \tilde{V}^{K3}}{2 (2\pi)^2} \right]
e^{4U_{1}}
- \frac12 \frac{[\textrm{Im}(y_{3})]^{2} }{\textrm{Im}(t)}\ . \label{sdef_pi}
\end{equation}
With this definition the gauge coupling \eqref{gaugecoupling} is
automatically the real part of a holomorphic function, i.e.\
\begin{equation} \label{gs}
g^{-2}= \textrm{Re}(i s) - \frac{1}{(2\pi)^2} \textrm{Re}\, \zeta (y_{3},t)\ , \quad
\zeta (y_{3},t) = \ln \vartheta_{1}(\sqrt{2 \pi} y_{3},t) + \ldots \ .
\end{equation}
The error one would have to make in order to ``create'' a
``rho-problem'', would be to omit the second term in (\ref{sdef_pi}) in
the definition of $\textrm{Im}(s)$, and instead define $\textrm{Im}(s)$ as the
breathing mode only:
\begin{equation}\label{bm}
-\textrm{Im}(s) :=  \left[ \frac{T_{3}h_{0} \tilde{V}^{K3}}{2 (2\pi)^2}
\right] e^{4U_{1}} \textrm{ (} \rightarrow \textrm{ would imply rho-problem)}\ .
\end{equation}
With such a definition, the  second term in  (\ref{gaugecoupling}) would make
$g^{-2}$ a function of the moduli $(s,y_3,t)$ that is manifestly not
the imaginary part of a holomorphic function.

While we have avoided any holomorphicity problems in \eqref{gs},
the result \eqref{gs} does not seem
to agree with the result of our 4D supergravity
approach of section \ref{secholomorphic}, where $g^{-2}$ was just $-\textrm{Im}(s)$ without
the extra  $\zeta$-term.
Instead, the correct interpretation
is that in order to reproduce the $\zeta$-term in $g^{-2}$, one has to
add another holomorphic contribution, $\delta \mathcal{F}$, to the
prepotential $\mathcal{F}$ of eq.\ (\ref{prep}). As the last two terms
in (\ref{gaugecoupling}) have a common origin in $\delta h$
they arise at the same order in $g_s$ and thus the cubic prepotential
does not represent an expansion in $g_s$. However, in a sense
(\ref{prep}) represents an expansion for large $u$ and $s$
(as $\zeta$ does not depend on these fields). However, this expansion
only holds in the prepotential. The low energy effective action
also contains derivatives of the prepotential and, thus, it is not
obvious under what circumstances a truncation to the cubic prepotential
actually leads to a consistent expansion of the effective action
for large $u$ and $s$ (cf.\ also
the discussion in section 3.3.2 of \cite{Berg:2004ek}).

We finally note that $-\textrm{Im}(s)$ is not the volume of $K3$
if branes are present. Instead, it is the full combination
$g^{-2}=-\textrm{Im}(s) - \frac{1}{(2\pi)^2} \textrm{Re}\, \zeta (y_{3},t)$
that should be identified with
the physical (warped) volume of $K3$.


\subsection{Stringy threshold corrections to D7-brane gauge coupling}
\label{stringy}

In the last section the dependence of the D7-brane gauge
coupling on the D3-brane scalars was determined using the Green's
functions method \cite{Giddings:2005ff, Baumann:2006th, Baumann:2007ah}.
However, it can alternatively be obtained as a 1-loop open string
threshold correction
\cite{Bachas:1996zt,Antoniadis:1999ge,Berg:2004ek,Akerblom:2007np}.
Even though it is not clear yet
how these world-sheet computations have to be adapted in the
presence of RR-fluxes, the D3-brane dependent part
of the threshold corrections
nicely reproduces the results of the supergravity calculations of
appendix \ref{10to4} (which are valid in the presence of
fluxes) and offer a useful alternative
view on these corrections. Conversely, the agreement between
the two results suggests that at least the D3-brane dependent part
of the world-sheet calculation is not modified considerably by the presence
of the fluxes.\footnote{Of course, as already mentioned in
footnote \ref{fluxprepot}, the fluxes modify the
global tadpole conditions and, thus, the number of D3-branes is
modified in general. This has an indirect influence on the result, because
the final formula for the D7-brane gauge coupling would contain
a sum over D3-branes in \eqref{sdef_pi} and \eqref{gs}.}

Let us discuss this dual approach in a bit more detail.
At tree level plus one-loop, the D7-brane gauge coupling has the form
\begin{equation}
\frac{1}{g^2}=\frac{1}{g_{\rm tree}^2}+\Delta(M,\bar M)\, ,
\end{equation}
where the threshold function $\Delta(M,\bar M)$ is a moduli
(open and closed string moduli) dependent function.
In general it has a
non-holomorphic term plus a holomorphic contribution:
\begin{equation}
\Delta(M,\bar  M)=\Delta_{\rm non-hol}(M,\bar M)
+{\rm Im}(\Delta_{\rm hol}(M))\, .
\end{equation}
$\Delta_{\rm non-hol}(M,\bar M)$ comes from the integration over
massless fields. It is
related to so-called infrared K\"ahler anomalies
(see
\cite{Derendinger:1991kr}
for details and \cite{Louis:1996ya} for a nice review on
gauge couplings in string theory).
$\Delta_{\rm hol}(M)$, on the other hand, is the Wilsonian
part of the threshold corrections
from integrating over the massive modes.

Note that the non-holomorphic term we discussed earlier in this
section in the context of the rho-problem is not part
of $\Delta_{\rm non-hol}(M,\bar M)$. Rather it is absorbed in the definition
of $s$ (cf.\ eq.\ \eqref{sdef_pi}), and in this way
it becomes part of the holomorphic
gauge kinetic function (holomorphic in the correctly defined variable $s$).
In order to calculate the
non-holomorphic pieces $\Delta_{\rm non-hol}(M,\bar M)$ one has to resort to a
direct open string calculation.

As already mentioned, $\Delta_{\rm hol}(M)$
arises from massive charged states running in the loop.
In the D3-D7 model under investigation, the massive
states leading to a dependence of the D7-brane gauge coupling
on the D3-brane scalars correspond to open strings
stretched between the D7- and the D3-branes. Suppose the D7-branes
wrap the K3, which
is in the $z_2,z_3$ directions of the internal space
with altogether three complex coordinates $z_i$.
Moreover, suppose that
the D7-branes are located at the point  $z_1=y_7$ inside the $T^2$.
On the other hand, the D3-branes are located at the
point $y_3$ inside $T^2$, and their  distance $d$ from
the D7-branes is hence given by
\begin{equation}
d=|y_3-y_7|\, .
\end{equation}
Furthermore, the mass of the lowest open string
states between the D3- and D7-branes is proportional to $d$:
\begin{equation}
\alpha' m_{D3-D7}^2 \sim |y_3-y_7|^2\, .\label{mass}
\end{equation}
This result can be used to compute the one-loop threshold
corrections in field theory
\cite{Kaplunovsky:1987rp},
where we first take
into account only one open string multiplet
(which is a hypermultiplet\footnote{When we say one hypermultiplet here, we actually mean a set of minimal hypermultiplets that form  an irreducible representation of the gauge group.} for the D3-D7 strings)
with mass given in eq.\ (\ref{mass}):
\begin{equation}\label{ftthres}
{\rm Im}(\Delta_{\rm hol}(M))=-\frac{b}{ 16 \pi^2}\ln (\alpha' m^2_{D3-D7})
=-\frac{b}{8 \pi^2}\ln |y_3-y_7|+ \ldots \, .
\end{equation}
Here $b = 2 T(r)$ for a hypermultiplet in the representation $r$
of the D7-brane gauge group, where $T(r)$ is the index of the corresponding
representation (a massive ${\cal N}=2$ vector multiplet
charged under the D7-brane gauge group would have contributed
$b=- 2 T({\rm adj})$).\footnote{As a reminder, some relevant indices
of $SU(N)$ representations are: $T(\Box) = 1/2,\, T({\rm adj}) = N,\,
T(\Box \hspace{-.5ex} \Box) = (N+2)/2,\,
T(\raisebox{2pt}{$\Box$} \raisebox{-3pt}{\hspace{-2.51ex} $\Box$})
= (N-2)/2$, cf.\ for instance (2.45) in \cite{Terning:2003th}. \label{indices}}

Now we want to include all massive open string
states of the tower whose lightest
members have the mass \eqref{mass}. They arise as winding states
from open strings beginning on the D7-brane and winding around the torus
before ending on the D3-brane. The string oscillator states instead do not
contribute, as they are non-BPS and
only short BPS multiplets contribute to the gauge couplings
\cite{Douglas:1996du,Bachas:1996zt}.
The inclusion of the massive states
can either be done by an explicit world sheet computation
\cite{Bachas:1996zt,Antoniadis:1999ge,Berg:2004ek,Akerblom:2007np}
or by taking into account the backreaction of the D3-branes on the internal
geometry as described above (and in appendix \ref{10to4}).
The world sheet calculation gives
\begin{equation}\label{sthres}
\Delta_{\rm hol}(M)=-\frac{i b}{8\pi^2} \ln
\vartheta_1(\sqrt{2 \pi} (y_3-y_7),t) + \ldots \, ,
\end{equation}
where the ellipsis stands for terms independent of $(y_{3}-y_{7})$.
This is compatible with \eqref{g3} from the Green's function method
because $b=1$ for a hypermultiplet
in the fundamental representation, which is the relevant case
for a D3-D7 string (between the stack of D7-branes and a single D3-brane).
To see whether these stringy threshold corrections agree
with the field theory result eq.\ (\ref{ftthres}),
we use the expansion \eqref{expandtheta1}
of the Jacobi-function $\vartheta_1(\sqrt{2 \pi} y,t)$
for small values of $y$, i.e.\
for small distances between the D3 and the D7-branes.
Then we obtain, in agreement with (\ref{ftthres}),
\begin{equation}
{\rm Im}(\Delta_{\rm hol}(M))=-\frac{b}{8\pi^2} \ln |y_3-y_7| +\dots \, .
\end{equation}

To obtain the final result for $\Delta_{\rm hol}(M)$, one still
has to sum over the different towers of the massive hypermultiplets,
which in the case at hand (i.e.\ if we are interested in the
D3-brane dependent terms) amounts to summing \eqref{sthres} over the
different D3-branes (including the orientifold images).

One comment is in order here. The factor $b$ appearing in
\eqref{sthres} is not to be confused with the ${\cal N}=2$ beta-function
coefficient of the theory, which is given by
\begin{equation} \label{bn2}
b^{{\cal N}=2} = 2 \Big(\sum_k T(r_k) - T({\rm adj})\Big)\ .
\end{equation}
Here the sum runs over the light (charged)
${\cal N}=2$ hypermultiplets with masses below the threshold
in the representation $r_k$ of the gauge group.
Instead, the $y_3$-dependent terms of the threshold corrections
all come from \emph{massive} D3-D7 strings, whose excitations
are always in hypermultiplets. Thus, they always contribute
with the same sign in the threshold corrections, and there is no way
of getting rid of them by modeling the spectrum appropriately (for instance
by choosing a spectrum with $b^{{\cal N}=2}=0$).
In other words, $y_3$-dependent terms of the form \eqref{sthres}
will always be present if there are massive D3-D7 strings.
Once the D3-brane reaches the D7-branes, the D3-D7 strings
contribute massless hypermultiplets to the spectrum and the corresponding
$b$ would contribute to $b^{{\cal N}=2}$ in \eqref{bn2}.
However, then also the corresponding contribution
to the threshold correction, i.e.\ \eqref{sthres},
disappears.


\subsection{Gaugino condensate superpotential}
\label{gcsupo}

In this section we would like
to discuss non-perturbative superpotentials from gaugino condensation, which
requires to break supersymmetry (spontaneously) to ${\cal N}=1$. Thus,
in order to comply with our notation from eq.\ \eqref{n1notation} and
\eqref{n2notation}, we define the ${\cal N}=1$ gauge kinetic function $f_{D7}$
as
\begin{equation} \label{fd7}
f_{D7}= i s - \frac{1}{8 \pi^2} \zeta(y_3-\mu,t)
- \frac{1}{8 \pi^2} \zeta(y_3+\mu,t) + \ldots \ ,
\end{equation}
so that the gauge coupling is given by the real part of $f_{D7}$.
In \eqref{fd7}, the second term with the function $\zeta$ comes from the symmetrization with respect to $\mathbb{Z}_{2}$ (cf. Appendix \ref{10to4}), we assumed the D7-branes to be fixed at $y_7=\mu$ and
we concentrate on a single D3-brane again (the dependence on the other
branes is in the ellipsis).

In order to ensure the appearance of a non-perturbative superpotential
one has to require that the quantity
\begin{equation} \label{c}
c = \sum_j T(r_j) - T({\rm adj})
\end{equation}
be negative. In \eqref{c}, the sum runs over the light (charged)
${\cal N}=1$ chiral multiplets in the representation $r_j$ of the gauge group.
In particular, no
adjoint matter is allowed in the light spectrum of the ${\cal N}=1$
gauge theory. We assume that the charged matter content of the D7-brane
gauge theory is such that it fulfills $c<0$,
for example by giving mass to unwanted matter via fluxes
\cite{Gorlich:2004qm,Cascales:2004qp,LMRS,Jockers:2005zy}.\footnote{The
two antisymmetric tensors which are present in the ${\cal N}=2$ theory
\cite{GP} are barely compatible with $c<0$, cf.\ footnote \ref{indices}.}

Then the non-perturbative superpotential due to gaugino condensation,
which stabilizes the volume of the K3 manifold, acquires the form
\begin{equation}
W=A_0 \exp\Big( {\frac{8 \pi^2 f}{c}}\Big) = \A e^{\frac{8 \pi^2}{c}
(i s - \frac{1}{8\pi^2} \zeta (y_{3}-\mu,t)- \frac{1}{ 8\pi^2} \zeta (y_{3}+\mu,t))}\ , \label{W}
\end{equation}
where $A_0$ might depend on any light charged matter fields
and $A$ incorporates in addition an overall factor independent
of $y_3$ coming from the ellipsis in \eqref{fd7}. Using the explicit
form of the string threshold corrections
eq.\ (\ref{gs}) we derive
\begin{eqnarray}\label{supo2}
W=A_0 \exp\Big(\frac{8 \pi^2 f(M}{ c}\Big) = A
\biggl(\vartheta_1\Big(\sqrt{2 \pi} (y_3+\mu),t\Big)
\vartheta_1\Big(\sqrt{2 \pi} (y_3-\mu),t\Big)\biggr)^\frac{- 1 }{ c}
e^{8 i \pi^2 s/c}.
\end{eqnarray}
For small values of $y_3-\mu$ (with $y_3+\mu$
staying finite) this becomes
\begin{eqnarray}\label{supo4}
W = \A\,
\biggl(\vartheta_1\Big(\sqrt{2 \pi} (y_3+\mu),t\Big)\biggr)^\frac{- 1}{ c}
\left((2 \pi)^{3/2} \eta(t)^3\right)^\frac{-1}{ c}
(y_3-\mu)^\frac{-1 }{ c} e^{8 i \pi^2 s/c}+ \ldots \ .
\end{eqnarray}
Let us compare this result with the superpotential which was obtained in
\cite{Baumann:2007ah} by studying the
embedding equations of the D7-branes  into the (warped) geometry of the deformed
conifold. Specifically, it is given as (adjusting their eq.\ (2.14) to our
notation and using $\tilde A$ instead of $A$ in order
to avoid confusion with the function $\A$ of eq.\ \eqref{W})
\begin{equation}
W=W_0 + \tilde A(z_\alpha)e^{-8 i \pi^2 s/N}\, . \label{supo3}
\end{equation}
Here, $z_{\alpha}$ $(\alpha=1,2,3)$ denotes the complex coordinates on $K3\times T^{2}/\mathbb{Z}_{2}$, $W_0$ is the flux superpotential, and the second term is,
as before, due to gaugino condensation on a stack of $N$ D7-branes.
The function $\tilde A(z_\alpha)$ is determined in terms of the D7 embedding function
$f(z_\alpha)$ (not to be confused with the
gauge kinetic function) as follows (see eq.\ (2.15) in \cite  {Baumann:2007ah}):\footnote{An independent argument for this form of the 
D3-brane dependence of the non-perturbative 
superpotential was given in section 6 of \cite{Koerber:2007xk}, using the 
results of \cite{Martucci:2006ij}.}
\begin{equation}\label{supob}
\tilde A(z_\alpha)=\tilde A_0\Biggl(\frac{f(z_\alpha)}{f(0)}\Biggr)^{1/N}\, .
\end{equation}
To compare this with the torus case, where one has
a stack of D7-branes, filling the $z_2$ and $z_3$
directions and being located at the complex point $\mu$
inside the first (complex) direction, one should
take an embedding function of the (Kuperstein) form
\cite{Kuperstein:2004hy}:
\begin{equation}
f(z_\alpha)=z_1-\mu = 0\, .
\end{equation}
Plugging this back into eq.\ (\ref{supob}), we see
that for $z_1\equiv y_3$ the
gaugino condensate superpotentials eq.\ (\ref{supo4})
and eq.\ (\ref{supo3}) are indeed both
proportional to a power of the D7-brane embedding function,
with the same exponent if we use
the value $c=-N$ for a pure $SU(N)$ SYM theory on the D7-branes.
However, the prefactor in the toroidal orientifold case still depends
on $y_3$, cf.\ \eqref{supo4}, and if the D7-branes were at
$\mu=0$, the exponent of $y_3$ in \eqref{supo4} would be twice as
large, i.e.\ $-2/c$. This is due to the symmetrization
under the orientifold action present in the toroidal orientifold case
(cf.\ the derivation in appendix \ref{d7gc}).

Moreover, using the gaugino condensate superpotential
eq.\ (\ref{supo2}), which contains the stringy threshold
correction function, and comparing it with eq.\ (\ref{supob}),
it is more natural to use the function
$\vartheta_1(\sqrt{2 \pi} (y_3-\mu),t)$ as embedding function as it
provides a modular covariant way to describe the position
of the D7-branes on the two-torus (or rather on $T^{2}/\mathbb{Z}_{2}$),
i.e.\
\begin{equation}
f(z)=\vartheta_1(z-\mu,t)=0 \quad \Longleftrightarrow
\quad z-\mu \in m + n t \quad  ,
\quad m,n \in \mathbb{Z} \, .
\label{embedding}\end{equation}
%


\section{Canonical field range of D3-brane coordinate}
\label{inflaton}

As pointed out in \cite{Lyth} (see also \cite{BM,BLS} for recent
discussions in the context of D-brane inflation models), the canonical
field range of the inflaton leads to important upper bounds on the
amount of tensor modes produced during inflation.
In D3/D7-brane inflation, the inflaton is related to the D3-brane
coordinate, $y_{3}$, on $T^{2}/\mathbb{Z}_{2}$, or, more precisely, to a
certain real curve in the complex $y_3$-plane. It is the maximal field
variation along such a real curve in units of the 4D Planck mass
which enters the bound on tensor modes and is therefore of great phenomenological interest. The purpose of this section is to point out that the \emph{kinematical}
field range of
the canonically normalized D3-brane coordinate can be much larger
than naively expected. More precisely, we will determine this field
range for the real part of the canonically normalized
D3-brane coordinate, $\phi\equiv\textrm{Re}(y_{3}^c)$,
and show that its kinematical range can be much larger than the 4D
Planck mass.\footnote{Whether $\textrm{Re}(y_{3}^c)$ is a good
inflaton candidate and whether
all possible values of $\textrm{Re}(y_{3}^c)$  really fall on a possible
inflaton trajectory are different questions that require a more careful
dynamical study of the full scalar potential. This is beyond the scope
of the present paper. We just mention again that, in general,
due to the quantum corrections discussed in the last section,
$\textrm{Re}(y_{3}^c)$ is not necessarily
preferred over $\textrm{Im}(y_{3}^c)$.
This would also follow from the $SL(2,\mathbb{Z})_{t}$ invariance of the
theory which we show in appendix \ref{dualsym} to be restored by 1-loop
effects, at least in the case without fluxes and gaugino
condensation. Even in that special case there is the question why the
real part $\textrm{Re}(y_{3}^c)$ should play a special role. This
$SL(2,\mathbb{Z})_{t}$ symmetry mixes $\textrm{Re}(y_{3}^c)$ and
$\textrm{Im}(y_{3}^c)$ and what is called $\textrm{Re}(y_{3}^c)$
in one $SL(2,\mathbb{Z})_{t}$ frame might become $\textrm{Im}(y_{3}^c)$
in another. The reason to favor $\textrm{Re}(y_{3}^c)$ in our discussion here
is the hope to find
a region in moduli space where (in a certain $SL(2,\mathbb{Z})_{t}$ frame)
the 1-loop corrections are small and, thus, $\textrm{Re}(y_{3}^c)$
would be the inflaton candidate due to its tree-level shift symmetry.}

\begin{figure}[t]
\begin{center}
  \resizebox{5.5cm}{!}{\psfig{figure=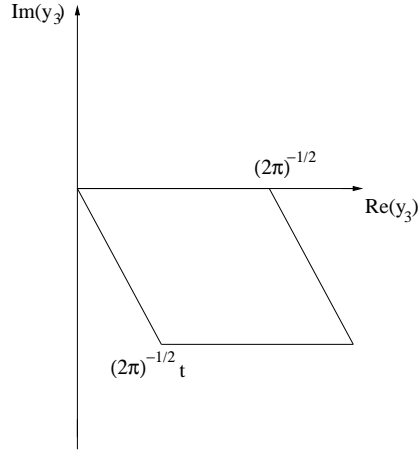,width=5cm}}
\caption{The torus has base length $(2 \pi)^{-1/2}$
(in dimensionless supergravity fields) and $t_2<0$.} \label{torusfig}
\end{center}
\end{figure}

Let us start with the range of the dimensionless 4D supergravity field
\begin{equation}
\textrm{Re} (y_{3})= y_3^1+\textrm{Re}(t)y_{3}^{2}\ ,
\end{equation}
where the coordinates refer to the decomposition $y_{3}=y_{3}^{1}+ty_{3}^{2}$.
In appendix \ref{10to4}, we show that the field range of the
component fields $y_{3}^{1,2}$ is from $0$ to
$(2 \pi)^{-1/2}$ (cf.\ eq.\ (\ref{range})). This implies that the field range of $\textrm{Re}(y_{3})$ is
\begin{equation}
\textrm{Re}(y_3)\in [0,(2 \pi)^{-1/2}(1+t_1))\ . \label{range1a}
\end{equation}
The full range can only be exploited if the field moves along
the diagonal of the parallelogram, cf.\ figure \ref{torusfig}.
If it moves along the base only, the $t_1$ factor of \eqref{range1a}
would be absent in the corresponding range.

We now have to convert this range of the dimensionless 4D supergravity field
to the field range of the corresponding canonically normalized field $\phi$.
Neglecting quantum corrections to the K\"ahler potential,
the kinetic term of $\textrm{Re}(y_3)$ can be read off from (\ref{kinscalars}):
\begin{equation}
M_P^2 \int d^4 x \sqrt{\det(\tilde{g}_{\mu\nu})} \tilde{g}^{\mu\nu}
\frac{\partial_{\mu}\textrm{Re}(y_{3})\partial_{\nu}
\textrm{Re}(y_{3})}{4\textrm{Im}(t)\textrm{Im}(s)
- 2[\textrm{Im}(y_{3})]^{2}}\ . \label{kinscalars2}
\end{equation}
The canonically normalized field, $\phi$, is therefore
\begin{equation} \label{canonphi}
\phi=\frac{M_{P}\textrm{Re}(y_3)}{\sqrt{2\textrm{Im}(t)\textrm{Im}(s) -[\textrm{Im}(y_{3})]^2}} \ .
\end{equation}
Using \eqref{sdef_pi}, as well as
\begin{equation}
T_3=\frac{1}{(2\pi)^3g_s(\alpha^{\prime})^2}\ ,
\end{equation}
the denominator becomes
\begin{equation}
\sqrt{2\textrm{Im}(t)\textrm{Im}(s) -[\textrm{Im}(y_{3})]^2}
=\sqrt{-\frac{\textrm{Vol}_{0}(K3)\textrm{Im}(t)}{(2\pi)^{5}
g_{s}(\alpha^{\prime})^2}}\ ,
\end{equation}
where
\begin{equation}
\textrm{Vol}_{0}(K3)\equiv h_{0} \tilde{V}^{K3}e^{4U_{1}}
\end{equation}
is the volume of $K3$ with respect to the zero mode, $h_0$, of the warp factor
and the minus sign under the square root is required because
$\textrm{Im}(t)<0$ in our conventions. We thus have
\begin{equation} \label{phi}
\phi=M_{P}\textrm{Re}(y_{3})\sqrt{-\frac{(2\pi)^{5}g_{s}
(\alpha^{\prime})^{2}}{\textrm{Vol}_{0}(K3)\textrm{Im}(t)}}\ ,
\end{equation}
or, using (\ref{range1a}),
\begin{equation}
\left( \frac{\phi_{max}}{M_P} \right) = \frac{1}{\sqrt{-t_2}}
\sqrt{\frac{(\alpha^{\prime})^{2}}{\textrm{Vol}_{0}(K3)}} \sqrt{g_s}
(2 \pi)^2\ (1 + t_1)\ .\label{range3}
\end{equation}
\mbox{} The corresponding range on $T^{2}/\mathbb{Z}_{2}$ would be
smaller by a factor of order $2$, but the main importance of \eqref{range3}
lies in its moduli dependence. The range
depends on the torus only via the complex structure, but it is
independent of the volume of $T^{2}/\mathbb{Z}_{2}$!
Using a rectangular torus (i.e.\ $t_1=0$) with base length $L_{1}$
along $\textrm{Re}(y_{3})$
and height $L_{2}$ along $\textrm{Im}(y_{3})$, this result is easy to understand:
The maximal distance for the dimensionful field along the base is proportional to $L_{1}$,
whereas the 4D Planck mass is proportional to
$\sqrt{\textrm{Vol}_0(K3) V^{T^{2}}}=\sqrt{\textrm{Vol}_0(K3) L_{1}L_{2}}$. Hence the canonically normalized
field has a range proportional to $\sqrt{L_{1}/L_{2}}=\sqrt{-1/t_{2}}$ in 4D
Planck units.

Using similar arguments, it is also easy to see that the maximal canonically
normalized field range along $\textrm{Im}(y^c_{3})$ is proportional to
$\sqrt{L_{2}/L_{1}}= \sqrt{-t_{2}} $ with otherwise identical numerical
factors as in (\ref{range3}), but without the $t_1$-term.

Let us now study how large the range of $\phi/M_{P}$ can be
within the regime of validity of the supergravity approximation.
In order to have a weakly coupled supergravity description, the two factors
$\sqrt{(\alpha^{\prime})^{2}/\textrm{Vol}_{0}(K3)}$ and $\sqrt{g_{s}}$
in (\ref{range3}) should both be smaller than one. This tends to suppress
the maximal range to values below the Planck scale. The factor $(2\pi)^2$
can compensate only part of this suppression. However, if we could choose
the factor $1/\sqrt{-t_{2}}$ sufficiently large, we could make the field
range larger than $M_{P}$. Furthermore, one might choose a large $t_1$, which also increases the field range.

Let us, for simplicity, again consider the case $t_{1}=0$.
The complex structure $t_{2}=-L_{2}/L_{1}$ then
gives the ratio of the lengths of the two sides of the rectangle that
defines the torus. Thus, eq. (\ref{range3})  implies that one could increase
the range of $\phi$ by simply making the torus asymmetrical, i.e., very long
and thin. Of course, we still have to require that the smaller length be
larger than the string length: $l_{s}\ll L_{2} \ll L_{1}$, but provided
these inequalities are respected, there is much freedom to dial a large
kinematical field range in 4D Planck units, and $\Delta \phi/M_{P}\gg 1$
becomes possible.\footnote{There is another constraint on the values of
$t$ arising from the requirement that Kaluza-Klein masses not be too light. This leads to upper bounds on $L_1$ and $L_2$.}

As, on the other hand, the range of $\textrm{Im}(y^c_{3})$
is proportional to $\sqrt{-t_{2}}$,
the range of either $\textrm{Re}(y^c_{3})$ or $\textrm{Im}(y^c_{3})$ can be
made very large by considering a very asymmetrical torus with either
$|t_2|\ll 1$ or $|t_2|\gg 1$ (depending on the $SL(2,\mathbb{Z})_{t}$
frame one uses). To our knowledge, this additional
freedom of having very different lengths for the torus sides
was neglected in the literature so far and, thus, the possibility of
having a large kinematical field range for the canonically normalized
scalar fields was not discussed.\footnote{We think that the complex
structure dependence of the range of the canonically normalized
open string fields should be more general
and also occur if the compactification space is a full-fledged
Calabi-Yau manifold.} However, it remains to be seen if the
potential (including the threshold corrections) can at the same time lead
to a stabilization of $t_2$ at such a small (or large) value and be fine-tuned to be
flat enough along the direction with a large kinematical field range.

Let us finally see how this result is consistent with
$SL(2,\mathbb{Z})_{t}$-invariance (again restricting to
a rectangular torus).
The inversion $\alpha:t\rightarrow \tilde{t}= -t^{-1}$ corresponds to
the exchange of the real and the imaginary part of
$y_{3}$, cf.\ \eqref{generaltrafo}
(plus an irrelevant conformal rescaling which drops out in the
complex structure). Thus, what was formerly the real part of $y_{3}$
now corresponds to
the imaginary direction of the transformed torus. The maximal range
of this imaginary part is proportional to $\sqrt{-\tilde{t}_{2}}=\sqrt{-1/t_{2}}$,
i.e., just the same as for the real part in the old $SL(2,\mathbb{Z})_{t}$-frame.
The only thing that has changed is that
this direction is now called the imaginary part of $y_3$.

The Dehn twist $\beta: t\rightarrow \tilde{t}=t+1$, finally, does not change the
imaginary part of $t$ and leaves $\textrm{Re}(y_{3})$ unchanged  (see
(\ref{xtrafo})): $\textrm{Re}(\tilde{y}_{3})=\tilde{y}_{3}^{1}
+\tilde{t}_{1}\tilde{y}_{3}^{2}=y_{3}^{1}=\textrm{Re}(y_{3})$. The canonical
field range therefore also stays the same.


\section{Some interesting brane configurations}
\label{braneconfigs}

Let us finally in this and the next
section sketch possible applications of our findings
for D3/D7-inflation, leaving a more detailed study for the future
\cite{pheno}. As mentioned in the introduction, one of the motivations
to revisit the D3/D7-inflationary scenario of \cite{Dasgupta:2002ew} was the
recent work of \cite{Bevis:2007gh}, where it was found that a spectral
index of $n_s \approx 1$ may be compatible with the CMB data, if cosmic strings
have a contribution of 11\% to the CMB. Such a spectral index and a viable
cosmic string tension arise naturally in D-term inflation with a small
coupling constant \cite{Kallosh:2003ux} and corresponds to the regime B mentioned in the introduction.

\begin{figure}[t]
\begin{center}
  \resizebox{10cm}{!}{\psfig{figure=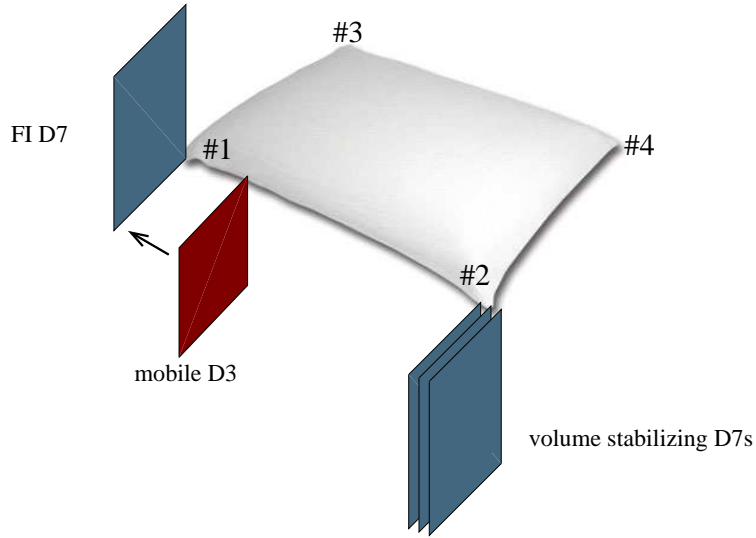}}
\caption{The simplest brane setup on the pillow.} \label{pillowfig}
\end{center}
\end{figure}

From the phenomenological point of view, the cleanest setup would look like
in figure \ref{pillowfig}. As reviewed in section \ref{d3d7inflation},
the $\mathbb{Z}_{2}$ operation \eqref{z2} has 4 fixed points on $T^{2}$ and
the space $T^{2}/\mathbb{Z}_{2}$ has the shape of a pillow with the
fixed points at the corners. We denote them by $\#1, \ldots , \#4$.
At one of the fixed points ($\#1$ in the figure) there is a single
D7-brane (called FI D7 in the figure)
on which a non-selfdual world-volume flux is turned on, breaking
supersymmetry. This induces an FI-term in the 4D action. A D3-brane in
the vicinity of this D7-brane is attracted by the resulting Coleman-Weinberg potential induced by the broken supersymmetry
(in the simplest setup, there is only
a single D3-brane, with the 3-brane tadpole
canceled by flux, cf.\ \eqref{tadpol}; in general,
there might be several D3-branes of course).
If there is no (strong) dependence of the F-term potential
on the D3-brane position, this leads to a phase of slow roll
D3/D7-brane inflation.

A possible source for such $y_3$-dependent F-terms are the F-terms
responsible for the stabilization of the other moduli. For the
$K3$ volume this stabilization might
proceed via gaugino condensation on a stack
of D7-branes wrapped around the $K3$. In order to avoid a destabilization
of the volume after inflation, this stack of D7-branes
should be placed away from the FI D7-brane, as the non-perturbative
superpotential becomes zero when the D3-brane hits the
gaugino condensate D7-branes \cite{Ganor:1996pe}.
In the simplest setup depicted in figure \ref{pillowfig},
there is a single stack of volume stabilizing D7-branes,
placed at fixed point $\#2$ (i.e.\ at $x=1/2 \times (2 \pi)^{-1/2}$),
to be explicit. In general, there might be
other D7-branes, for instance at the other fixed points.

The hope would be now that one can, by an appropriate choice of fluxes, fine-tune the complex structure of the
torus in such a way that the dependence of the non-perturbative
superpotential on the D3-brane position (arising via the threshold
corrections to the gauge coupling of the gauge theory on the stack
of D7-branes at fixed point $\#2$) is small. In fact, we have argued that in
the context of our  $T^{2}/\mathbb{Z}_{2}$ geometry the stabilizing superpotential
depends on the position of the D3-brane as suggested by equation \eqref{W} (for some
positive constant $a$)
\begin{equation} \label{Wcorrected}
W_{np} = A \Big( \vartheta_1( \sqrt{2\pi}y_3-1/2, t ) \vartheta_1( \sqrt{2\pi}y_3+1/2, t ) \Big)^{-1/c}  e^{-ias}\ .
\end{equation}
In this notation $\sqrt{2\pi}y_3 = 1/2$ would correspond to the position of the D3-brane
coinciding with the stack of the stabilizing D7's at the fixed point \#2. Since
\begin{equation}
\vartheta_1( w, t )_{w\rightarrow 0} \rightarrow 0\ ,
\end{equation}
eq.\ \eqref{Wcorrected} is in agreement with the fact that the volume
would be destabilized if
the mobile D3 hits the D7-brane stack at the fixed point \#2.
However, this can be prevented if the D3-brane is close to the
FI D7-brane (i.e.\ for small $y_3$), due to the interbrane
attraction between the two. One
can use \eqref{1/2} in order to expand
the non-perturbative superpotential \eqref{Wcorrected} as a function of $y_3$ and finds
\begin{equation} \label{Wcorrected1}
W_{np} = A \Big(1 - \Delta (t) \, y_3^2 + \ldots \Big) e^{-ias}\ ,
\end{equation}
where we absorbed the overall factor of $\vartheta_2(0,t)$ into the prefactor $A$.
To lowest order, the quantum corrections to the superpotential are quadratic in $y_3$ and
the coefficient of this quadratic term is the product of a parameter $\Upsilon$ and a function $\delta (t)$,
\begin{equation} \label{constant}
 \Delta (t)= \Upsilon \delta(t)\ ,
\end{equation}
where
\begin{equation} \label{constant2}
 \Upsilon = -\frac{2 \pi^3}{3c}
\end{equation}
depends on the constant $c$ defined in \eqref{c} (which, as a reminder, has to be negative
for a non-perturbative superpotential to be present). 
Note that $\Upsilon$ is not very small.
Finally, the dependence on the complex structure modulus $t$
which can be stabilized by the choice of the bulk fluxes is encoded in the function $\delta(t)$
\begin{equation} \label{delta}
\delta(t)  := [E_2(t) + \vartheta_3^4(0,t) + \vartheta_4^4(0,t)] \ .
\end{equation}
In order to realize the original D-term inflation scenario, it would 
be sufficient (but not necessary, see below) to fine-tune the function 
$|\delta(t)|$ to small values.

\begin{figure}[t]
\begin{center}
  \resizebox{6cm}{!}{\psfig{figure=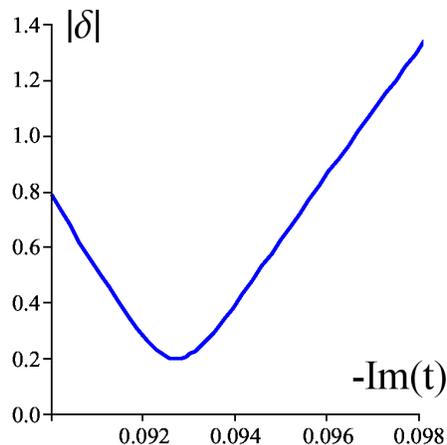}}
\caption{The function $|\delta|$ with $\delta$ given by \eqref{delta} for ${\rm Re}(t)=0.26$.}
\label{deltafig}
\end{center}
\end{figure}

However, this turns out to be pretty difficult to achieve.
We only performed a preliminary analysis of the function $\delta$
given in \eqref{delta} and leave a more detailed investigation
for the future \cite{pheno}. Here we only give one example in
figure \ref{deltafig}, which shows $|\delta|$ as a function of
$-t_2$ for $t_1=0.26$ (a value which we found to allow for relatively small
$|\delta|$).

Of course, also the situation where $|\delta|$ is not very small might
be interesting for inflation. A large value for $|\delta|$ is not
necessarily a problem but might lead to interesting modifications
of the pure D-term scenario. It is certainly conceivable that a small
slow roll parameter $\eta$ is possible even for relatively large values
of $|\delta|$, at least for particular values of the complex structure.
After all, $\delta$ is just a parameter in the superpotential and
$\eta$ is determined from the full potential (including also the
D-term potential). We will come back to this point in section
\ref{cosmology}.

It  might seem counterintuitive at first sight that the corrections
in \eqref{Wcorrected} do not become small, even if one increases the
distance between the D3-brane and the stack of volume stabilizing D7-branes
by increasing the volume of $T^{2}/\mathbb{Z}_{2}$.
However, the correction is completely independent of the volume of
$T^{2}/\mathbb{Z}_{2}$. In fact, it has to be independent, as the corresponding modulus is a
member of an ${\cal N}=2$ hypermultiplet and, thus, cannot appear
in the gauge coupling.

%

%
%

We end this section by mentioning some further caveats and open
questions on the way to a successful model of D3/D7-brane inflation
on $K3\times T^{2}/\mathbb{Z}_{2}$. The simple brane setup shown in
figure \ref{pillowfig} is not close to the orientifold point at which
4 D7-branes are on top of each O7-plane and for which the 7-brane tadpole is canceled
locally. Only this orientifold limit would lead to a dilaton that is constant along the internal
directions.\footnote{Its constant value still is a modulus in the external directions and has to be stabilized by fluxes, of course.} Thus, one is in the realm of F-theory \cite{Vafa:1996xn}
and one has to worry about the backreaction
on the geometry and the dilaton and about the fact that the O-planes
split into pairs of $(p,q)$ 7-branes at finite coupling \cite{Sen:1996vd}.
It has been argued though in \cite{Sen:1997gv},
and more recently in \cite{Braun:2008ua}, that there is a limit (called
the ``weak coupling limit'' in \cite{Sen:1997gv}) in which the backreaction
on the imaginary part of the dilaton (i.e.\ $g_s$) and the geometry can be
very small and a description using the toroidal orientifold with D7-branes
at arbitrary positions is valid
(of course, there is still the monodromy for the real part of the dilaton,
i.e.\ the RR-scalar, when going around groups of branes and O-planes
with net 7-brane charge). It is an interesting question though, how to
stabilize the moduli in this region of moduli space, for instance by fluxes.
Part of this question
would be to understand more concretely, which fluxes might stabilize the
D7-branes at the positions shown in figure \ref{pillowfig}. This issue becomes
complicated by the fact that only close to the orientifold point  there is a
clean distinction between open and closed string moduli/fluxes  \cite{LMRS}.\footnote{We thank P.\ Mayr for discussions on this
point.}

Some of these complications could be avoided by staying close to the
orientifold point, i.e.\ by distributing the D7-branes in groups of
four on top of (or close to) the O7-planes. This configuration can be
stabilized by closed string fluxes
\cite{Tripathy:2002qw,Gorlich:2004qm,Cascales:2004qp,LMRS}.\footnote{In
\cite{D'Auria:2004qv} it was found that the D7-brane coordinates are fixed at
$y_7=0$, cf.\  \eqref{conditions}. However, as was stressed in \cite{LMRS} and as we
also mentioned in section \ref{withoutd3}, supergravity alone
does not contain enough information to interpret this condition. Additional
information from string theory is required which determines the reference points
from which the distances of the individual D7-branes are measured. In \cite{LMRS},
this additional information is encoded in the choice of basis of 2-forms of the upper
$K3$, denoted $\widetilde{K3}$ in our section \ref{withoutd3}.} If one switches on a worldvolume flux on the brane stack at fixed point \# 1 in order to attract a nearby D3-brane to that point,
the corresponding gauge group on that stack would be broken to a smaller group involving Abelian factors. If this smaller gauge group no longer allows for gaugino condensation, the volume stabilization will again come from the other brane stacks at the other fixed points. It remains to be seen whether this could lead to
an interesting phenomenology. One issue one would have to reconsider would be the production of cosmic strings, which might be semilocal. We plan to come back to some of these issues in the future
\cite{pheno}.


\section{Towards D3/D7-brane cosmology}
\label{cosmology}
The previous sections analyzed string theory corrections to the D3/D7-brane
inflation model. In 4D this model is effectively described by
D-term inflation (with the $\N=2$ supersymmetry relation $\lambda^2=2g^{2}$,
where $\lambda$ is   the trilinear coupling between the waterfall fields and the inflaton in the superpotential,  and $g$ is the
$U(1)$ coupling) \cite{Binetruy:1996xj,Kallosh:2003ux,Dvali:2003zh,Binetruy:2004hh},
for which we want to study the impact of the corrections now. To set the stage, we
start with a brief review of the version of D-term inflation relevant for the D3/D7-brane model
and add in a second step the stringy corrections. We will see that to leading order
the stringy corrections add an inflaton mass term to the potential, with
interesting cosmological implications.


\subsection{Basic model of D3/D7-brane inflation}
\label{basic}

The potential of D-term inflation in the near de Sitter valley where
inflationary perturbations are generated   is given by a constant term and the
Coleman-Weinberg term:
\begin{equation}
V =\frac {g^{2}\xi^{2}}{ 2}\left(1+\frac{g^{2}}{ 16 \pi^{2}} U(x)\right) \ ,
\label{D3D7CW}
\end{equation}
where $x\equiv \frac{\phi}{ \sqrt\xi}$ and
\begin{equation}
U(x)= (x^2+1)^2 \ln (x^2+1) + (x^2-1)^2 \ln (x^2-1)
- 2x^4 \ln (x^2) - 4\ln 2 \ .
\label{U}
\end{equation}
The last term is added to account for the normalization condition $U(1) = 0$,
but it can be ignored in our subsequent calculations. Indeed, in the approximation
which we are going to use, the corrections to the potential do not affect much its
value, $V \approx  g^{2}\xi^{2}/2$, but these corrections are fully responsible for
the value of its derivative $V'$, which does not depend on the last term in (\ref{U}).
Furthermore, we remind the reader of the origin of the Coleman-Weinberg potential. 
It arises as a 1-loop correction, with the hypermultiplets of the waterfall
fields running in the loop. This correction only appears after 
supersymmetry breaking by the FI-term $\xi$, 
which leads to a mass splitting of the waterfall fields.
The occurrence of the Coleman-Weinberg potential is independent of the string 
theoretical threshold corrections
discussed in section \ref{qc} (which are present even if supersymmetry 
is not broken). 

Density perturbations on the scale of the  present cosmological horizon
have been produced at $\phi \approx \phi_{N}$ with $N$ in the range of about $50$
to $60$, and their
amplitude is proportional to $\frac{V^{3/2}}{ V'  }$ at that time. In our estimates we will
use $N = 60$ for definiteness.
From the WMAP5 observational data \cite{Komatsu:2008hk} it is known that the
amplitude of adiabatic perturbations, in the absence of any contribution from
cosmic strings, constrains the inflaton potential at the beginning of the last 60 e-foldings so that
\begin{equation}
\frac{V^{3/2}}{ V'} \approx
5.4 \times 10^{-4}\ .
\label{deltaWMAP3}
\end{equation}
In WMAP3 the corresponding value was $5.3 \times 10^{-4}  $. If one assumes that  there
is a contribution of 11\% from cosmic strings at $l=10$, the corresponding contribution
from scalar perturbations becomes \footnote{ We are grateful to M. Hindmarsh for
pointing out to us that the regime where cosmic strings contribute 11\% to the CMB
temperature power spectrum corresponds, according to \cite{Bevis:2007gh}, to an
approximately 15\% decrease of the square of the amplitude of inflationary curvature
perturbations, using WMAP3 data. }
\begin{equation}
  \frac{V^{3/2}}{ V'} \approx
4.9 \times 10^{-4}\ .
\label{deltaStrings}
\end{equation}
In the following, when we will be interested in the regime without cosmic strings,
we will use $\frac{V^{3/2}}{ V'} \approx
5.4 \times 10^{-4} $, and with cosmic strings we will use $\frac{V^{3/2}}{ V'}\approx
4.9 \times 10^{-4}$. (However, our final results are not very sensitive to the choice
of either one of these two numbers.)

At the end of D-term inflation Abelian local (or semilocal) BPS cosmic strings can be  produced with a tension related to the FI term $\xi$ as \cite{Dvali:2003zh}
\begin{equation}
\mu= 2\pi \xi \ .
\end{equation}
This means that in units $M_P^2= \frac{1}{8\pi G}=1$, used in this section,
the dimensionless string tension is given by
\begin{equation}
G\mu= \frac{ \xi }{ 4}\ .
\end{equation}
In case we consider a contribution of cosmic strings at the level of 11\% as in \cite{Bevis:2007gh}, which is achieved for $G\mu = 7\times 10^{-7}$, $\xi = 2.8\times 10^{-6}$, we have to take into account that the amplitude of  perturbations  includes also a contribution from cosmic strings and use  eq.\ (\ref{deltaStrings}) for inflationary fluctuations.

Two phenomenological regimes of D-term inflation were discussed in \cite{Binetruy:1996xj,Kallosh:2003ux,Dvali:2003zh,Binetruy:2004hh} and in its stringy version in \cite{Dasgupta:2004dw}, however without account of the stringy corrections studied in the present paper.

\textbf{Regime A:}  If $g\geq 2\times 10^{-3}$, the last 60
e-foldings of inflation start far away from the bifurcation point
where the local de Sitter minimum turns into a de Sitter maximum.
For $x\gg 1$
\begin{equation}
V = \frac{g^{2}\xi^{2}}{ 2}\left(1+\frac{g^{2}}{ 4\pi^{2}} \ln  \frac{\phi}{ \sqrt\xi}\right) \ .
\label{D3D7potential}\end{equation}
From Friedmann's equation
$
H^2 = \left(\frac{\dot a}{ a}\right)^2 = V/3 \approx \frac{g^2\xi^2}{ 6},
$
where $a(t)$ is the scale factor of the universe, one infers that one
has an approximately constant Hubble parameter $H =
g \xi/\sqrt 6$.
This leads to inflation with
$
a(t) = a(0)\, \exp{ {\frac{g \xi\,  t}{\sqrt 6} }}
$.
During the slow-roll regime the field $\phi$ obeys the equation
$3H\dot \phi = -V'(\phi)$
which gives
$
\phi^2(t) = \phi^2(0) - \frac{g^3\xi\, t}{ 2\sqrt 6 \pi^2} $.
Note that, in this section, the complex structure field $t=t_0$ is fixed, and the letter $t$
is used for the time-variable in the 4D FRW geometry as  is customary
in cosmology.

One can find the value of the
field $\phi_{{}_N}$ such that the universe inflates $e^N$ times when the
field rolls from $\phi_{{}_N}$ until it reaches the bifurcation point
$\phi = \phi_c= \sqrt \xi$:
\begin{equation}\label{infl3}
\phi_{{}_N}^2  = {\phi_c}^2  +\frac {g^2 N}{ 2 \pi^2} = {\xi}  + \frac{g^2 N}{ 2 \pi^2}\ .
\end{equation}
In our model with $N \approx 60$ and vanishing contribution from
cosmic strings, this yields
\begin{equation}\label{smallga}
\frac{V^{3/2}}{ V'}  =  \frac{2\sqrt 2\pi^2 \xi}{  g}~ \phi_{{}_N} \approx
5.4 \times 10^{-4}\ .
\end{equation}
In this regime, one has $\frac{g^2 N}{ 2 \pi^2} \gg
{\xi}$ and, consequently, 
$\phi_{{}_N} \approx \frac{g
\sqrt{N}}{ \sqrt 2 \pi }$. For $N \approx 60$, eq.\ \eqref{smallga} implies that
\begin{equation}\label{COBE2}
\xi  \approx 1.1 \times  10^{-5}  \ .
\end{equation}
 The spectrum of perturbations in
this model is nearly flat. It is characterized  by the spectral index
\begin{equation}\label{COBE4}
n_s = 1
  -{3}\left(\frac{V'}{V}\right)^2 + 2\frac{V''}{V} \approx 1-\frac{1}{N} \approx 0.98 \ .
\end{equation}
The problem here is that the tension of the cosmic strings
produced after inflation in this model is given by
\begin{equation}
G\mu  =  \frac{ \xi}{ 4} \approx   2.8 \times 10^{-6} \ .
\end{equation}
This is significantly higher than  the current bound on the cosmic string tension.

An attempt to avoid local strings and replace them with semilocal ones was made in
\cite{Urrestilla:2004eh,Dasgupta:2004dw}, where it was found that for the $SU(2)$ case
\begin{equation}
G\mu =\frac {\xi}{ 4} \approx 3.7\times 10^{-6}\ , \qquad n_s=0.98\ .\label{slnumbers}
\end{equation}
According to \cite{Urrestilla:2007sf} this may be a marginally viable solution. With
better data in the future and a more detailed numerical investigation of the semilocal
strings, this regime of D-term inflation may be either confirmed or falsified.

\textbf{Regime B:} If $g \ll 2\times  10^{-3}$, one can study the model assuming
that $x-1$ is small and  using the  potential as given in equations (\ref{D3D7CW}),
(\ref{U}). In this case, near $x=1$, we find,  at leading order,
\begin{equation}
\frac{\partial V}{ \partial \phi}=\frac{g^4 \xi^{3/2} \ln 2}{ 4\pi^2}
\end{equation}
and
\begin{equation}\label{smallg2}
\frac{V^{3/2}}{ V'  }= \frac{\sqrt 2\pi^2}{ \ln 2 \,g}\xi
^{3/2} \approx
4.9 \times 10^{-4} \ ,
\end{equation}
where (by choosing the value \eqref{deltaStrings})
we assumed already that there is a contribution to the CMB
from cosmic strings. \eqref{smallg2} then implies that
\begin{equation}
 \qquad   {\xi } \approx 8.4\times 10^{-4}~ g^{2/3} \ .
\end{equation}

Solving the scalar field equation together with the Friedman equation one finds that
\begin{equation}
\phi_N= \phi_c +\frac{g^2\ln 2}{ 2\pi^2 \sqrt{\xi}} N\ .
\end{equation}
This means that at the beginning of the last $N$ e-foldings
\begin{equation}
x-1=  \frac{g^2\ln 2}{ 2\pi^2 {\xi}} N \ .
\label{x-1}
\end{equation}
Now let us estimate the spectral index
\begin{equation}\label{COBE4a}
n_s \approx  1+\frac{g^2}{ 2\pi^2 \xi}  [2\ln(x^2-1) + 4 \ln 2]\approx 1+\frac{g^2}{ \pi^2 \xi}  [\ln(x-1) + 3 \ln 2]\ ,
\end{equation}
 where we neglected terms of order $g^{4}$, as we always do for $n_{s}$.
We can evaluate the spectral index using eq.\ (\ref{x-1}) and find
\begin{equation} \label{ns-1}
n_s -1 \approx \frac{g^2}{ \pi^2 \xi}  [\ln(\frac{g^2\ln 2}{ 2\pi^2 {\xi}} N) + 3 \ln 2]\ .
\end{equation}
For $ g \ll 2 \times 10^{-3}$ one finds a practically flat spectrum,
\begin{equation}
n_{s} \approx 1  \ ,
\end{equation}
which is a distinguishing feature of this class of models.

To be precise we have also solved the FRW equations numerically for specific  values of parameters associated with the fit to CMB data in \cite{Bevis:2007gh} by Hindmarsh et.al. Namely, we take
\begin{equation}
G\mu = 7\times 10^{-7}\ , \qquad \xi = 2.8\times 10^{-6}\ .
\end{equation}
To fit the data of the level of fluctuations we need
\begin{equation}
g\approx  2.2 \times 10^{-4}
\end{equation}
and this leads to the spectral index
\begin{equation}
n_s= 0.997\ .
\end{equation}
If we had used instead the approximate solution presented  in \eqref{ns-1},
these numbers would have been only slightly different.
This confirms that the approximation used above is valid and gives
approximately the same value of the spectral index as a full numerical solution.
In case that this model with 11\% of cosmic strings and $n_s\approx 1$
were confirmed by future data, it would be a simple version of D3/D7-brane inflation
in the regime of very small couplings \cite{Kallosh:2003ux} of the associated D-term
hybrid inflation. This is the case when  quantum corrections are associated with
FI terms generating Coleman-Weinberg terms and other  stringy quantum
corrections  are negligibly small. 

In view of the proposal in
\cite{Bevis:2007gh, Pogosian:2008am} it will be very interesting to follow
the new data which may confirm or falsify this model and to study the level 
of non-Gaussianity of perturbations of the metric produced by cosmic strings 
in this scenario.


\subsection{New features of the model with stringy corrections}

In this section we will start to analyze how the basic picture of 
D-term inflation reviewed in the last section changes if moduli stabilization 
is taken into account. This requires the presence of 
an F-term potential in addition to the D-term and Coleman-Weinberg 
potential discussed in the last section. It 
arises from a superpotential $W=W_{flux}+W_{np}$ that 
gets contributions both from fluxes and non-perturbative effects. 
A part of the closed string moduli is 
stabilized by three-form fluxes and gets high masses, 
in particular the dilaton and the
complex structure of the torus. It has been shown in \cite{Aspinwall:2005ad}
that the remaining closed string moduli can be fixed by non-perturbative 
contributions to the potential, which also depend on the D3-brane coordinates. 
Furthermore, fluxes can stabilize the D7-brane position moduli, cf.\
\cite{Gorlich:2004qm,Cascales:2004qp,LMRS,Jockers:2005zy}.
Ideally one would like to simply treat all fields except the 
inflaton candidate as
already stabilized by these effects and solely 
focus on the dynamics of the inflaton, i.e.\ the 
D3-brane position along $T^2/\mathbb{Z}_{2}$. 
However, the stabilization of the moduli fields in general 
interferes with the dynamics of the inflaton. The field for which 
this interference is expected to be strongest, is the $K3$-volume because 
it is the only modulus that is stabilized 
by a superpotential which directly depends on the 
inflaton candidate (cf.\ footnote \ref{independent}). 
Thus, we have to consider its minimization more carefully in order to determine 
the dynamics of the inflaton. 

The superpotential at the minimum of all moduli other than $y_3$ and $s$ is 
given by $W = W_0 +W_{np}$, where $W_{np}$ is now only the part of the 
non-perturbative superpotential depending on $s$, and $W_0$ includes the 
flux superpotential as well as
further non-perturbative contributions fixing the other K\"ahler moduli. 
Moreover, close to the FI D7-brane, $W_{np}$ can be expanded as discussed 
in section \ref{braneconfigs}.
For a non-perturbative superpotential of the form $W_{np} = A \big(1 - \Delta (t_0) \,
y_3^2 + \ldots \big) e^{-i a s}\ ,$ where $t_0$ is the value of the complex
structure fixed by the choice of fluxes, one can compute the F-term
potential following a similar computation in Appendix F of \cite{KKLMMT}
(using the relation $is = \rho$ between our variable and the one employed there).
The difference between the KKLMMT model and our
D3/D7-model (without account of the
threshold corrections to the K\"ahler potential)
is the absence of the Hubble square contribution to the
mass term for the canonically normalized real part
\begin{equation} \label{phican}
\phi={\rm Re}(y_3)/\sqrt{2t_2s_2}
\end{equation}
of the D3-brane coordinate,\footnote{This canonical normalization factor is based on
\eqref{canonphi}, neglecting the $[{\rm Im}\, y_3]^2$ term, which is valid to the order $\phi^2$ at which we are working.} i.e.\ there is no contribution
$m_\phi^2=2H^2$ as in eq.\ (F.7) of \cite{KKLMMT}.
This is due to its shift symmetry in the D3/D7-model
without quantum corrections. For $\phi$ to be the
inflaton candidate, we make the simplifying assumption that ${\rm Im}(y_3)$ is
fixed (at zero). This would have to be justified in a
more complete treatment, as we said before.

Performing now a similar calculation as
the one described in appendix F of \cite{KKLMMT}, leads to an F-term
potential which is similar to the second and third terms of their
eq.\ (F.7). More precisely, we modified the calculation
in some details. For example, we relaxed the assumptions of
a real $A,W_0$ and $s$ in the minimum. Moreover, we solved the
condition $D_sW=0$ not for $W_0$ (which we take to be
constant after fixing the moduli other than $\phi$ and $s$), but for
$s$. This is because we are interested in a potential for
$\phi$ only and, as was
discussed in \cite{Baumann:2007np,Baumann:2007ah}, minimizing the F-term potential
with respect to the volume modulus $s$ leads to a $\phi$-dependence
of the value of $s$ in its minimum.\footnote{Strictly speaking, integrating out $s$ would require solving $\partial_{s}V=0$, where $V$ denotes the total scalar potential. As this would necessitate a much more elaborate analysis, which is beyond the scope of the present paper, we restrict ourselves to solving
the simpler condition $D_s W=0$, which actually need not be a bad approximation.
The main purpose of the present analysis is to demonstrate some general features
 of the inflaton mass term, in particular its tunability to small values. This is all we use in the cosmology analysis of this section. }
This
is obvious from
\begin{equation}
D_{s}W|_{{\rm Im}(y_{3})=0}=0 \Longleftrightarrow W_{0}
=A(1- {\rm Re}(y_{3})^{2}\Delta)e^{-ias}[2as_{2}-1]\ .\label{constraint}
\end{equation}
Solving this for $s$ leads to an implicit dependence of $s$
in its minimum on $[{\rm Re}(y_{3})]^{2}$ (note that
\eqref{constraint} only depends on ${\rm Re}(y_{3})$ quadratically)
and, thus, we can expand, for small ${\rm Re}(y_{3})$,
\begin{equation} \label{s}
s=\tilde{s} +i\lambda [{\rm Re}(y_{3})]^{2}
+\mathcal{O}([{\rm Re}(y_{3})]^{4})
\end{equation}
with some constants $\tilde{s}$ and $\lambda$ (the $i$ in front of $\lambda$
is for convenience and, with a slight abuse of notation,
we denote the value of $s$ in its minimum still by the same symbol, i.e.\ $s$).
The constant piece, $\tilde{s}$, is implicitly determined by (\ref{constraint})
in the limit ${\rm Re}(y_{3})=0$, where it becomes
\begin{equation}
W_{0}=Ae^{-ia\tilde{s}}[2a\tilde{s}_{2}-1]\ . \label{W0}
\end{equation}
This equation defines $\tilde{s}$ and can be used to eliminate everywhere $W_{0}$
in favor of the constant $\tilde{s}$. The constant $\lambda$ can be determined
by expanding (\ref{constraint}) in ${\rm Re}(y_{3})$ and  comparing
coefficients (which results in
${\rm Re}(\lambda)= (2a\tilde{s}_{2}-1){\rm Re}(\Delta)/[(2a\tilde{s}_{2}+1)a]$ and
${\rm Im}(\lambda)= {\rm Im}(\Delta)/a$). Plugging this into \eqref{s} and switching
to the canonically normalized field \eqref{phican}, one finally
obtains
\begin{equation}
V_{F}=\frac{|Ae^{-ia\tilde{s}}|^{2}\tilde{s}_{2}}{2u_{2}}\Big[\frac{3 a^{2}}{t_{2}} - 2 \phi^{2}\Big(3 a\textrm{Re}(\Delta) + 4t_2 |\Delta|^{2}\Big)\Big] +\mathcal{O}(\phi^{4}).\label{paper2}
\end{equation}
%
%
Strictly speaking, there is an additional overall constant prefactor 
which arises from the hypermultiplet sector. As mentioned after 
\eqref{KP} the K\"ahler potential has two parts, one originating from the 
${\cal N}=2$ vector multiplets and one from the ${\cal N}=2$ hypermultiplets. 
If the hypermultiplet scalars are fixed, the K\"ahler potential becomes 
$K=K_{vector}+K_{fix}$ with $K_{fix}$ the contribution of the fixed 
hypermultiplet scalars. 
Inserting this into the $e^K$ prefactor of the F-term potential leads to the 
mentioned overall constant factor.
Of course, this 
can be absorbed in the factor $|A|^2$ and we did so in order not 
to overload the notation. 

\begin{figure}[t]
\begin{center}
  \resizebox{8cm}{!}{\psfig{figure=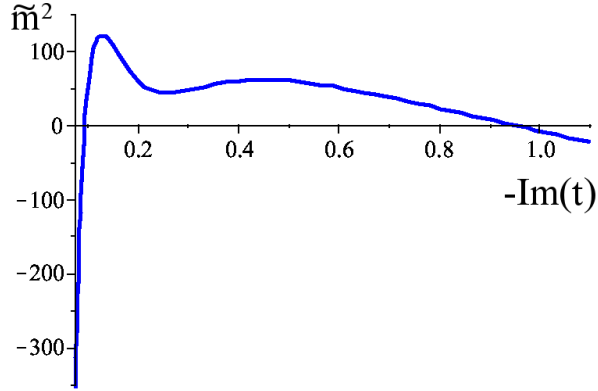}}
\caption{The function $\tilde{m}^2$ given in \eqref{tildem} as a function of
$-{\rm Im}(t)$ for ${\rm Re}(t)=0.26$ (we chose the
values $a = 8\pi^2/10$ and $\Upsilon=2\pi^3/30$, i.e.\ $c=-10$).}
\label{mfig}
\end{center}
\end{figure}

In order to ensure that the cosmological constant is almost
zero after inflation, the first term in \eqref{paper2} (a negative
contribution to the vacuum energy), has to be canceled. This might
require an additional uplifting mechanism, a discussion of which is
beyond the scope of this paper.\footnote{The uplifting might complicate
the story in several ways. First, the uplifting potential
might also depend on $\phi$. Even if it did not do so explicitly, a
possible dependence on $s$ would introduce an implicit
dependence according to \eqref{s}, which would modify the
mass term \eqref{msquare} (some preliminary ideas on
how to avoid this problem can be found in appendix
\ref{appmarco};  see also \cite{Brax:2006yq} for
a related discussion ). Moreover, if the uplifting
mechanism proceeded via anti D3-branes or additional
D7-branes with non-selfdual world volume flux, this might
modify the Coleman-Weinberg potential for $\phi$. \label{upliftcaveat}}
Here we just assume that the $\phi$-independent contribution to the
F-term potential is canceled after inflation ends.
In that case, the correction due to the F-term potential arising from
stringy corrections to the superpotential takes the
form\footnote{There is a further caveat here. We did not explicitly
take into account threshold corrections to the K\"ahler potential.
These would modify the mass term, but the formulas get a bit messy.
In any case, these corrections are suppressed in the weak coupling limit
(i.e.\ for large $|u_2|$).
Nevertheless, they should be taken into account in a more
detailed analysis and they might lead to interesting additional
possibilities to fine-tune the mass parameter.
We think however, that the qualitative features of the mass squared term
would not change, neither by this effect nor by the one mentioned in footnote
\ref{upliftcaveat},
i.e.\ we still expect that it can take both signs and that it can be fine-tuned
to small values via a fine-tuning of the complex structure.}
\begin{equation} \label{msquare}
V_{F}= -\frac {m^2}{2}  \phi^2 \ , \qquad
m^2= \frac{2|A|^2 \tilde{s}_2 e^{2 a \tilde{s}_2}}{u_2} \left[
3 a {\rm Re} (\Delta) + 4 t_2 |\Delta|^2\right]\ .
\end{equation}
The function $m^2$ of \eqref{msquare} gets a strong suppression from
the exponential prefactor (note that $\tilde{s}_2$ is negative in our conventions and that
$|\tilde{s}_{2}|$ has to be considerably larger than one in the supergravity
regime). Furthermore, also $|u_2|$ is large in the weak coupling limit. In addition, $m^{2}$
depends on the complex structure
and is thus tunable via a choice of fluxes. Note that even though $t_2,u_2$ and $\tilde{s}_2$ are all negative in our
conventions, $m^2$ is not necessarily positive, because ${\rm Re} (\Delta)$ can have either sign.
In fig.\ \ref{mfig}, we plot the function
\begin{equation} \label{tildem}
\tilde{m}^2 \equiv 3 a {\rm Re} (\Delta) + 4 t_2 |\Delta|^2
\end{equation}
for $a = 8\pi^2/10$ and $\Upsilon=2\pi^3/30$ (cf.\ \eqref{constant2}, where we
set $c=-10$ to be explicit) as a function
of $-t_2$ for the sample value $t_1=0.26$ (which is the same value we
used in section \ref{braneconfigs}). As $\tilde{m}^2=\gamma m^2$ with $\gamma>0$,
the vanishing of $\tilde{m}^2$ means also a vanishing of $m^2$.
It is thus plausible that $m^2$ can be made
small and positive. We will assume this in the following, and this is
all we will make use of in the remainder of the section; the explicit form of 
$m^2$ given in \eqref{msquare}, which should get modified by the various 
effects mentioned in the footnotes, is not relevant for the following discussion.

Thus, the whole D3/D7-brane inflation model potential at small $\phi$ (i.e.\
in the regime where inflationary perturbations are generated)
in the notation of \cite{Kallosh:2003ux}, in Planck units, and with
account of stringy corrections from the stabilizing $F$-term as
explained above, is
\begin{equation}
V =\frac {g^{2}\xi^{2}}{ 2}\left(1+\frac {g^{2}}{ 16 \pi^{2}} U\Big(\frac{\phi}{\sqrt{\xi}}\Big)\right) -\frac {m^2}{ 2}\phi^{2} \ , \label{D3D7potentialcorrFull}
\end{equation}
where $U(x)$ is given in \eqref{U}.\footnote{We should mention that, even though we
take the $y_3$-dependent threshold corrections to the gauge coupling in
$W_{np}$ into account, we simplified the analysis at the present stage by ignoring
any $y_3$-dependence of the FI D7-brane gauge
coupling $g$. The latter could originate either from $y_3$-dependent
stringy threshold corrections or from an ${\rm Im}(s)$-dependence of $g$ which,
at the critical point of $s$, induces a dependence on $y_3$, along the lines of
\eqref{s}.}
This should be compared with eq.\ (\ref{D3D7CW}) where the quadratic
term due to stringy corrections was absent.  In (\ref{D3D7potentialcorrFull}), we ignore the
higher orders in $\phi$.

Now we would like to outline the new possibilities,\footnote{Interesting modifications of the supersymmetric hybrid inflation models have been considered in the literature. In \cite{Seto:2005qg,BasteroGil:2006cm} corrections to the \K~potential  were studied which  tend to suppress the spectral index and the cosmic strings contribution. In \cite{Brax:2006yq} a class of quantum corrections to D-term inflation was studied which is due to moduli stabilization and the uplifting procedure  (however, the superpotential was taken to be independent
of the inflaton, in contrast to the situation at hand). In both cases the particular regime of very small coupling and $n_s\approx 1$ was not studied, particularly since these papers came out before \cite{Bevis:2007gh}.}
which are present in the updated version of the D3/D7-brane inflation model when, in addition to the world-volume flux on the D7-brane at fixed point \#1, there is also an attraction\footnote{ We are assuming here that $m^2$ is positive. For negative $m^{2}$, the D3-brane would of course be repelled by the
volume stabilizing D7-branes. } of the mobile D3 towards the fixed point \#2, at which the stack of stabilizing D7's is placed, as shown in Fig.\ 2. Both effects break the shift symmetry: the first one is responsible for the $U(x)$-term in the potential due to the effective FI term $\xi$, the second one is responsible for the negative quadratic inflaton term. In absence of a quadratic term both the height of the potential as well as the deviation from the flatness are due to the gauge coupling $g$ and the FI term $\xi$.

Now we have one more parameter in the problem, since the D7's at fixed point \#2  attract the D3 away from its main motion towards the FI D7 at fixed point \#1, shown in Figure 2. This gives a clear stringy interpretation of each term in the potential of the D3/D7 model in
eq.~(\ref{D3D7potentialcorrFull}) and the simplified version in eq.~(\ref{D3D7potentialcorr}), see below.

Figure \ref{Ramond} illustrates the behavior of the potential in
eq.~(\ref{D3D7potentialcorrFull}) for various values of the constant
$m^2$. The left part of the plot corresponds to the region near the bifurcation
point $\phi = \sqrt{\xi}$ where inflation ends (i.e.\ $\phi/\sqrt\xi = 1$,
as in the graph). The upper curve corresponds to $m^{2} = 0$ and the
lower curves correspond (from top to bottom) to increasing values for
$m^{2}$.
In the lowest case the D3-brane always moves towards the
stack of volume stabilizing D7-branes
at fixed point \#2, which is not the regime that we want as it leads
to a destabilization of the volume.\footnote{Taking into account also the higher powers of $\phi$ in the F-term potential, there could in principle also be a local minimum
for finite $\phi$, which could prevent the D3-brane from eventually reaching the volume stabilizing D7-branes.}

\begin{figure}[tb]
\centering{\includegraphics[height=5cm]{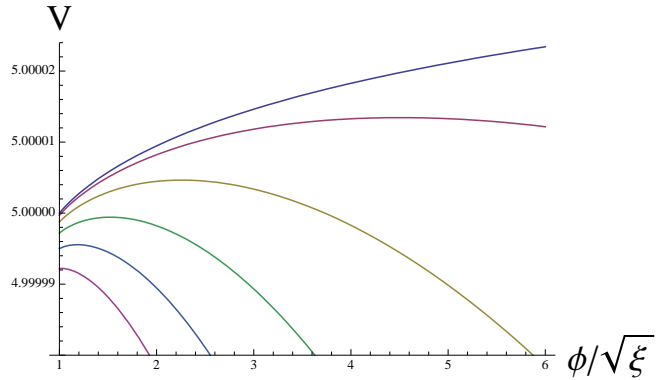}}
 \caption{The inflaton potential $V(\phi)$ including the negative mass term
$-\frac{m^2}{2}\phi^2$, which results from stringy corrections. The
potential is shown in units of $10^{{-17}}$ for a particular case
$g= 10^{-2}$, $\xi = 10^{{-6}}$, for several different values of $m^{2}$.
The upper line corresponds to the case $m^{2} = 0$, i.e.\ to the theory
without stringy corrections. For all sufficiently small $m^{2}$, the
potential acquires a maximum, which allows the regime of eternal inflation
starting from the top of the potential.}
  \label{Ramond}
\end{figure}

For all sufficiently small values of $m^{2}$, the potential has a maximum at some value of $\phi$ (which is supposed to be small, so that eq.\ (\ref{Wcorrected1}) is valid). If we start near the maximum, the potential near the top is approximately quadratic. {\it Inflation near this maximum is eternal } (it is always eternal if inflation occurs near the top of the inflationary potential) \cite{Vilenkin:1983xq}.

The most important fact is that to the left of the maximum the potential is
less steep than the original logarithmic potential, i.e.\ $V'$ is smaller.

Now let us remember that the amplitude of density perturbations is
proportional to  $V^{3/2}/V'$. Note that the corrections practically do
not change $V$, their main role is to change $V'$, which may become
significantly smaller (for example, $V'$ vanishes at the maximum). As a
result, the corrections have a tendency to increase the amplitude of
inflationary perturbations. Meanwhile, they do not directly affect the
string tension, so they do not affect the contribution of cosmic
strings to the density perturbations.

Thus, we find a novel possibility to suppress the cosmic string contribution
as compared to inflationary perturbations. As we will see, this can be done
simultaneously with allowing the spectral index taking a broad range of
values in the range $n_{s}\leq 1$.

To analyze these possibilities, let us investigate the inflationary regime
in the model including stringy corrections.


\subsubsection{Regime A: Inflation far away from the bifurcation point,
$\phi^2\gg \xi$}
\label{regimeA}

For $x\gg 1$ the Coleman-Weinberg potential simplifies and we get
\begin{equation}
V = \frac{g^{2}\xi^{2}}{ 2}\left(1+\frac {g^{2}}{4\pi^{2}} \ln\frac {\phi}{ \sqrt\xi}\right) - \frac{m^2}{ 2}\phi^{2} \ .
\label{D3D7potentialcorr}\end{equation}
In this case an analytic solution of the cosmological evolution is available.
The equations of motion for the field $\phi$ during inflation can be written as follows:
\begin{equation}
3H\dot\phi = -V' = - \frac{g^{4}\xi^{2}}{ 8\pi^{2}\phi} +m^{2}\phi \ .
\end{equation}
The potential has a maximum at
\begin{equation}  \label{max}
\phi_{*}^{2} = \frac{g^{4}\xi^{2}}{ 8\pi^{2} m^{2}}\ .
\end{equation}
 As we would like this maximum to be at $\phi_{*}^{2} > \xi$, we see that
for typical values of $\xi$ and $g$ the value of $m^2$ has to be very small.

The leading contribution to $V$ is given by $\frac{g^{2}\xi^{2}}{ 2}$.
Therefore, the Hubble constant during inflation remains approximately constant,
\begin{equation}
H = \sqrt\frac{V}{ 3} \approx \frac{g\xi }{ \sqrt 6} \ .
\end{equation}
The total number of e-folds of inflation is equal to $N = Ht$, where $t$ is the time since the beginning of inflation. Using this relation, one can represent the equation of motion as follows:
\begin{equation}
V \frac{d\phi}{ dN}  \approx \frac{g^{2}\xi^{2}}{ 2}  \frac{d\phi}{ dN} = -\frac{g^{4}\xi^{2}}{8\pi^{2}\phi} +m^{2}\phi \ .
\end{equation}
The solution to this equation can be written in the following form:
\begin{equation}
\phi_{*}^{2}-\phi^{2} = (\phi_{*}^{2}-\phi_{N}^{2})\  \ e^{2m^{2}N/V}  \ .
\end{equation}
Here $\phi_{N}$ is the initial value of the field starting from which the universe experiences N e-folds of inflation until the field reaches the point $\phi$.

Inflation ends when the field $\phi$ reaches the bifurcation point, $\phi  = \sqrt\xi$, which implies that the total number of e-folds of inflation is determined by the relation
\begin{equation}
N = \frac{V}{ 2m^{2}} \ln\left(\frac{\phi_{*}^{2}-\xi}{ \phi_{*}^{2}-\phi_{N}^{2}}\right)   \ .
\end{equation}
Equivalently, one can find $\phi_{N}$,
\begin{equation}\label{phin}
\phi_{N}^{2} = \phi_{*}^{2}-(\phi_{*}^{2}-\xi) \, e^{-\alpha N} \ ,
\end{equation}
where we have introduced the following notation
\be
\alpha \equiv g^{2}/2\pi^{2}\phi_{*}^{2}=\frac{4m^2}{ g^2\xi^2}\ .
\end{equation}
Using the fact that $\phi_{*}^2\gg \xi$ we find
\begin{equation}\label{phinS}
\phi_{N}^{2} = \phi_{*}^{2}(1- e^{-\alpha N} )\ .
\end{equation}

For comparison with observations, we will need the following expression for
$V^{3/2}/V'$ at $\phi = \phi_N$:
\begin{equation}\label{pertgeneral}
\frac{V^{3/2}}{ V'  }= \frac{2\sqrt 2\,\pi^{2}\, \xi\, \phi_{N}\,
\phi_{*}^{2}}{ g (\phi^{2}_{*} -\phi_{N}^{2})}\approx 5.4 \times 10^{{-4}} \ .
\end{equation}
As discussed at the beginning of section \ref{basic}, this quantity should be equal to
$5.4 \times 10^{{-4}}$ only  if inflationary perturbations are fully responsible for the
CMB anisotropy. This number may be a few percent smaller if strings give some contribution
to the CMB anisotropy.
We will also need an expression for the spectral index
\begin{equation}\label{spectralindex}
1-n_{s} = \frac{g^{2}}{ 2\pi^{2}}\left[\frac{1}{\phi_{N}^{2}}+\frac{1}{ \phi_{*}^{2}}\right] \ .
\end{equation}
The value of $\phi_{*}$ is a function of the parameters $g^{2}$, $\xi$ and $m^{2}$, see eq.\ (\ref{max}). In the limit $m^{2}\to 0$ one returns to the previously studied case without stringy corrections. This corresponds to the regime where $\phi_{*} \to \infty$.
Indeed, one can easily check that one can obtain the results for the case without the string theory corrections by taking the limit $\phi_{*}\to \infty$ in the  expressions  (\ref{phin}), (\ref{pertgeneral}) and (\ref{spectralindex}).
In the opposite limit, when $m^{2}$ is too large, the potential has a maximum at $\phi < \xi$, which does not allow for any inflationary regime.

Since the analytic solution is known we may try to extract the most important properties of this model concerning the string tension and the spectral index.
From eq.\ (\ref{pertgeneral}) we have the value of $\xi$ as follows
\be
\xi =  \frac{2.7 \times 10^{-4}\, \sqrt{ \alpha} \,  e^{-\alpha N}}{\pi  \sqrt {1 - e^{-\alpha N}}}\ ,
\end{equation}
where $N$ can be in the range of 50 to 60. As we already mentioned,  in our estimates we will use, for definiteness, $N = 60$. One can also find an expression for the spectral index
\begin{equation}
n_s=1-\alpha \left (1+\frac{1}{ 1 - e^{-\alpha N}}\right)\ .
\end{equation}
In particular, in the limit $\alpha \rightarrow 0$, i.e.\ in the absence of stringy corrections, we get back to $\xi=10^{-5}$ and $n_s=0.98$ for $N=60$. In general, a wider range of values for $\xi$ and $n_s$ is possible.

For semilocal strings with $p$ Higgs multiplets one finds
\begin{equation}
\xi =  \frac{2.7 \times 10^{-4}\, p \sqrt{ \alpha} \,  e^{-\alpha N p}}{ \pi  \sqrt {1 - e^{-\alpha N p}}}
\end{equation}
and
\begin{equation}
n_s=1-p\, \alpha \left (1+\frac{1}{1 - e^{-\alpha N p}}\right)\ .
\end{equation}
In particular, in the limit $\alpha \rightarrow 0$
we get back to $\xi=1.5 \times 10^{-5}$ and
$n_s=0.98$ for $N=60$ as in the case without stringy corrections (cf. eq. (\ref{slnumbers})).
Again, the
dependence on $\alpha$ allows for more possibilities for the
values of $\xi$ and $n_s$.
 \begin{figure}[tb]
\centering{\includegraphics[height=5.2cm]{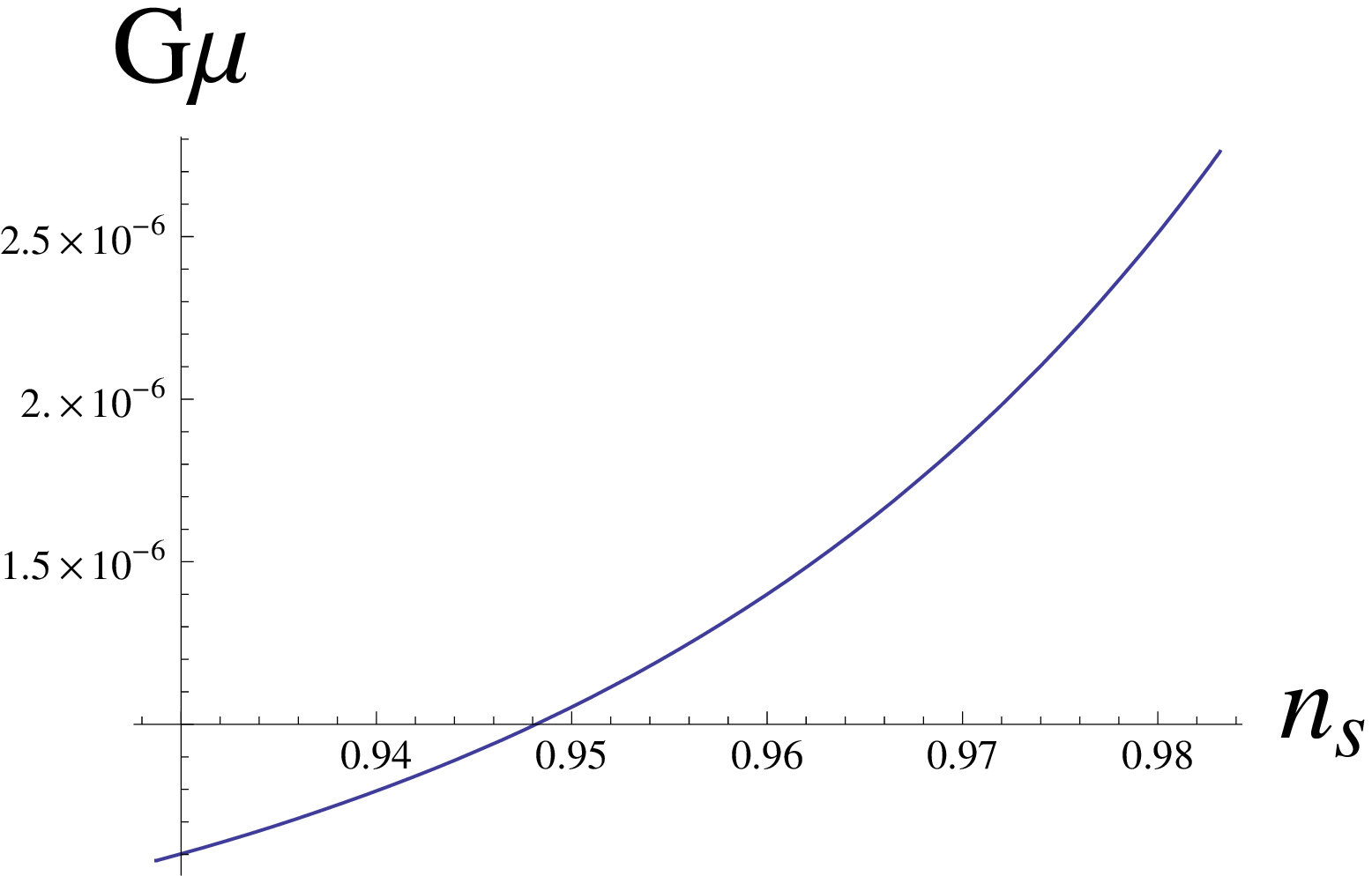}   \hskip 0.3cm   \includegraphics[height=5.2cm]{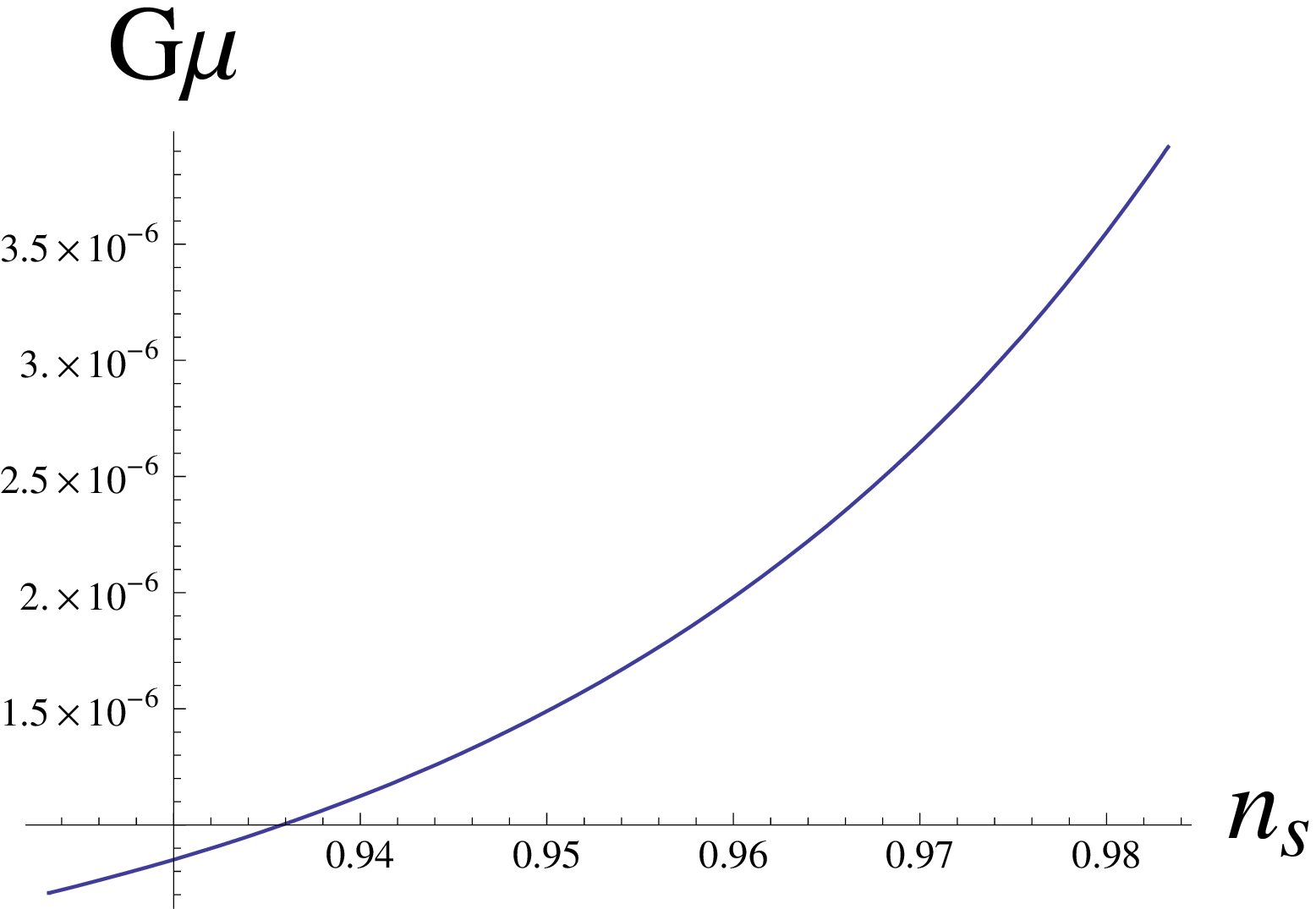} }
 \caption{A parametric plot for the cosmic string tension versus the spectral index. On the left figure we plot the case of local cosmic strings, on the right figure the case of semilocal ones with $p=2$.}
  \label{parametric}
\end{figure}

For both cases we made parametric plots for the values of the string tension versus the spectral index, see Fig. \ref{parametric}.
One can try to compare the results of these models to \cite{Bevis:2007gh} for local Abelian strings or for the semilocal ones in \cite{Urrestilla:2007sf}.
In both cases, inclusion of string theory corrections allows to obtain a broad spectrum of values of $n_{s}$. It also allows to reduce the string contribution to the amplitude of perturbations, but only for relatively small values of $n_{s}$. As a result, it is hard to make the theory of local strings compatible with observations in this regime. The situation with semi-local strings is similar, but it is a bit better; a further analysis of the cosmological constraint on semilocal strings along the lines of Ref.  \cite{Urrestilla:2007sf} would help to reach a final conclusion in this respect.

We should note that the potential advantage of this regime is that it can occur for
relatively large $g$, and the results do not depend on $\phi_{*}$ for
$\phi_{*} \gg \sqrt \xi$. However, at very large $\phi$ one may need to take into
account the string theory corrections to the potential proportional to higher powers
of $\phi$.  Fine-tuning the shape of the full potential may allow to decrease the
cosmic string contribution for realistic values of $n_s$.
We leave this possibility for further investigation.


\subsubsection{Regime B: Inflation near the bifurcation point, $\phi^2\approx \xi$}
\label{near}

For $x\approx 1$ we have to use the complete potential in eq.~(\ref{D3D7potentialcorrFull}).
Leaving a full analysis of the model for the future, in the regime B with stringy corrections we will consider the case when  the quadratic term becomes significant and leads to formation of the maximum of the potential very close to the bifurcation point. It is interesting to see what kind of cosmology one should expect here. A qualitative analysis of the potential suggests that one can try to suppress the cosmic string tension simultaneously with suppressing the value of the spectral index. This would be an attempt to explain the CMB data without cosmic strings.

Let us remember that, as we have found in Section 7.1, if one ignores string theory corrections (i.e. considers the case $m^{2} = 0$), one can have the cosmic string contribution at the level of 11\% for $\xi = 2.8\times 10^{-6}$, $g = 2.15 \times 10^{{-4}}$, with $n_{s} = 0.997$. According to \cite{Bevis:2007gh}, this regime is consistent with observations.

The contribution of cosmic strings to the amplitude of the CMB perturbations, in the limit when this contribution is small, is proportional to $\xi$. Therefore the contribution of the cosmic strings will become negligibly small if, for example, instead of $\xi = 2.8\times 10^{-6}$ one considers the model with $\xi = 2.8\times 10^{-7}$. If, as before, one takes $m^{2} = 0$, and solves numerically the FRW equations for $\xi= 2.8\times 10^{-7}$, $G\mu= 7\times 10^{-8}$ with $g=5.7\times 10^{-6}$, one finds a correct amplitude of perturbations  and the spectral index $n_s=0.9999$. Once again, it is  close to 1,  but in this case it is disfavored by the WMAP5 data due to the absence of the cosmic string contribution.

 On the other hand, by a slight increase of $m^{2}$ one can easily suppress cosmic strings and simultaneously  reduce $n_{s}$ down to $0.95 - 0.97$ or even further. Consider, e.g.,  a case with  $\xi= 5\times 10^{-7}$, $G\mu= 1.2 \times 10^{-7}$, $g=4\times 10^{-4}$ and $m^2=1.65\times 10^{-22}$. \footnote{ Note that the value $g=4\times 10^{-4}$ is actually not that small compared to $g=2  \times 10^{-3}$, i.e., this case could also be considered as being in the intermediate region between regimes A and B. Our calculations in this paragraph, however, take this into account in that they do not use
 the simplifying approximations that define regime B, but instead use a  numerical solution of the full FRW equation. }   In this case the contribution of cosmic strings to the CMB anisotropy is negligible. The potential has a maximum at $x=1.08$. The 60 e-foldings take place when the system evolves from the position at $x=1.065$, not far from the maximum of the potential. One finds that the value of fluctuations computed via $V^{3/2}/V'$ at this point is in agreement with the data. Moreover, the value of the spectral index turns out to be $n_s= 0.945$. By a slight change of parameters, one can  obtain somewhat bigger or smaller values of $n_{s}$.

 Thus, by a proper choice of the parameters of our D3/D7 model in which the stringy corrections from the stack of D7's are taken into account one can control the amplitude of the perturbations while allowing the spectral index and the contribution of the cosmic strings to take a broad range of values consistent with the existing cosmological observations. Moreover, as we already mentioned in the previous subsection, inflation in this simple model is eternal.

In this paper we have restricted ourselves to the two regimes A and B,
either far away from the bifurcation point or very close to it and we
made a first analysis of the corresponding cosmology. A more detailed
and thorough study is postponed to the future. In fact, the intermediate
regimes may also be interesting. It would also be worthwhile
studying the situation
where the string corrections can not just be approximated
by a quadratic term. This might be particularly relevant in any
situation where the initial position of the mobile
D3-brane is at large distance from the FI D7-brane. As
observed in section \ref{inflaton}, this might be possible
on an asymmetric torus.


\subsubsection{The issue of fine-tuning}
\label{fine}

One should note that our model, just as many other models of string inflation, requires significant fine-tuning. In particular, in order to construct the model with an 11\% contribution of cosmic strings and $n_s\approx 1$ one must take a very small value of the coupling constant $g$ and of the parameter $m^{2}$, which parametrizes the strength of the stringy corrections quadratic in $\phi$ (note, however, that due to the exponential prefactor in (\ref{msquare}), the fine-tuning issue for $m^{2}$ might not be such a severe problem in the present context).  Moreover, if $m^{2}$ is extremely small (or if  $\phi$ is of order one), one may additionally need to fine-tune stringy corrections which are proportional to $\phi^{4}$, etc.\ (a more detailed analysis of this issue can be found in appendix \ref{appquartic}). On the other hand, many features of our model do not depend on the fact that stringy corrections were quadratic in $\phi$. The main idea was that these corrections may decrease the value of $V'$ near the bifurcation point, and therefore increase the relative contribution of inflationary perturbations as compared to perturbations produced by cosmic strings.

There are two other considerations which may render the required fine-tuning a bit more
natural. First of all, suppose that the term $-m^{2}\phi^{2}/2$ is unsuppressed and big.
This would imply that the mobile D3-brane would run towards the stabilizing D7-branes (unless there is a local minimum of the potential at finite distance due to the higher powers of $\phi$),  which
would destabilize our 4D world and make our universe 10 dimensional. Life as we know it
cannot exist in such a universe. Therefore, the most natural regime consistent with the
existence of our life would be the regime where $m^{2}$ is smaller than some critical
value, so that the stabilizing maximum of the potential at $\phi > \sqrt\xi$ still exists.
On the other hand, it would seem unnatural for $m^{2}$ to be much smaller than this value.
In other words, we must have a maximum of the potential at $\phi > \sqrt\xi$, but it would
be unnatural to have it at $\phi \gg \sqrt\xi$. This singles out the regime which we
discussed in section \ref{near} of our paper.

An interesting aspect of our construction is related to the possibility of a slow roll eternal inflation in string theory. Although the slow roll eternal inflation appears in most of the models of modular inflation, see e.g.\ \cite{BlancoPillado:2004ns,Linde:2007jn}, there  was a conjecture that eternal inflation is generically absent in the brane inflationary scenario \cite{Chen:2006hs}. From our results it follows, however, that in the D3/D7 model with stringy corrections inflation is eternal.  A similar conclusion was reached in \cite{Linde:2007jn} with respect to the recent version of the KKLMMT scenario proposed in \cite{Baumann:2007np,Baumann:2007ah}. Slow roll eternal inflation  also occurs in the model proposed in \cite{Iizuka:2004ct}, and it may be possible in the model of ref.\ \cite{Silverstein:2008sg}. The issue of the proper choice of the probability measure in models involving eternal inflation is still unsettled, but it is interesting that by using a certain class of probability measures one may conclude that the existence of  eternal  inflation and a long stage of slow roll inflation increases the probability of inflation  in the landscape even if it requires fine-tuning, see e.g.\ \cite{Linde:2007jn} for a recent discussion.


\subsubsection{Tensor-to-scalar ratio}

The relative magnitude of tensor modes is described by the tensor-to-scalar ratio $r$. In slow-roll inflation $r$ is directly related to the potential via
\begin{equation}
V^{1/4} = 3.3 \times 10^{16}\, r^{1/4}\ \text{GeV} \; .
\end{equation}
In $M_P^2=1$ units the potential to leading order is $V\approx \frac{g^2\xi^2}{2}$ and we obtain
\begin{equation}
r \approx 1.5 \times 10^7\, g^2\,\xi^2 \; .
\end{equation}
With a cosmic string contribution of less than 11\%, we must have $\xi < 2.8 \times 10^{-6}$, which implies
\begin{equation}
r < 10^{-4}\, g^2 \; .
\end{equation}
The projected experimental sensitivity for the next decade lies in the regime\, $r\gtrsim 0.01-0.001$. For $g<1$, the level of tensor modes in our model is  below this bound.


\subsubsection{On reheating after inflation and entropy fluctuations}
Until now, we discussed the dynamical evolution of the inflaton field, but we did not consider the evolution of the fields $\phi_{\pm}$ which are responsible for  spontaneous symmetry breaking after inflation, as well as for the Coleman-Weinberg corrections (\ref{U}) to the inflaton potential.
The evolution of the fields $\phi_{\pm}$ becomes important at the stage of reheating after inflation. In addition, these fields could contribute to isocurvature (entropy) perturbations because of their long-wavelength quantum fluctuations during inflation. We are going to discuss these two issues briefly.

During inflation $\phi_\pm=0$. For these values, the masses of the fields $\phi_{\pm}$ 
depend on the inflaton field according to \cite{Binetruy:1996xj}:
\be \label{mpm}
m_{\pm}^{2} = g^{2}(\phi^{2} \mp \xi) \ .
\ee
After the inflaton field reaches the bifurcation point $\phi^{2} = \xi$, the mass squared of the field $\phi_{+}$ becomes negative, and the process of spontaneous symmetry breaking begins due to the tachyonic instability with respect to the growth of fluctuations of the field $\phi_{+}$. 

This process is rather nontrivial because of the combination of two different effects. First of all, the tachyonic mass $m_{+}^{2}$ is time-dependent (cf.\ eq.\ \eqref{mpm}).  Moreover, for small $g$, the fields $\phi$ and $\phi_{+}$ after a short period of adjustment  `fall down from the cliff'   along the straight line  $ \phi_{+}/\sqrt 2 + \phi = \sqrt\xi$ \cite{BasteroGil:1999fz,Dasgupta:2004dw}. One can easily check that the potential along this line is not quadratic but cubic with respect to the deviation from the bifurcation point. For both of these reasons, one cannot describe the growth of tachyonic fluctuations by the simple rule $\delta \phi_{+} \sim e^{g\sqrt\xi\, t}$, as it is done, e.g., in \cite{Brandenberger:2008if}.

Another feature of this process is even more important. Originally, many people visualized the process of spontaneous symmetry breaking as a rolling of a homogeneous scalar field down to the minimum of the potential, and a subsequent long stage of oscillations of this field with an amplitude which slowly decreases due to particle production. However, the theory of spontaneous symmetry breaking and reheating in hybrid inflation  \cite{Felder:2000hj,Felder:2001kt} and in new inflation \cite{Desroche:2005yt} shows that this process typically occurs in an entirely different way.

For example, during the tachyonic instability in the process of spontaneous symmetry breaking in the model with a quadratic tachyonic potential (e.g.\ in the Higgs model), all modes with momenta $k < |m_{+}|$  exponentially grow, $\delta \phi_{+} \sim e^{\sqrt{|m_{+}^{2}|-k^{2}}\, t}$. The rate of growth only weakly depends on $k$ for $k \ll |m_{+}|$. As a result, at the moment when the growing field reaches the minimum of the potential, it does not look like a homogeneous oscillating classical field, but like a collection of colliding classical waves of the scalar field with typical momenta $k \sim |m_{+}|$. Analytical investigation of this issue accompanied by lattice simulations shows that the field distribution typically experiences just a single oscillation before it relaxes near the minimum of the potential \cite{Felder:2001kt}. Similar results are valid for a cubic potential \cite{Felder:2001kt}, and for spontaneous symmetry breaking during reheating in hybrid inflation \cite{Felder:2000hj}. This effect precludes the process of a narrow parametric resonance described in \cite{Brandenberger:2008if}. For a more detailed discussion  of reheating in D3/D7 hybrid inflation see ref.\ \cite{Dasgupta:2004dw}.

As we already mentioned, there is another potentially important effect associated with the field $\phi_{+}$: If this field has a small mass during inflation, $m_{+}^{2} \ll H^{2}$, then inflationary fluctuations of this field can be generated. These fluctuations, under certain conditions, may contribute to the total amplitude of metric perturbations after inflation, and may alter the initial conditions for reheating. This could affect some details of the theory of reheating in D3/D7 hybrid inflation \cite{Brandenberger:2008if}. 

At first glance this could seem a reasonable possibility  because $m_{+}^{2}$ vanishes near the bifurcation point $\phi^{2} = \xi$. Let us, however, consider this issue more accurately.  

The condition required for generation of fluctuations of the field $\phi_{+}$ can be written as follows: 
\be
m_{+}^{2} =  g^{2}(\phi^{2} - \xi) \ll H^{2} = g^{2}\xi^{2}/6\ .
\ee
 This condition is satisfied for
\be
\phi^{2} - \xi \ll \xi^{2}/6 \ .
\ee
Consider for simplicity the basic inflationary regime (\ref{infl3}), ignoring for a moment string theory corrections. In this case the condition given above, combined with (\ref{infl3}), yields
\be\label{N60}
g \ll \sqrt{\frac{\pi^{2} }{3N}}\ \xi  \sim \xi/4 \ ,
\ee
where we took $N \sim 60$.
If one is interested in perturbations produced during the last e-fold of inflation (i.e. take $N = 1$), the constraint changes a bit,
\be\label{N1}
g \ll     2\xi  \ .
\ee

However, in all   versions of the D3/D7 model studied in our paper the constant  $g$ is much greater than  $2\xi$, so the conditions (\ref{N60}), (\ref{N1}) are not satisfied. Therefore, $|m_{+}|$ is much greater than $H$ everywhere except in the immediate vicinity of the bifurcation point. A similar result is valid for all other regimes of D3/D7 inflation which we studied in our paper, including string theory corrections.

 We conclude that in all versions of the D3/D7 scenario studied in our paper reheating occurs not in the narrow resonance regime, but in the regime of  tachyonic preheating described in \cite{Felder:2000hj,Dasgupta:2004dw}, and no long-wavelength entropy perturbations related to inflationary fluctuations of the field $\phi_{+}$ are generated in this scenario. Such perturbations may become possible if one considers anomalously small values of $g$, or introduces additional light degrees of freedom to the D3/D7 model discussed above.


\section{Conclusions and outlook}
\label{concl}
In this paper, we started a careful re-investigation of the
D3/D7 inflationary model on $K3\times T^{2}/\mathbb{Z}_{2}$ with a focus on the effects of quantum corrections and volume stabilization.
As described in the introduction, this model provides a valuable laboratory for probing, in a well-controlled setting, various features that are shared by many of the contemporary string theoretical
models of inflation. In particular we discussed the tension between moduli stabilization and slow-roll inflation and  the possible role of cosmic strings. As for the latter, we pointed out that a particular parameter regime
(regime B of the introduction), in which the model has 11\% of cosmic
strings and $n_s=1$  might be  interesting  given the recent analysis of data in \cite{Bevis:2007gh,Pogosian:2008am}.

Starting from the observation \cite{Hsu:2003cy} that the theory has, at the level of $\mathcal{N}=2$ supergravity with a cubic prepotential, an inflaton shift symmetry that could protect the inflaton mass from large corrections due to volume stabilization, we took a closer look at the fate of that shift symmetry when various effects  are taken into account.

The first of these effects we studied is the spontaneous partial supersymmetry breaking $\mathcal{N}=2 \rightarrow \mathcal{N}=1$ induced by the bulk three-form fluxes.
We verified in several relevant cases that the partial supersymmetry breaking
comes along with the emergence of holomorphicity of the gauge couplings.   Furthermore, we observed how the shift symmetry along the real part of the D3-brane coordinate $y_3$ in the $\mathcal{N}=2$ theory (based on the cubic prepotential) is passed down to the resulting effective $\mathcal{N}=1$ supergravity, where it is encountered as a shift symmetry of the D7-brane gauge coupling and the K\"{a}hler potential.

The $\mathcal{N}=1$ theory so obtained, however, is only an
approximation to the full effective action, and further corrections
break the shift symmetry.
In order to quantify these corrections and their effect on the shift
symmetry, we investigated which quantum corrections are already
captured by the $\mathcal{N}=1$ theory that descends from the
$\mathcal{N}=2$ theory with the cubic prepotential. Using the
Green's function method of refs.\
\cite{Giddings:2005ff,Baumann:2006th,Baumann:2007ah} we established
the proper relation between the 4D and 10D variables and identified
the part of the (closed string dual of the) open string one-loop
corrections that is captured by the cubic prepotential. The remaining
part it does not capture was found to reproduce results
from direct open string computations \cite{Berg:2004ek}, even though
the latter
were not performed in a background with RR-fluxes. This confirms
earlier observations \cite{Giddings:2005ff,Baumann:2006th} and
quantifies
the violations of the inflaton shift symmetry due to the presence
of the stack of the volume stabilizing D7-branes.

As a result of the breaking of the shift symmetry, the
real part of $y_3$ is not
necessarily preferred over the imaginary part as a flat
direction in the scalar potential, and it becomes, in general,
a matter of fine-tuning the threshold corrections (e.g., by dialing
a suitable vev of the complex structure, $t$, of the torus) in order
to preserve the original flatness along $\textrm{Re}(y_{3})$.
Whether and in which cases this is actually possible is a complicated question which we
only started to discuss.

We found that there is a $t$-dependence of the range of the canonically normalized D3-brane field, which
allows for an unexpected enhancement of its range
for asymmetrical tori. At present, the question whether the possible
range of the canonically normalized inflaton
(candidate) is smaller or larger than $M_P$
seems to depend on the geometric properties of the compactification space.
 For example, in the warped  throat geometry there is a bound \cite{BM} which allows
only small fields. On the other hand, in the recent work
 \cite{Silverstein:2008sg} it was found that large fields are possible due to the
monodromy of D-branes when the compactification is performed on Nil manifolds with
negative curvature in type IIA string theory. In our compactification of type IIB
string theory on a $K3\times T^{2}/\mathbb{Z}_{2}$
orientifold, we also find the possibility of
large values for the canonical field, under the condition that  the torus geometry is
very asymmetrical. This is possible due to the special feature of this model that the
volume of the torus is in a hypermultiplet, decoupled from the vector multiplets.
It is interesting that in  \cite{Silverstein:2008sg} the asymmetry of the manifold
was also important for the existence of the large field range. It would be worthwhile 
to analyze whether the large kinematical field range in the D3/D7-model can indeed
be realized dynamically in a stage of slow roll inflation. Until now we 
only looked into the phenomenology of the model in the small field regime (cf.\ section
\ref{cosmology}).

We also gave an outlook on a more concrete study of the most promising brane setups
and listed a number of additional effects that might be important for a complete analysis.
We believe that this model provides a versatile and controllable testing ground for
many of the features discussed in string theoretical models of inflation. The first steps towards a realistic cosmology have been made in
the context of a brane configuration presented in Fig.\ 2 where the
mobile D3-brane is attracted towards the FI D7-brane at one of the
fixed points of the pillow. Due to the effective FI terms (world-volume
fluxes on D7) there is a small breaking of the shift symmetry
providing a standard D-term hybrid inflation attractive
potential for the D3-brane in the direction of the FI D7. A new
feature of the model discussed in section \ref{cosmology} is the presence of
the stabilizing D7-branes at
another fixed point of the pillow: we find that it takes part in the
breaking of the shift symmetry.  We considered the case where it
effectively adds a negative mass squared term to the standard Coleman-Weinberg potential
controlled by the FI term in D-term inflation.  This mass term may be tuned
by the choice of fluxes. The two effects together allow for more flexibility in
adjusting the values of the cosmic string tension and the spectral index.

 In the case that the attraction from the stack of stabilizing D7-branes is negligible,  one finds a simple model consistent with \cite{Bevis:2007gh} with an 11\% contribution of cosmic strings and $n_s\approx 1$. There are several phenomenological models based on supergravity where this regime may occur, see e.g.\ \cite{Kallosh:2003ux}, but we are unaware of any other version of a string inflation model which has this property. On the other hand, by a slight increase of the string theory corrections one can strongly suppress the cosmic string contribution and reduce the spectral index $n_{s}$ from 1 down to 0.95 - 0.97, which provides a good fit to the WMAP5 data in the absence of cosmic strings.

\ 


{\bf \large Note Added:}

Recently the authors of \cite{Burgess:2008ir} also revisited the D3/D7-model and found another possible realization of D3/D7-brane inflation that has some similarities with the racetrack scenario \cite{BlancoPillado:2004ns}. They also analyzed the D-term inflationary scenario studied in our paper but did not find any sets of parameters which would lead to the stage of slow roll inflation of the kind we were describing in section \ref{cosmology} and in particular in \ref{near}. The situation can be explained as follows.

One could easily find the regime similar to the standard D-term inflation by suppressing the string theory corrections \eqref{paper2}. This could be achieved, for example, by making the overall factor $Ae^{-ias_2}$ very small. However, this term is responsible for the depth of the F-term minimum and, thus, a large enough value is important for volume stabilization; one cannot make it too small without vacuum destabilization during inflation \cite{Kallosh:2004yh}. The second possibility is to keep the overall factor large, but expand the $\phi$-dependent part of $V_F$ for small values of $\phi$, i.e.\ close to the bifurcation point. This was the strategy pursued in our paper. Ideally one would like to have a situation where the quadratic term in $\phi$ is the dominant contribution albeit with a sufficiently small mass to allow for a flat inflaton potential. For an extremely small mass term, however, higher powers of $\phi$ may also become relevant, with the $\phi^4$-term being the most important one. Thus, as was emphasized in section \ref{fine}, one must verify that the higher powers in $\phi$, in particular the $\phi^{4}$-term, are indeed subleading. Whereas we were focusing on the inflationary regime near the bifurcation point with $\phi \sim 10^{-3}$, the authors of \cite{Burgess:2008ir} concentrated on the regime with relatively large $\phi$, where the terms $\phi^{4}$ are generically much more dangerous. They did not find any parameters for which one would simultaneously have a deep enough F-term minimum and a small quadratic and even smaller quartic term.

However, the authors of \cite{Burgess:2008ir} emphasized that their search was broad but not exhaustive.  In the present version of the paper we added Appendix \ref{appquartic} in which we show that, for the regime B discussed in section \ref{near}, there are indeed points in the moduli space of the complex structure, where both the quadratic and the quartic term of the inflaton F-term potential can be made arbitrarily small simultaneously. At the same time the volume of the K3 can be in a range that leads to a hierarchy between the depth of the F-term potential and the height of the inflaton potential. This ensures that the volume is not destabilized during inflation. We conclude that there is no inconsistency between our results and the results of \cite{Burgess:2008ir}. Rather, the analyzed regimes are different, and having the inflaton very close to the bifurcation point can open up new possibilities for D3/D7 inflation. A first step in this direction was made in section \ref{cosmology}. 



\

\begin{center}
{\bf Acknowledgments}
\end{center}

We would like to thank Ana Achucarro, Richard Battye, Marcus Berg, Ralph Blumenhagen,
Dick Bond, Massimo Bianchi, Cliff Burgess, Jim Cline, Aurelien Fraisse, Amihay Hanany, Arthur Hebecker,
Mark Hindmarsh, Shamit Kachru, Igor Klebanov, Lev Kofman, Daniel Krefl,
Peter Mayr, Liam McAllister, Viatcheslav Mukhanov, Rob Myers,
Hans-Peter Nilles, Lyman Page, Enrico Pajer, Joe Polchinski, Marieke Postma, Misao Sasaki,
Eva Silverstein,  David Spergel,  Stephan Stieberger, Mario Trigiante,
Sandip Trivedi, Angel Uranga, Vitaly Vanchurin and Antoine Van Proeyen
for helpful discussions or email correspondence. This work is supported
in part by the European Community's Human Potential Programme under
contract MRTN-CT-2004-005104 ``Constituents,
fundamental forces and symmetries of the universe'', the Excellence Cluster
``The Origin and the Structure of the Universe'' in Munich and the
Transregional Collaborative Research Centre TRR 33 ``The Dark Universe''.
The work of M.~H. and M.~Z. is supported by the German
Research Foundation (DFG) within the Emmy-Noether-Program (grant numbers:
HA 3448/3-1 and ZA 279/1-2). The work of R.~K. and A.~L. is supported by
the National Science Foundation grant 0244728, by the Alexander Von
Humboldt Award.  They are grateful for the hospitality extended to
them at LMU, Munich and at YITP, Kyoto where parts of this work were performed.


\section{APPENDIX}

\begin{appendix}

\section{Theta functions}
\label{thetaf}

Due to the mostly minus signature convention for the metric that we
use throughout (which implies negative values for the
complex structure modulus $t_2$), we use a different definition of
the theta functions compared to the usual one which is defined in the
upper half plane of the complex structure modulus
(cf.\ chapter 7 of \cite{Polchinski:1998rq}).
In particular we use
\begin{equation}
\thba{a}{b} (\nu,t) = \sum_{n=-\infty}^{\infty} \exp \Big[ - \pi i (n+a)^2 t - 2 \pi i (n + a)(\nu + b) \Big]\ ,
\end{equation}
which is related to the usual one by complex conjugation
(and subsequent renaming $\bar \nu \rightarrow \nu$ and
$\bar t \rightarrow t$ in order to comply with our definition of
$\nu$ and $t$ using the mostly minus convention).
This relation to the usual definition of the theta function has
to be taken into account when comparing with formulas from
chapter 7 of \cite{Polchinski:1998rq} and appendix A of
\cite{Kiritsis:1997hj}, for instance. To translate their formulas
involving theta (or eta) functions to
our conventions, one has to take the complex conjugate of the
corresponding formula
and afterward rename $\bar \nu \rightarrow \nu$ and
$\bar t \rightarrow t$.

For $\vartheta_1 = - \thba{1/2}{1/2}$ this implies the
$SL(2,\mathbb{Z})$ transformation
\begin{eqnarray}
\vartheta_1(\nu,t+1) &=& \exp(-i \pi/ 4) \vartheta_1(\nu,t)\ , \\
\vartheta_1(\nu/t,-1/t) &=& i (i t)^{1/2} \exp(-i \pi \nu^2/t) \vartheta_1(\nu,t) \label{newsl2z}\ .
\end{eqnarray}
Moreover, the Dedekind eta function in our conventions has the transformation
properties
\begin{eqnarray}
\eta(t+1) &=& \exp(-i \pi/ 12) \eta(t)\ , \\
\eta(-1/t) &=& (i t)^{1/2} \eta(t) \label{newsl2zeta}\ .
\end{eqnarray}

Let us now review the periodicity of the theta functions
(in particular of $\vartheta_1$) concerning their first argument.
Using formulas (7.2.32a), (7.2.32b) and (7.2.36) of Polchinski, one
can determine\footnote{This is also consistent with formula (A.13) of
\cite{Kiritsis:1997hj}, if one takes into account that the $a$ and $b$
of Kiritsis are minus two times the ones of Polchinski.}
\begin{eqnarray} \label{period}
{\thba ab} (\nu+1,t) & = & e^{-2 \pi i a} {\thba ab} (\nu,t)\ , \non
{\thba ab} (\nu+t,t) & = & e^{\pi i t + 2 \pi i (\nu + b)}
{\thba ab} (\nu,t)\ .
\end{eqnarray}
Specializing this to the case of $\vartheta_1$, i.e.\ $a=b=1/2$, one obtains
\begin{equation} \label{theta1period}
\vartheta_1(\nu+1+t,t) = - \thba{1/2}{1/2}(\nu+1+t,t)
= - e^{2 \pi i \nu + \pi i t} \thba{1/2}{1/2}(\nu,t)
= e^{2 \pi i \nu + \pi i t} \vartheta_1(\nu,t)\ ,
\end{equation}
where the minus sign in the definition of $\vartheta_1$
is conventional and we again follow the conventions of Polchinski (cf.\
(7.2.37d)).

One can now use equations \eqref{period} and
\eqref{theta1period}, the fact that $\vartheta_1$ is
an odd function of its first argument (i.e.\ $\vartheta_1(-\nu,t) =
- \vartheta_1(\nu,t)$) and the expansion
\begin{equation} \label{expandtheta1}
\vartheta_1(\nu,t) = 2 \pi \eta(t)^3 \nu + {\cal O}(\nu^3)
\end{equation}
for small $\nu$, in order to expand $\vartheta_1$ around the points
$1+t, 1$ and $t$ with the result
\begin{eqnarray} \label{expand}
\vartheta_1(1+t-\nu,t)
& = & - e^{-2 \pi i \nu + \pi i t} \vartheta_1(\nu,t)
= - 2 \pi \eta(t)^3 e^{\pi i t} \nu + {\cal O}(\nu^2)\ , \non
\vartheta_1(1-\nu,t)
& = & \vartheta_1(\nu,t) = 2 \pi \eta(t)^3 \nu + {\cal O}(\nu^3)\ , \\
\vartheta_1(t-\nu,t)
& = & e^{-2 \pi i \nu + \pi i t} \vartheta_1(\nu,t) =
2 \pi \eta(t)^3 e^{\pi i t} \nu + {\cal O}(\nu^2)\ . \nonumber
\end{eqnarray}

We also give the expansion around
the points $1/2$, $t/2$ and $1/2+t/2$, which can be determined
from (A.13) and (A.25) - (A.27) of \cite{Kiritsis:1997hj}. This
leads to
\begin{eqnarray}
&& \vartheta_1(1/2+t/2-\nu,t)
= e^{-\pi i \nu + \pi i t/4} \vartheta_3(\nu,t) \\
&& \hspace{.5cm} = e^{\pi i t/4} \vartheta_3(0,t)
\Big(1 - i \pi \nu -\frac12 \pi^2 \nu^2 -
\frac{\pi^2}{6} (E_2(t) + \vartheta_2^4(0,t) - \vartheta_4^4(0,t)) \nu^2 \Big)
+ {\cal O}(\nu^3)\ , \non
&& \vartheta_1(1/2-\nu,t)
= \vartheta_2(\nu,t) \label{1/2} \\
&& \hspace{.5cm} = \vartheta_2(0,t) \Big(1 -
\frac{\pi^2}{6} (E_2(t) + \vartheta_3^4(0,t) + \vartheta_4^4(0,t)) \nu^2 \Big)
+ {\cal O}(\nu^3)\ ,  \non
&& \vartheta_1(t/2-\nu,t)
= - i e^{- \pi i \nu + \pi i t/4} \vartheta_4(\nu,t)\\
&& \hspace{.5cm} = - i e^{\pi i t/4} \vartheta_4(0,t)
\Big(1 - i \pi \nu -\frac12 \pi^2 \nu^2 -
\frac{\pi^2}{6} (E_2(t) - \vartheta_2^4(0,t) - \vartheta_3^4(0,t)) \nu^2 \Big)
+ {\cal O}(\nu^3)\ , \nonumber
\end{eqnarray}
where $E_2(t)$ is the second Eisenstein series.


\section{Emergence of holomorphicity after
partial SUSY breaking:  }
\label{Apphol}
In this appendix, we discuss a few more representative
examples for the restoration of the holomorphicity of the gauge couplings under partial supersymmetry breaking. We also give an example for the case of unbroken $\mathcal{N}=2$ supersymmetry, where holomorphicity in
general does not arise. The case numbers here refer
to the classification in section \ref{secholomorphic3}. The last example
refers to another class of $\mathcal{N}=1$ vacua with more
general values for the stabilized moduli fields (i.e stabilized vevs of $t$ and $u$ different from $-i$).

\

\subsection{ Case 2: $\mathcal{N}=2  \rightarrow  \mathcal{N}=1$ for
$g_0,g_1,g_2,g_3 \neq 0$}
\label{Apphol1}

\

We now consider the case when $A_{\mu}^{0,1}$ gauge the two
spacelike directions
$C^{m=1,2}$ and $A_{\mu}^{2,3}$ gauge the timelike
directions $C^{a=1,2}$. According to the classification of \cite{Tripathy:2002qw}, which we reviewed in section \ref{secholomorphic}, this
is expected to yield an $\mathcal{N}=1$ vacuum as well.
The Killing spinor equations again lead to  equations
(\ref{conditions}) for the moduli $(t,u,y_7)$. In this
case, the holomorphicity of the kinetic matrix simply
follows from the holomorphicity of case 1 discussed in section \ref{secholomorphic3},
as we just have to restrict the attention to
 the surviving gauge couplings $\mathcal{N}_{44,45,55}$,
given that the other vectors $A_{\mu}^{0,1,2,3}$ all get
masses comparable to  the second (i.e., the massive)
gravitino. This leads to the following gauge couplings
\begin{eqnarray}
 \mathcal{N}_{44}&=&-\bar{s} \label{coupling1b}\ ,\\
\mathcal{N}_{45}&=&0\ ,\\
\mathcal{N}_{55}&=& -i\ ,
\end{eqnarray}
 which are again all antiholomorphic, as desired.


We finally note that the above vev for the axion-dilaton $u$ is phenomenologically not very attractive, because it corresponds to  a large string coupling. In Appendix \ref{Apphol3},
we give a short summary of more general gaugings that may lead to more general values for the moduli $u,t$ \cite{D'Auria:2004td}.

\

\subsection{ Case 3: Unbroken  $ \mathcal{N}=2 $ SUSY for
$g_2,g_3\neq 0$}\label{Apphol2}

\

In order to illustrate that the emerging holomorphicity of the gauge kinetic matrix
is really non-trivial and requires the partial breaking $\mathcal{N}=2 \rightarrow \mathcal{N}=1$, we briefly consider also the case when $\mathcal{A}_{\mu}^{2}$ and $A_{\mu}^{3}$ gauge two timelike directions $C^{a=1}$ and $C^{a=2}$, and nothing else is gauged. According to \cite{Tripathy:2002qw}, this should lead to $\mathcal{N}=2$ vacua, and the constraints on the moduli are found to be \cite{D'Auria:2004qv}:
\begin{eqnarray}
t&=&u\ , \\
1+t^2&=&\frac{(y_7)^{2}}{2} .
\end{eqnarray}
Plugging these values into the kinetic matrix for the remaining vector fields,
$A_{\mu}^{0,1,4,5}$, one finds highly non-holomorphic dependencies on the surviving moduli. As an illustration, the real and imaginary parts of the component $\mathcal{N}_{45}$ are given by
\begin{eqnarray}
\textrm{Re}\mathcal{N}_{45}&=& f(u,t,y_7) \textrm{Im}(y_3)\\
\textrm{Im} \mathcal{N}_{45}&=& g(u,t,y_7) \textrm{Im}(y_3),
\end{eqnarray}
with two complicated functions $f$ and $g$. This is not an antiholomorphic function of $y_3$,
as long as $f$ and $g$ are nonvanishing.



\

\subsection{$\mathcal{N}=1$ vacua with generic stabilized $t, u$ values}
\label{Apphol3}

\

The gauged supergravity vacua described in Section \ref{secholomorphic3} and the above cases 2 and 3 are the simplest ones, but they may not be very useful for physical applications (if $u=-i$).  As mentioned above, they involve large dilaton vevs corresponding to a string coupling of order one, but they also allow only for a quadratic torus without any freedom to dial its complex structure.

\mbox{}From the stringy analysis in \cite{Tripathy:2002qw}, however, we know that also other stabilized values of $t$ and $u$ are allowed. In fact, the analysis of \cite{Tripathy:2002qw} (see also \cite{LMRS}) leads to a discretuum of possible vevs of $t$ and $u$ that depend on the chosen background fluxes (cf., e.g., eqs. (4.4) and (4.5) in \cite{Tripathy:2002qw}, where their $\phi$ and $\tau$
correspond to our $u$ and $t$).

It was shown in \cite{D'Auria:2004td} how these more general vevs of $u$ and $t$  (the condition $y_7=0$ remains valid in the analysis in \cite{D'Auria:2004td}) can also be obtained in the 4D gauged supergravity framework.
More precisely, the authors of \cite{D'Auria:2004td} show how to obtain arbitrary vevs for $t$ and $u$,
\begin{equation}
t=a_t-i e^{2\lambda_t}\ , \qquad u=a_u-i e^{2\lambda_u},
\end{equation}
where $a_t, a_u, \lambda_t, \lambda_u$ are generic real numbers.\footnote{A discretuum of possible vevs is obtained in supergravity only if one takes into account the flux quantization conditions, which translate to suitably quantized gauge parameters in the gauged supergravity theory.} To this end, one has to go to a different symplectic duality frame by means of a particular type of  $SU(1,1)_t \times SU(1,1)_u$
isometry. The specific form of the transformation matrix is given in
\cite{D'Auria:2004td}, in eqs.\ (35)-(38).
To achieve such a stabilization at generic points in the
moduli space for $t$ and $u$, one needs the presence of all four charges (cf. section \ref{secholomorphic3}),
\begin{eqnarray}
 g_{0}&:=&q_{0}^{m=1}\ , \nonumber\\
   g_{1}&:=&q_{1}^{m=2}\ , \nonumber\\
   g_{2}&:=&q_{2}^{a=1}\ , \nonumber\\
   g_{3}&:=&q_{3}^{a=2}\ ,
 \end{eqnarray}
which have to be rotated to
\begin{equation}
q_{\Lambda}^{\prime \, I}=  (A_\Lambda{}^\Sigma)(a_t, a_u, \lambda_t, \lambda_u)
q_{\Sigma}^{I}\ .
\label{general}
\end{equation}
Here, as indicated, the matrix $(A_\Lambda{}^\Sigma)$  depends on the desired vevs $(a_t, a_u, \lambda_t, \lambda_u)$.
 The explicit form of
the matrix $(A_\Lambda ^\Sigma)(a_t, a_u, \lambda_t, \lambda_u)$
is given in eq.\ (38) of \cite{D'Auria:2004td}. After this transformation of the charges $q_{\Lambda}^{I}$, there are
$4\times 4$ gauge parameters, instead of the original four, which corresponds to a more general choice of fluxes. Of course, in order to really correspond to a real string theory setup, one has to make sure that these new fluxes satisfy all consistency conditions (such as quantization and tadpole conditions).

What remains true for all the above gaugings is that $y_7=0$ is still a requirement for unbroken $\mathcal{N}=1$ supersymmetry. Surprisingly, this is sufficient to render the surviving gauge couplings $\mathcal{N}_{44,45,55}$ antiholomorphic (note that these gauge couplings are not changed by the symplectic rotation, as that rotation only changes the $(0,1,2,3)$-directions). To demonstrate how non-trivial this emergence of holomorphicity is, we present here the non-holomorphic gauge couplings in the 4,5 directions before the $\mathcal{N}=1$ condition is inserted \cite{D'Auria:2004qv}:
\begin{eqnarray}
 \mathcal{N}_{44}&=& -s_1 -\frac{4u_2 (y_7)_1 (y_{7})_2 (y_3)_{2}^{2}}{(-2t_2 u_2+(y_7)_{2}^{2})^2}\nonumber\\ + && \!\!\!\!\!\!\!\!\!\!\!\!\!\!\! i \frac{s_2(2t_2u_2-(y_7)_2^2)(2t_2u_2+2(y_7)_1^2+(y_7)_2^2)+2u_2(-(y_7)_1+(y_7)_2)((y_7)_1+(y_7)_2)(y_3)_2^2}{(-2t_2u_2+(y_7)_2^2)^2}\ , \nonumber\\
 \mathcal{N}_{45}&=& \frac{2u_2(y_7)_1(y_3)_2}{-2t_2u_2+(y_7)_2^2}  +i \frac{2u_2(y_7)_2(y_3)_2}{2t_2u_2-(y_7)_{2}^{2}}\ , \nonumber\\
 \mathcal{N}_{55} &=& -\bar{u}\ .
 \end{eqnarray}
Here, the subscripts 1 and 2 denote, respectively, the real and imaginary parts of the moduli.

Clearly these are complicated non-holomorphic functions of the moduli. At $(y_7)_1=(y_7)_2=0$ these three entries of the coupling matrix do not depend on $t, y_{3}$ and collapse to
\begin{eqnarray}
 \mathcal{N}_{44}&=&-\bar{s}\\
\mathcal{N}_{45}&=&0\\
\mathcal{N}_{55}&=& - (a_u+i e^{2\lambda_u}).
\end{eqnarray}
Hence, also in this case, the couplings become (anti-)holomorphic, and the shift symmetry of the D7-brane gauge coupling is preserved.


\section{From $D=10$ to $D=4$}
\label{10to4}

In this appendix we would like to identify the definition
of the variable ${\rm Im}(s)$ (cf.\ \eqref{sdef1}) in terms
of 10-dimensional quantities. To this end, we consider the D7-brane
gauge coupling and the D3-brane kinetic term arising from a
dimensional reduction of the D7-brane and D3-brane DBI
actions using the metric \eqref{metric1}.
In our analysis we follow closely \cite{Giddings:2005ff}
and, in particular, \cite{Baumann:2006th,Burgess:2006cb,Baumann:2007ah}.

\subsection{D7-brane gauge coupling}
\label{d7gc}

For the warp factor we consider the split \eqref{hexp}
and, for the moment, we will
 concentrate on its dependence on a single D3-brane with coordinates
 $(\hat y_3^m,\hat y_3^i)$ on $K3\times T^2/\mathbb{Z}_{2}$
(keeping the positions of the other
 D3- and D7-branes fixed):
\begin{equation} \label{deltah}
 \delta h = \delta h (\hat y_3^m,\hat y_3^i;x^m,x^i)\ .
 \end{equation}
In order to distinguish the D3-brane coordinates from the
(dimensionless) supergravity coordinate $y_3$
of the last chapter, we used the notation with a hat.
The relation between the dimensionful and the dimensionless
coordinates involves a conversion factor, $\C$, of length
dimension one,
\begin{equation}
\hat{y}_{3}=\C y_{3}\ . \label{conversion}
\end{equation}
The $\delta h$ of \eqref{deltah} integrates to zero over
the whole internal manifold (cf. eq. (64) in
\cite{Baumann:2006th}) and hence does not contribute to the 4D Planck mass
(cf. eq. (\ref{kappa4d}) below, where only the zero mode
$h_{0}$ enters $\kappa_{4}^{2}$).

The Einstein equation implies a Poisson equation for the perturbation
$\delta h$, which incorporates the D3-brane as a  point source
and includes the background charge density $\rho_{bg}$
necessary to satisfy the
Gauss law constraint on a compact manifold.
Let us first consider the resulting equation on the covering
torus (times $K3$) and later
consider its symmetrization under the orbifold involution.
On $K3\times T^2$,
the Poisson equation is
\begin{equation}
\Big( \nabla^{2}_{x^m} + \nabla^{2}_{x^i}  \Big)\,\,
\delta h = C \Big[ \rho_{bg}(x^m,x^i) -
 \frac{\delta^{(2)}(\hat y^{i}_{3}-x^i)}{\sqrt{g^{T^2}(x^i)}}
 \frac{\delta^{(4)}(\hat y_{3}^{m}-x^m)}{\sqrt{g^{K3}(x^m)}} \Big]\ .
\label{L1}
\end{equation}
Here, (cf.\  the expression  below eq. (67) in \cite{Baumann:2006th})
 \begin{equation} \label{C}
 C=  2\kappa_{10}^2T_3= (2\pi)^4g_s(\alpha^{\prime})^{2}\ ,
 \end{equation}
 where
 \begin{eqnarray}
 \kappa_{10}^{2}&=& \frac{1}{2}(2\pi)^{7} g_{s}^2 (\alpha^{\prime})^{4}
\label{kappa10d}
 \end{eqnarray}
 is the $10$D gravitational coupling,
 and the Laplace operators are with respect to the metrics
$g_{mn}$ and $g_{ij}$,
 which do not include the warp factor, but do contain the
breathing modes (the same
 holds for the determinants  $g^{T^2}(x^i)$ and $g^{K3}(x^m)$).
 For consistency, the background charge density  has to satisfy
 \begin{equation}
 \int_{K3}d^4 x^{m}\sqrt{g^{K3}}\int_{T^2}d^2 x^i \sqrt{g^{T^2}}
\rho_{bg} =1\ .
 \end{equation}
 Integrating both sides of (\ref{L1}) over $\int_{K3}\sqrt{g^{K3}}d^4x^{m}$,
 one obtains a two-dimensional Laplace equation,
\begin{equation}
\nabla_{x^i}^{2} \hat{\delta h} = C \Big[ \hat{\rho}_{bg}(x^i)
- \frac{\delta^{(2)}(\hat y_{3}^{i} -x^{i})}{\sqrt{g^{T^2}(x^{i})}} \Big]\ ,
\label{L3}
\end{equation}
 where
 \begin{eqnarray}
 \hat{\delta h}(\hat y_3^i;x^i) & \equiv& \int_{K3} d^4 x^m \sqrt{g^{K3}}
\delta h\ , \label{hatdh}\\
 \hat{\rho}_{bg}(x^{i})&\equiv&  \int_{K3}d^4 x^m \sqrt{g^{K3}} \rho_{bg}\ .
 \end{eqnarray}
Following now similar steps as the ones  that lead
from eq.\ (18) to eq.\ (22) in \cite{Baumann:2006th} (the numbers refer to
the version on spires), one derives
\begin{equation} \label{lapltorus}
\nabla_{\hat y_3^{i}}^{2} \hat{\delta h} = C \Big[ \frac{1}{V^{T^2}}
- \frac{\delta^{(2)}(\hat y_{3}^{i} -x^{i})}{\sqrt{g^{T^2}(\hat y_3^{i})}}
\Big]\ .
\end{equation}
This can be transformed to a Poisson equation in complex
dimensionless coordinates\footnote{In
order not to overload the notation, we use the same,
unhatted, notation $x$ also for the
dimensionless torus coordinates. Only for the D3-brane
position moduli we make the distinction
between dimensionful and dimensionless explicit.
Depending on whether a formula contains
$y$ or $\hat y$, the occurring $x$ is also dimensionless
or dimensionful, respectively.}
\begin{equation}
y_3 = y_3^1 + t y_3^2\ , \quad x = x^1 + t x^2\ ,
\end{equation}
leading to
\begin{equation} \label{laplcomplex}
\partial_y \partial_{\bar y} \, \hat{\delta h} = - C \Big[ \frac{\C^2}{4 \lambda^2 t_2} + \frac12 \delta^{(2)}(y_{3} -x) \Big]\ .
\end{equation}
Here we assumed that the dimensionful $x^1$ and $x^2$ (and thus also $\hat y_3^1$
and $\hat y_3^2$) range from $0$ to $\lambda$, where $\lambda$ denotes an a priori
arbitrary parameter of dimension length (this $\lambda$ is, of course,
not the same as the one used in \eqref{s};
we will come back to its value in the next section,
or rather to the value
of the dimensionless parameter $\lambda/\C$).
Thus, the volume of the torus appearing in \eqref{lapltorus} is given by
\begin{equation}
V^{T^2} = \sqrt{g^{T^2}} \lambda^2\ .
\end{equation}
The two-dimensional Laplace equation \eqref{laplcomplex} is very similar to
the one in eq.\ (7.2.1) of Polchinski's book \cite{Polchinski:1998rq}. The
different sign of the first term on the
right hand side is due to the different metric signature (leading
to negative values for $t_2$ in our case, cf.\ fn.\ \ref{fnconventions}).
The solution is, therefore,
\begin{equation}
\hat{\delta h} = - \frac{C \C^2}{2 \lambda^2} \frac{[\textrm{Im}(y_{3}-x)]^{2} }{\textrm{Im}(t)}
- \frac{C}{4 \pi} \ln \left| \vartheta_1 \Big( \frac{\C (y_{3}-x)}{\lambda} \Big| t \Big) \right|^2\ + \ldots\ ,
\label{hatexp}
\end{equation}
where the ellipsis stands for terms not depending on $y_3$. The additional terms
are necessary to render the
integral of $\delta h$ over the whole internal space vanishing and $\lambda/\C$ is the
range of the dimensionless variables $y_3$ and $x$.
Using
\begin{equation}
\frac{C}{2 \pi} = \frac{1}{T_3}
\end{equation}
and defining the holomorphic function
\begin{equation} \label{zeta}
\zt (y_{3},t) = \ln \vartheta_{1}(\tfrac{\C}{\lambda} y_{3},t)\ ,
\end{equation}
$\hat{\delta h}$ takes the form
\begin{equation} \label{hatexp2}
\hat{\delta h} = - \frac{\pi \C^2}{T_3 \lambda^2} \frac{[\textrm{Im}(y_{3}-x)]^{2} }{\textrm{Im}(t)}
- \frac{1}{T_3} \textrm{Im}\, (i \zt(y_3-x,t))+ \ldots \ .
\end{equation}

We are now ready to express the D7-brane gauge coupling in
terms of the above higher-dimensional quantities. To this end,
we insert (\ref{hexp}) into (\ref{couplingg}) and obtain
\begin{equation}
g^{-2}=  \left[ \frac{T_{3}h_{0} \tilde{V}^{K3}}{2(2\pi)^2} \right]
 e^{4U_{1}} + \frac{T_{3}}{2(2\pi)^{2}}
\hat{\delta h}(y_{3}^i,x^i=y_{7}^i) +\ldots \ , \label{g2}
\end{equation}
where, as indicated, $\hat{\delta h}$ (cf.\  eq.\ (\ref{hatdh})) is evaluated at the
point $x^i=y_{7}^{i}$ on the two-torus where the D7-brane is located and the ellipsis
stands for terms depending on the positions of the other D3- and D7-branes that we ignored
in the above and terms that do not depend on any brane positions at all, but just on $t$.\footnote{The
other D-branes contribute to the background
charge $\rho_{{\rm bg}}$ in \eqref{L3}. One could take them into account more explicitly by
adding several delta function sources on the right hand side.
This would lead to a sum of terms like \eqref{hatexp2}.
At the orbifold point of $K3$ and without fluxes  one could calculate these contributions
using CFT methods \cite{Berg:2004ek}. This also shows that there are additional
contributions to the 1-loop gauge coupling which do not depend on any
D-brane moduli and, thus, can not be obtained by the present Green's function method.
Moreover, if the K3 is at the orbifold point and the D3-branes are at fixed points of the $K3$ orbifold there are
contributions from exchanges of twisted strings which are also not captured by the
Green's function method. We assume in the present discussion that
we are at a position in moduli space where these additional corrections can be
neglected (for example, the D3-branes could be away from the fixed points
of the $K3$ orbifold).}

Inserting now (\ref{hatexp2}) into (\ref{g2}) and symmetrizing with respect to
the orbifold involution $x \rightarrow -x$ (following the appendix of \cite{Antoniadis:1996vw},
in particular their eq.\ (A.3); cf.\ also \cite{Burgess:2006cb}), we obtain
\begin{eqnarray}
g^{-2}&=& \left[ \frac{T_{3}h_{0} \tilde{V}^{K3}}{2 (2\pi)^2} \right]
e^{4U_{1}} - \frac{\C^2}{8 \pi \lambda^2}
\Big( \frac{[\textrm{Im}(y_{3}-y_7)]^{2} }{\textrm{Im}(t)}
+ \frac{[\textrm{Im}(y_{3}+y_7)]^{2} }{\textrm{Im}(t)}
\Big) \nonumber \\
&&- \frac{1}{2 (2\pi)^2} \Big( \textrm{Im}(i \zt(y_3-y_7,t))
+ \textrm{Im}(i \zt(y_3+y_7,t))
\Big) + \ldots \nonumber \\
&=&\left[ \frac{T_{3}h_{0} \tilde{V}^{K3}}{2 (2\pi)^2} \right] e^{4U_{1}} - \frac{\C^2}{4 \pi \lambda^2}
\Big( \frac{[\textrm{Im}(y_{3})]^{2} }{\textrm{Im}(t)} + \frac{[\textrm{Im}(y_7)]^{2} }{\textrm{Im}(t)}
\Big) \nonumber \\
&& - \frac{1}{2 (2\pi)^2} \Big( \textrm{Im}(i \zt(y_3-y_7,t)) + \textrm{Im}(i \zt(y_3+y_7,t))
\Big) + \ldots\ .\label{gexp2}
\end{eqnarray}

In order to proceed we have to have some information about the additional terms
left out in \eqref{gexp2}, coming from the other branes.
It was noticed before (cf.\
\cite{Berg:2005ja} and
footnote 1 in \cite{Antoniadis:1997gu})
that the overall scalars $\sum_r y_3^r$ and $\sum_k y_7^k$ are special (here
$r$ and $k$ denote the different (stacks of) D3- and D7-branes, respectively).
They correspond to massive scalars (in the T-dual picture with D9/D5-branes they would
correspond to Wilson-line moduli of anomalous 6-dimensional
vectors \cite{Antoniadis:1997gu}).
In \cite{Antoniadis:1997gu} they were just set to zero and also in
\cite{Berg:2005ja} it was noticed that they had to be set to zero in order
to allow to express the K\"ahler potential in terms of a prepotential. Imposing
these conditions in the open string 1-loop calculation of \cite{Berg:2005ja},
i.e.\footnote{Of course this condition only has to
hold modulo a lattice vector of the covering torus.}
\begin{equation} \label{centreofmass}
\sum_r y_3^r=0\ , \quad \sum_k y_7^k=0\ ,
\end{equation}
leads to (non-Abelian) D7-brane gauge couplings whose only dependence
on the D7-brane scalars is via the arguments of $\zt$, i.e.\ the terms linear and quadratic in
$y_7$ arising from the $[{\rm Im} (y)]^2$-terms (including those from the other D-branes
hidden in the ellipsis of \eqref{gexp2}) cancel out (cf.\ eqs.\ (43) and (44) in \cite{Berg:2005ja}).
We will assume in the following that the same would happen in the case at hand if all
branes were properly taken into account. In any case, in the next subsection we will
consider the kinetic term of the D3-brane scalars for vanishing $y_7$ only. Thus, strictly
speaking we can only make a strong statement about that case anyway.

Now, defining\footnote{We stress again that the term $\sim [{\rm Im}(y_3)]^2$ would
be replaced by a sum $\sum_r [{\rm Im}(y_3^r)]^2$ over the D3-brane coordinates if one takes into
account the other D3-branes as well.}
\begin{equation}
-\textrm{Im}(s) := \left[ \frac{T_{3}h_{0} \tilde{V}^{K3}}{2 (2\pi)^2} \right]  e^{4U_{1}} - \frac{\C^2}{4\pi \lambda^2} \frac{[\textrm{Im}(y_{3})]^{2} }{\textrm{Im}(t)}\ , \label{sdef}
\end{equation}
the gauge coupling is given by a holomorphic function
of the moduli $(s,y_3,y_7,t)$,
\begin{equation}
g^{-2}=\textrm{Im}  \left[ - s - \frac{i}{2 (2\pi)^2}  \Big(\zt (y_{3}-y_7,t) + \zt (y_{3}+y_7,t) \Big)  \right]  +\ldots \ .\label{g3}
\end{equation}
We will verify in the next section that the definition \eqref{sdef} is indeed
consistent with the expression (\ref{KP}) for the K\"{a}hler potential.


\subsection{K\"{a}hler potential}
\label{seckaehler}

In order to verify the consistency of \eqref{sdef}
with the K\"{a}hler potential (\ref{KP}), we perform a reduction of the
D3-brane DBI action and read off the kinetic term for the D3-brane
position modulus $y_3$ on $T^{2}/\mathbb{Z}_{2}$, again following closely
\cite{Baumann:2007ah}.
As a byproduct, we determine the
field range of the variable $y_{3}$, which is important for the application to
inflation in section \ref{inflaton}. For simplicity, we again
only consider the case of a single
D3-brane. Moreover, we set $y_7^k = 0$.

For the dimensional reduction of the D3-brane DBI action, we again use the
metric \eqref{metric1}.
We parametrize the fiducial metric $\tilde{g}_{ij}$ on the torus as
\begin{equation}
\tilde{g}_{ij}dx^idx^j=\frac{\tilde{V}^{T^2}}{\lambda^{2}t_{2}}dxd\bar{x}\ . \label{torusfid}
\end{equation}
Remembering that on the covering torus $x^1$ and $x^2$ range from $0$ to $\lambda$, it is
easy to check that $\tilde{V}^{T^2}$ is indeed
the fiducial volume of the two-torus with respect to the metric
$\tilde{g}_{ij}$:
\begin{equation}
\tilde{V}^{T^2}=\int_0^\lambda d x^1  \int_0^\lambda dx^2 \sqrt{\det \tilde{g}_{ij}}\ .
\end{equation}
%
As in the last section, we will use
\begin{equation}
\hat{y}_{3}= \hat{y}_3^1+t\hat{y}_3^2
\end{equation}
to denote the D3-brane position on $T^2/\mathbb{Z}_{2}$.
Inserting now the metric (\ref{metric1}) into the DBI action,
one obtains a kinetic term for the D3-brane scalars,
\begin{equation}
 -\frac{T_3}{2} \int d^4x \sqrt{\det{G^{4D}}}\ G^{4D,\, \mu \nu}(G_{mn}^{K3}\partial_{\mu}\hat{y}_3^m \partial_{\nu}\hat{y}_3^n + G_{ij}^{T^{2}} \partial_{\mu}\hat{y}_{3}^{i}\partial_{\nu}\hat{y}_{3}^{j})\ .
\end{equation}
Ignoring the D3-brane positions, $\hat{y}_{3}^{m}$, on $K3$ and using
eqs.\ (\ref{metric1}), (\ref{conversion}) and (\ref{torusfid}), this becomes
\begin{equation}
- T_3 \int d^4x \sqrt{\det{\tilde{g}_{\mu \nu}}}\ e^{-4 U_1}\frac{\tilde{V}^{T^2} \C^2}{2 t_2\lambda^{2}}
\tilde{g}^{\mu \nu}\,
\partial_\mu  y_{3} \partial_\nu \bar y_{3}\ . \label{kineticD3}
\end{equation}
Note that the warp factor has dropped out of the kinetic term.

The consistency of this result
with the gauged supergravity approach of Section \ref{secholomorphic}
requires that the kinetic term for $y_3$ should be reproduced by the kinetic term
derived from the K\"{a}hler potential
\begin{equation}
K=-\ln [(t - \bar t) (s - \bar s) - \frac12 (y_{3} - \bar y_{3})^2] -\ln [u - \bar u]\ ,
\end{equation}
which is, up to irrelevant additive constants, the K\"ahler potential
following from \eqref{prep} for  $y_7^k = 0$.
The above K\"{a}hler potential leads to
\begin{equation} \label{kyy}
K_{y_3 \bar y_3} = - \frac{1}{(t - \bar t) (s - \bar s) - \frac12 (y_{3} - \bar y_{3})^2} + \ldots \ ,
\end{equation}
where the ellipsis stands for a term proportional to $\textrm{Im}(y_3)^2$ which
is suppressed by an additional power of the denominator appearing
in \eqref{kyy} (and which arises from the annulus level instead of the
disk level we are interested in when comparing with the DBI action).
Neglecting that term, the kinetic term for $y_{3}$ then becomes
\begin{equation}
\frac{1}{\kappa^{2}_{4}} \int d^4 x \sqrt{\det(\tilde{g}_{\mu\nu})}
\tilde{g}^{\mu\nu} \frac{\partial_{\mu}y_{3}\partial_{\nu}
\bar{y}_{3}}{4\textrm{Im}(t)\textrm{Im}(s) - 2[\textrm{Im}(y_{3})]^{2}}\ , \label{kinscalars}
\end{equation}
where $\tilde{g}_{\mu\nu}$ denotes the 4D Einstein frame metric and
\begin{equation}
\kappa_{4}^{2} = M_P^{-2} =  \frac{2 \kappa_{10}^{2}}{h_{0}\tilde{V}^{K3} \tilde{V}^{T^2}} \label{kappa4d}
\end{equation}
is the $4$D gravitational coupling.
The extra factor of $2$ arises due to the
orientifolding ($\tilde{V}^{T^2/\mathbb{Z}_2}=\tilde{V}^{T^2}/2$).

Comparing \eqref{kineticD3} with \eqref{kinscalars} we read off
\begin{equation}
- e^{4 U_1}  \frac{2 t_2\lambda^{2}}{T_{3} \tilde{V}^{T^2}\C^{2}} =
\kappa_{4}^{2} [4 \textrm{Im}(t) \textrm{Im}(s) - 2 [\textrm{Im}(y_3)]^2]
\end{equation}
or
\begin{equation} \label{s2}
- \textrm{Im}(s) = \left[ \frac12 T_{3}^{-1}\kappa_{4}^{-2} \frac{\lambda^{2}}{\tilde{V}^{T^2}\C^{2}} \right] e^{4 U_1} - \frac12 \frac{[\textrm{Im}(y_{3})]^2}{\textrm{Im(t)}}\ .
\end{equation}
This indicates that $\textrm{Im}(s)$ is basically
given by the breathing mode of $K3$ corrected by
a quadratic term in the D3-brane scalars. The latter is (the non-harmonic)
part of the
backreaction of the D3-branes as calculated in our equation (\ref{sdef}).

For this to be true, however, we should now compare (\ref{s2}) with our equation
(\ref{sdef}) from the gauge coupling calculation (for $y_7=0$).
To this end, we need the following identity
 \begin{equation}
 T_{3}^{-1}\kappa_{4}^{-2}= \frac{T_{3}h_{0}\tilde{V}^{K3}\tilde{V}^{T^2}}{2 \pi}\ .
 \end{equation}
Equation (\ref{s2}) then becomes
\begin{equation}
-\textrm{Im}(s)= \left[ \frac{T_3 h_{0}\tilde{V}^{K3}\lambda^{2}}{4 \pi\C^{2}}  \right] e^{4U_1} - \frac12 \frac{[\textrm{Im}(y_{3})]^2}{\textrm{Im}(t)}\ . \label{sresult}
\end{equation}
Comparing this with the expression (\ref{sdef}),
\begin{equation}
-\textrm{Im}(s) =  \left[ \frac{T_{3}h_{0} \tilde{V}^{K3}}{2(2\pi)^2} \right]
e^{4U_{1}} - \frac{\C^2}{4\pi \lambda^2}
\frac{[\textrm{Im}(y_{3})]^{2} }{\textrm{Im}(t)} \label{sdef4}
\end{equation}
we see that we need to impose the relation
\begin{equation}
 \frac{\lambda^{2}}{\C^{2}}=\frac{1}{2 \pi}\ . \label{lambdac}
 \end{equation}
This then means that the range of the dimensionless supergravity fields $y_{3}^{1}$ and $y_3^{2}$ is
\begin{equation} \label{range}
y_{3}^{1,2}\in[0,\sqrt{\frac{1}{2 \pi}})\ .
\end{equation}
%


\section{Duality symmetries}
\label{dualsym}

In appendix \ref{10to4}, we saw that the 4D,
$\mathcal{N}=2$ supergravity theory that derives from the cubic
prepotential (\ref{prep}) encodes part of the D3-brane backreaction
effects, such as, e.g., the quadratic term
in $\textrm{Im}(y_{3})$ in the D7-coupling (\ref{gexp2}), which, in (\ref{prep}),
is included implicitly via the definition of the variable $s$ in eq.\
(\ref{sdef}), cf.\ also \cite{Antoniadis:1996vw,Berg:2004ek,Giddings:2005ff,Baumann:2006th}.
Other backreaction  effects, however, such as the last
two terms in (\ref{gexp2}), are \emph{not} captured by the cubic prepotential
(\ref{prep}). Nevertheless, taken all by itself, the
theory based on (\ref{prep}) still describes a
perfectly consistent 4D, $\mathcal{N}=2$ supergravity theory, so one might
hope that it can be used as an expansion for large $s$ and $u$. It is the
purpose of this section to show that,
while the restriction to (\ref{prep}) is consistent
with $\mathcal{N}=2$ supersymmetry,
it is \emph{not} consistent with the stringy duality symmetries.


\subsection{10D perspective}
Type IIB string theory on $K3\times T^{2}/\mathbb{Z}_{2}$
features a number of discrete duality symmetries. In this section we discuss
three duality groups, which we denote by $SL(2,\mathbb{Z})_{s}$,
$SL(2,\mathbb{Z})_{t}$ and $SL(2,\mathbb{Z})_{u}$, respectively.

The group $SL(2,\mathbb{Z})_u$ is just the usual IIB S-duality group
relating strong and weak string coupling, whereas $SL(2,\mathbb{Z})_{s}$
corresponds to a T-duality group associated with the size of $K3$.
The group $SL(2,\mathbb{Z})_t$, finally,
describes modular transformations of the two-torus, i.e.\ conformal
transformations that preserve its complex structure.
Physical quantities should not depend on how the torus is parametrized
and should thus be consistent with the $SL(2,\mathbb{Z})_t$-duality
group. In the following, we will focus on this discrete
reparametrization invariance of the torus.

Parametrizing the complex plane by $x \in\mathbb{C}$, a two-torus
with complex structure $t=t_1+it_2\in \mathbb{C}$  is defined in the
usual way via the identification
\begin{equation}
x\sim x+ m \sqrt{2 \pi}^{-1}\ , \qquad  x \sim x +nt\sqrt{2 \pi}^{-1}\
, \qquad m,n\in \mathbb{Z}\ ,
\end{equation}
where we use the periodicity $\sqrt{2 \pi}^{-1}$ in agreement
with our findings of the last section, cf.\ \eqref{range}.
A convenient way to label points on the torus is via the real coordinates
$(x^{1},x^{2})\in [0,\sqrt{2 \pi}^{-1})\times [0,\sqrt{2 \pi}^{-1})$
defined by
\begin{equation}
x=x^1+tx^2 \qquad \Leftrightarrow \qquad \textrm{Re}(x)=x^1+t_1 x^2\ , \qquad
\textrm{Im}(x)=t_{2}x^{2}\ .
\end{equation}
As is well known, tori whose complex structure differ by an
$SL(2,\mathbb{Z})_t$ transformation according to
\begin{equation}
t\rightarrow \tilde{t}=\frac{at+b}{ct+d}\ , \qquad \left(
\begin{array}{cc}a&b\\c&d \end{array} \right) \in SL(2,\mathbb{Z})_{t}
\end{equation}
are conformally equivalent. The group $SL(2,\mathbb{Z})_t$
is generated by the inversion
\begin{equation}
\alpha: t \rightarrow \tilde{t}=-\frac{1}{t}\
\end{equation}
and the Dehn twist
\begin{equation}
\beta: t \rightarrow \tilde{t}=t+1\ .
\end{equation}
It acts on the coordinates $x$ of a point on the torus as
\begin{equation} \label{xtrafo}
x \rightarrow \tilde{x}=\frac{1}{ct+d}x
\end{equation}
or, equivalently,
\begin{equation}
\left( \begin{array}{c}x^1\\x^2 \end{array} \right) =\left( \begin{array}{cc} d & b \\
c & a \end{array} \right) \left(\begin{array}{c} \tilde{x}^{1}\\
\tilde{x}^{2} \end{array} \right) . \label{generaltrafo}
\end{equation}

The transformations of $SL(2,\mathbb{Z})_{t}$ do not change the shape of the torus
and just correspond to the infinitely many equivalent ways to
parametrize one and the same torus. Physical quantities should
accordingly be $SL(2,\mathbb{Z})_{t}$-invariant (at least as long as
the backreaction leaves the torus intact, i.e.\ as long as the branes and
fluxes only modify the warp factor). We will now verify
to what extent this is true for the low energy effective action that
derives from the cubic prepotential (\ref{prep}).


\subsection{No branes, only D3-branes, only D7-branes}
In the absence of branes, the cubic prepotential (\ref{prep})
reduces to $\mathcal{F}=stu$, which describes the scalar manifold
\begin{equation}
\mathcal{M}_{V}\cong \left(\frac{SU(1,1)}{U(1)}\right)_s \times
\left(\frac{SU(1,1)}{U(1)}\right)_t \times \left(\frac{SU(1,1)}{U(1)}\right)_u
\end{equation}
with the isometry group $Iso(M_{V})\cong SU(1,1)_{s}\times
SU(1,1)_{t} \times SU(1,1)_{u}$, which contains the stringy duality group
$SL(2,\mathbb{Z})_s \times SL(2,\mathbb{Z})_t \times SL(2,\mathbb{Z})_u$
as an obvious subgroup. These duality symmetries act on one
of the scalars $s,t,u$ by fractional linear transformations
and, at least in the case without branes, leave the other two scalars invariant:
\begin{equation}
SL(2,\mathbb{Z})_{t}: \qquad s\rightarrow s\ ,
\qquad t\rightarrow \frac{at+b}{ct+d}\ , \qquad u\rightarrow u\ , \label{SLt}
\end{equation}
and analogously for $SL(2,\mathbb{Z})_{s,u}$.

At the level of the 4D, $\mathcal{N}=2$ supergravity theory, all
$SL(2,\mathbb{Z})$-dualities
are realized as particular, symplectic transformations of the symplectic
vector $(X^{\Lambda},F_{\Sigma})$ $(\Lambda,\Sigma=0,1,2,3)$ followed
by a possible K\"{a}hler transformation. Using the symplectic section
\eqref{newbasis} (with all brane coordinates omitted), \
$SL(2,\mathbb{Z})_t$ acts as
\begin{equation}
\left( \begin{array}{c} X^{\Lambda} \\ F_{\Sigma} \end{array} \right) \rightarrow
e^{-f} \left( \begin{array}{cc} S^{\Lambda}{}_{\Gamma} & 0 \\
0 & (S^{-1 \, T})_{\Sigma}{}^{\Omega} \end{array} \right) \left( \begin{array}{c}
X^{\Gamma} \\ F_{\Omega} \end{array} \right) , \label{trasym}
\end{equation}
with
\begin{eqnarray}
e^{-f}&=&\frac{1}{(ct+d)}\\
S^{\Lambda}{}_{\Sigma} &=& \frac{1}{2} \left( \begin{array}{cccc} (a+d)
& (-c+b) & (a-d) & (c+b)\\
(-b+c) & (a+d) & (b+c) & (-a+d)\\
(-d+a) & (c+b) & (d+a) & (-c+b)\\
(b+c) & (-a+d) & (-b+c) & (a+d) \end{array} \right) , \label{Sdef1}
\end{eqnarray}
where $f$ describes a K\"{a}hler transformation,
\begin{equation}
K\rightarrow K+f+\bar{f}\ .
\end{equation}
%


If one includes also the D3-branes,  the simple transformation property (\ref{SLt}) of $SL(2,\mathbb{Z})_t$ no longer holds and has to be generalized to also
 involve the D3-brane coordinates. Moreover, these brane coordinates
transform  themselves under $SL(2,\mathbb{Z})_t$, as they are nothing but
coordinates on the torus, whose (complex structure-preserving)
reparametrizations we are considering. For simplicity, we will only consider
one D3-brane coordinate $y_{3}$. The prepotential
\begin{equation}
\mathcal{F}=stu-\frac{1}{2}uy_3^2=u[st-\frac{1}{2}y_3^2]
\end{equation}
then describes the scalar manifold
\begin{equation}
\mathcal{M}_{V}\cong \left(\frac{SU(1,1)}{U(1)}\right)_u
\times \frac{SO(2,3)}{SO(2)\times SO(3)}\ .
\end{equation}
The duality groups $SL(2,\mathbb{Z})_{s,t}$ are now embedded in the
isometry group, $SO(2,3)$, of the second factor, and hence  still do
not act on $u$. However, $SL(2,\mathbb{Z})_{t}$ may now act on $s$
as well as on $y_3$, as the factorization property
of the $(t,s)$-part of the moduli space is lost. To find the new action
of $SL(2,\mathbb{Z})_{t}$ on $s$ and $y_{3}$, one first uses the upper
part of the transformation of the symplectic vector,
\begin{equation}
X^{\Lambda}\rightarrow e^{-f} \tilde{S}^{\Lambda}{}_{\Sigma} X^{\Sigma}\ ,
\end{equation}
where $\tilde{S}^{\Lambda}{}_{\Sigma}$ is now a $(5\times 5)$-matrix,
and $f$ again parametrizes a possible K\"{a}hler transformation.
This is consistent with $t\rightarrow \frac{at+b}{ct+d}$
and $u\rightarrow u$ if
\begin{eqnarray}
f&=&\ln (ct+d)\ , \\
\tilde{S} &=&\left( \begin{array}{cc} S & 0\\
0 & 1 \end{array} \right)\ ,
\end{eqnarray}
with $S$ as in (\ref{Sdef1}).
Using $X^5=y_3$, this in turn implies
\begin{equation} \label{sl2zony3}
y_3 \rightarrow \frac{y_3}{ct+d}\ .
\end{equation}
To infer the transformation property of $s$, one uses that the lower part,
$F_{\Lambda}$, of the symplectic vector transforms as (cf.\ eq.\ (\ref{trasym}))
\begin{equation}
F_{\Lambda}\rightarrow e^{-f} ({\tilde{S}}^{-1 \, T})_{\Lambda}{}^{\Sigma} F_{\Sigma}\ ,
\end{equation}
with the same $f$ and $\tilde{S}$ as above. Using
\begin{equation}
S^{-1 \, T}=  \frac{1}{2} \left( \begin{array}{cccc} (a+d) & (-c+b) & (-a+d) & (-c-b)\\
(-b+c) & (a+d) & (-b-c) & (a-d)\\
(d-a) & (-c-b) & (d+a) & (-c+b)\\
(-b-c) & (a-d) & (-b+c) & (a+d) \end{array} \right),
\end{equation}
this then fixes
\begin{equation}
s\rightarrow s-\frac{(y_{3})^2}{2} \frac{c}{ct+d}\ .
\end{equation}
Note that this is completely analogous to the heterotic theories
considered in \cite{dWKLL} (eq.\ (4.15) in that paper) and is
consistent with the definition (\ref{sdef}).


If we now consider the theory with D7-branes but without D3-branes,
we encounter a similar situation as in
the previous subsection. Indeed, the full cubic prepotential is
formally invariant under the simultaneous exchange
of $y_7 \leftrightarrow y_3$ and $s\leftrightarrow u$, so
(assuming for simplicity one D7-brane only) one now finds that
the scalar manifold factorizes into
\begin{equation}
\mathcal{M}_{V}\cong \left(\frac{SU(1,1)}{U(1)}\right)_s \times
\frac{SO(2,3)}{SO(2)\times SO(3)}
\end{equation}
and $SL(2,\mathbb{Z})_{t}$ acts as
\begin{equation}
 s\rightarrow s, \qquad t\rightarrow \frac{at+b}{ct+d}\ , \qquad u
\rightarrow u -\frac{(y_{7})^2}{2} \frac{c}{ct+d}\ , \qquad y_7 \rightarrow
\frac{y_7}{ct+d}\ . \label{D7trafo}
\end{equation}
Just as in the previous subsection, this action can be embedded as
a symplectic transformation on the section (\ref{newbasis}) via
\begin{equation}
X^{\Lambda}\rightarrow e^{-f} \tilde{S}^{\Lambda}{}_{\Sigma} X^{\Sigma}\ ,
\qquad F_{\Lambda}\rightarrow e^{-f} ({\tilde{S}}^{-1 \, T})_{\Lambda}{}^{\Sigma}
F_{\Sigma}\ , \label{sympltrafo3}
\end{equation}
where $\tilde{S}^{\Lambda}{}_{\Sigma}$ is now a $(5\times 5)$-matrix
with the fifth coordinate corresponding to $X^4=y_7$, and we again have
\begin{eqnarray}
f&=&\ln (ct+d)\ ,  \label{fK} \\
\tilde{S} &=&\left( \begin{array}{cc} S & 0\\
0 & 1 \end{array} \right), \label{sM}
\end{eqnarray}
with $S$ as in (\ref{Sdef1}).


\subsection{Both D3- and D7-branes included}

If both D3- and D7-branes are included, we face a problem: As shown in
\cite{D'Auria:2004qv}, the scalar manifold described by the cubic prepotential
(\ref{prep}) is no longer a symmetric (although still homogeneous)
space, and  some of the previously discussed isometries are broken.
In particular, the group $SU(1,1)_{t} $ is broken to a two-dimensional
subgroup that only contains the rescalings and shifts of $t$, but no
longer the inversion $t\rightarrow -t^{-1}$ (cf.\ eq.\ (41) in \cite{D'Auria:2004qv}).
This means that only those $SL(2,\mathbb{Z})_{t}$ transformations that
have $c=0$ are isometries of the
scalar manifold described by (\ref{prep}) if both D3- and D7-branes are
present.\footnote{On the slice $y_{7}=0$, $SL(2,\mathbb{Z})_{t}$ is presumably
still an isometry.}


\subsubsection{K\"{a}hler potential}\label{TKP}

That there are problems with the inversion of $t$ can be seen
in many different ways. The K\"{a}hler potential, for instance,
is easily seen to be invariant under the combined $SL(2,\mathbb{Z})_{t}$
transformations
\begin{eqnarray} \label{transform}
 s\rightarrow s-\frac{(y_{3})^2}{2} \frac{c}{ct+d}\ , \qquad t\rightarrow
\frac{at+b}{ct+d}\ , \qquad u\rightarrow u -\frac{(y_{7})^2}{2} \frac{c}{ct+d}\ ,
\qquad y_{3(7)} \rightarrow \frac{y_{3(7)}}{ct+d}\nonumber \\ \label{fulltrafo}
\end{eqnarray}
provided we restrict these transformations to the case $c=0$.
Conversely, if we take the inversion $\alpha: t\rightarrow -t^{-1}$
corresponding to $b=-1$ and $c=1$, and assume, for simplicity, a
rectangular torus with $t=it_2$ ($t_2 < 0$
with our signature of the metric), we obtain
\begin{eqnarray}
 K\rightarrow \tilde{K}&=&-\textrm{log}\left[-\frac{8}{t_{2}^{2}}
\left(s_2 t_2 u_2 -\frac{1}{2}u_2(\textrm{Im}(y_{3}))^2 -\frac{1}{2}
s_{2}\textrm{Im}(y_{7}))^{2} \right. \right. \nonumber \\
&& \left. \left.  +\frac{1}{4t_{2}}
\left\{(\textrm{Im}(y_{3}))^{2}(\textrm{Im}(y_{7}))^{2}
- (\textrm{Re}(y_{3}))^{2}(\textrm{Re}(y_{7}))^{2}
\right\}\right)\right] . \label{Ktrafo1}
\end{eqnarray}
Obviously, due to the last line in (\ref{Ktrafo1}), we cannot write this
as $K+f +\bar{f}$ for some holomorphic function $f$.

Let us try to understand better the origin of the non-invariance of the
K\"ahler potential. This also hints towards a way to make the
K\"ahler potential $SL(2,\mathbb{Z})_{t}$ invariant again. First note that the
breaking of $SL(2,\mathbb{Z})_{t}$ invariance of \eqref{KP} arises from the
necessity to add a (counter-) term to the K\"ahler potential resulting
from dimensional reduction, in order to make it consistent
with ${\cal N}=2$ supersymmetry. This issue was discussed in
\cite{Ferrara:1996wv,Antoniadis:1996vw} and \cite{D'Auria:2004qv}.
Let us follow the discussion of
\cite{D'Auria:2004qv} here. Without taking the counterterm into account, the
K\"ahler potential would be (cf.\ their formula (32))
\begin{equation} \label{kwithout}
K_{(0)} = - \ln \Big[(s-\bar{s}) (t - \bar{t})
- \tfrac{1}{2} (y_3 - \bar{y}_3)^2
\Big] - \ln \Big[(u-\bar{u}) (t - \bar{t}) - \tfrac{1}{2} (y_7 - \bar{y}_7)^2
\Big] + \ln (t - \bar{t})\ .
\end{equation}
It is straightforward to check that this transforms under \eqref{transform}
according to
\begin{equation} \label{k0trans}
K_{(0)} \rightarrow K_{(0)} + \ln (c t + d)
+ \ln (c \bar{t} + d) \ .
\end{equation}
However, as stressed in \cite{D'Auria:2004qv}, the K\"ahler potential \eqref{kwithout}
is not consistent with ${\cal N}=2$ supersymmetry. It only becomes
so after adding a term whose presence can also be understood from
$D=6$ anomaly cancellation \cite{Ferrara:1996wv} (by first performing
a reduction on K3 to $D=6$). The zeroth order
K\"ahler potential can be written as
\begin{equation} \label{k0with}
K_{(0)} = - \ln \Big[(s-\bar{s}) (t - \bar{t}) (u-\bar{u})
- \tfrac{1}{2} (u-\bar{u}) (y_3 - \bar{y}_3)^2 -
\tfrac{1}{2} (s-\bar{s}) (y_7 - \bar{y}_7)^2 + \frac{(y_3 - \bar{y}_3)^2
(y_7 - \bar{y}_7)^2}{4 (t - \bar{t})} \Big]
\end{equation}
and the anomaly counterterm amounts to subtracting the last term in the
argument of the logarithm. It arises at 1-loop level from the D3-D7 annulus and indeed one
can verify its appearance (in the case without fluxes)
by calculating that diagram. However, there are many further terms
arising at 1-loop and only when taking all of them into account, one can
hope to obtain an $SL(2,\mathbb{Z})_{t}$ invariant result.

Let us make this more precise. In \cite{Berg:2005ja}, the 1-loop
corrections to the K\"ahler potential were calculated for the
T-dual case with D9- and D5-branes and with vanishing Wilson lines
on the D5-branes. It was found that the result can be combined with the
``tree level'' result \eqref{KP} by shifting the argument of the
logarithm, i.e.\ (translated to the D3/D7-system)
\begin{equation}  \label{k1loop}
K = - \ln \Big[(s-\bar{s}) (t - \bar{t}) (u-\bar{u})
- \tfrac{1}{2} (u-\bar{u}) (y_3 - \bar{y}_3)^2 - \tilde{c} (t - \bar{t})
{\cal E}_2 (y_3, t)
\Big]\ ,
\end{equation}
where ${\cal E}_2 (y_3, t)$ is a combination of generalized Eisenstein
series that can be found in formula (2.67) of
\cite{Berg:2005ja}.\footnote{Note that the authors of \cite{Berg:2005ja}
use a different normalization for the open string scalars, leading
to a different range than \eqref{range}. Thus, in the conventions
used in the present paper, the factor in the exponent of (2.65) of
\cite{Berg:2005ja} would be $(2 \pi)^{3/2}$ instead of
$2 \pi$ and also the coefficient $\tilde{c}$ might differ from the
value found in \cite{Berg:2005ja}, because the theory
with D3/D7-branes is locally oriented and, therefore, the
tension of the branes does not involve the factor of
$1/\sqrt{2}$ that one finds in unoriented theories, see for instance
\cite{Bachas:1998rg}.}
One can generalize the calculation to the
D3/D7-brane picture with non-vanishing D3- and D7-brane
scalars.\footnote{As discussed in section \ref{braneconfigs},
if one wants to avoid issues of large backreaction one would
assume that the D7-branes are close to the O7-planes, in groups
of four. In this appendix, we measure all the D7-brane positions
with the origin $y_7=0$ as the reference point! This does not mean
of course that they are all close to $y_7=0$. Rather, four of them
could be of the form $y_7=1/2+ \delta y_7$ and similar for the other
fixed points.}
The resulting K\"ahler potential is, up to 1-loop order,
\begin{equation} \label{koutside}
K = K_{(0)} + \tilde{c} \frac{\tilde{\cal E}_2 (y_3,y_7, t)}{(s-\bar{s})
(u-\bar{u})}\ ,
\end{equation}
where
\begin{eqnarray} \label{e37}
\tilde{\cal E}_2 (y_3,y_7, t)  & = &
- \sum_{r,s} N_r N_s \big[  E_2(y_3^r - y_3^s,t) +
     E_2(-y_3^r + y_3^s,t) \nonumber \\[-.3cm]
&& \hspace{2.5cm}
    - E_2(y_3^r + y_3^s,t)
    - E_2(-y_3^r - y_3^s,t) \big] \nonumber \\[.3cm]
&& - \sum_{k,l} N_k N_l \big[  E_2(y_7^k - y_7^l,t) +
     E_2(-y_7^k + y_7^l,t) \nonumber \\[-.3cm]
&& \hspace{2.5cm}
    -  E_2(y_7^k + y_7^l,t)
    - E_2(-y_7^k - y_7^l,t) \big] \nonumber \\[.3cm]
&& + ~ \sum_{r,k} N_r N_k \big[  E_2(y_3^r - y_7^k,t) +
     E_2(-y_3^r + y_7^k,t) \big] \nonumber \\
&& - ~ \sum_{r} N_r \big[ E_2(2y_3^r,t) +
     E_2(-2y_3^r,t) \big] \nonumber \\
&& - ~ \sum_{k} N_k \big[ E_2(2y_7^k,t) +
     E_2(-2y_7^k,t) \big] \nonumber \\
&& + ~24\, E_2(0,t)\ .
\end{eqnarray}
Here $r$ labels the different stacks of D3-branes (with $N_r$ members
each) and $k$ the stacks of D7-branes (with $N_k$ members each) and the
various contributions come from the D3/D3 annuli, the D7/D7-annuli, the
D3/D7-annuli, the D3-M\"obius strip, the D7-M\"obius strip and the Klein
bottle. The function $E_2$ is a generalized non-holomorphic Eisenstein
series, whose concrete form will be given below, cf.\ \eqref{eisen}, after the
introduction of rescaled variables $Y$, cf.\ \eqref{Y}.

It should be possible to express the K\"ahler potential
in terms of a holomorphic prepotential according to
\begin{equation} \label{kprepot}
K = -\ln \Big( 2 {\cal F} - 2 \bar{\cal F} - \sum_I
(\phi_I - \bar{\phi}_I) (\partial_{\phi_I} {\cal F}
+ \partial_{\bar{\phi}_I} \bar{\cal F}) \Big)\ ,
\end{equation}
where $\phi_I$ denotes all of the moduli fields.
It turns out that this can indeed be done, if one includes
the correction term of \eqref{koutside} in the argument of the
logarithm, i.e.\
\begin{eqnarray} \label{kinside}
K &=& - \ln \Big[(s-\bar{s}) (t - \bar{t}) (u-\bar{u})
- \tfrac{1}{2} (u-\bar{u}) (y_3 - \bar{y}_3)^2 -
\tfrac{1}{2} (s-\bar{s}) (y_7 - \bar{y}_7)^2 \nonumber \\
&& \hspace{1cm} ~+ \frac{(y_3 - \bar{y}_3)^2
(y_7 - \bar{y}_7)^2}{4 (t - \bar{t})} - \tilde{c} (t - \bar t)
\tilde{\cal E}_2 (y_3,y_7, t)\Big]\ .
\end{eqnarray}
At 1-loop order the K\"ahler potentials \eqref{koutside} and
\eqref{kinside} coincide and it turns out that it can be brought
into the form \eqref{kprepot} (for the right choice of $\tilde{c}$).
Thus we expect \eqref{kinside} to be the full perturbative
K\"ahler potential for the moduli, because, if the K\"ahler potential
\eqref{koutside} got higher order corrections that cannot be
obtained from expanding \eqref{kinside}, this would lead to a higher order
perturbative correction to the prepotential, which is known to be absent.

Let us check now that \eqref{kinside} is of the form \eqref{kprepot}
and in addition also $SL(2,\mathbb{Z})_{t}$ invariant (up to a K\"ahler
transformation). To this end,
recall formulas (2.81) and (2.82) of \cite{Berg:2005ja}, i.e.\

\begin{equation} \label{tEh}
(t - \bar t) E_2(Y,t) = - \frac{4 i \pi^4}{3} \frac{Y_2^4}{t_2} + 2 h - 2\bar h
 - ( t - \bar t) (\partial_t h + \partial_{\bar t} \bar h)
 - ( Y - \bar{Y}) (\partial_{Y} h
+ \partial_{\bar{Y}} \bar h)
\end{equation}
with
\begin{eqnarray}
h(Y,t) &=&
\frac{\pi^4}{2} \Big[ \frac{1}{90} t^3  - \frac13 t Y^2 + \frac23 Y^3 \Big]
+ \frac{i \pi}{2} Li_3(e^{2\pi i Y})
\non
&&
~~~~~~~
+ \frac{i \pi}{2} \sum_{m>0} \Big[ Li_3(e^{2 \pi i (m t - Y)}) + Li_3(e^{2 \pi i (m t + Y)}) \Big]\ ,
\end{eqnarray}
where we introduced the rescaled variable
\begin{equation} \label{Y}
Y = \sqrt{2 \pi} y
\end{equation}
(which has the same periodicity as the variable used in \cite{Berg:2005ja})
in order to avoid unusual powers of $2 \pi$ at various places.

The first term on the right hand side of \eqref{tEh} is a potential
obstacle to writing the K\"ahler potential in the form \eqref{kprepot}.
Using \eqref{tEh} in \eqref{e37}, these quartic
terms are of three different kinds, those that involve only D3-brane
scalars, those involving only D7-brane scalars and those that
mix the two. Concretely, one finds that all terms proportional to $(y_3^r)^4$,
$(y_7^k)^4$, $(y_3^r)^2 (y_3^s)^2$ and $(y_7^k)^2 (y_7^l)^2$
vanish after performing the sums in \eqref{e37}. The only terms which are
not vanishing are those proportional to $y_3^r (y_3^s)^3$,
$y_7^k (y_7^l)^3$, $y_3^r (y_7^l)^3$,
$(y_3^r)^3 y_7^l$ and $(y_3^r)^2 (y_7^l)^2$. The last one is
special, as all the other ones are linear in one of the open string fields
and cubic in the respective other one. Thus, employing
the constraint \eqref{centreofmass} again,
the only quartic term that survives summation over
all $r$ and $k$ is the one proportional to
\begin{equation}
\sum_{r,k} \frac{(y_3^r - \bar{y}_3^r)^2
(y_7^k - \bar{y}^k_7)^2}{(t - \bar{t})}\ ,
\end{equation}
which is exactly of the form of the counterterms that had been
advocated to restore ${\cal N} = 2$ supersymmetry. Thus, the constant
$\tilde{c}$ has to be such that it cancels the last term in the
first line of \eqref{kinside} (where the summation over $r$ and $k$
is implicit).

Following \cite{Berg:2005ja} it is also not hard to see that the
complete K\"ahler potential is now $SL(2,\mathbb{Z})_{t}$ invariant.
As already mentioned in \eqref{k0trans}, the
first line of the argument of the logarithm \eqref{kinside} transforms under
\eqref{transform} by a multiplicative factor $|c t + d|^{-2}$. The same is
true for $(t-\bar t)$. Thus, we only have to check $SL(2,\mathbb{Z})_{t}$
invariance of the function $\tilde {\cal E}_2 (y_3,y_7, t)$ in order to
prove that also for the full K\"ahler potential we have
\begin{equation} \label{kfulltrans}
K \rightarrow K + \ln (c t+d)
+ \ln (c \bar{t} + d) \ .
\end{equation}
To check $SL(2,\mathbb{Z})_{t}$ invariance of $\tilde {\cal E}_2 (y_3,y_7, t)$
it is enough to do so for
\begin{eqnarray} \label{eisen}
E_2 ( Y, t) &=& \sum_{\vec n=(n,m)^T} \!\!\!\!\!\! ' \; \; \;
\frac{t_2^2}{|n+mt|^{4}}
\exp\left[ 2\pi i \frac{Y (n+m\bar t) - \bar{Y}
(n+mt)}{t-\bar t} \right]\nonumber \\
&=& \sum_{\vec n=(n,m)^T} \!\!\!\!\!\! ' \; \; \;
\frac{t_2^2}{|n+mt|^{4}} e^{2 \pi i (-m Y_1 + n Y_2)}\ .
\end{eqnarray}
It is straightforward to check
\begin{equation}
E_2 ( \tilde Y, \tilde t) = E_2 ( Y, t)\ ,
\end{equation}
where $\tilde Y$ and $\tilde t$ arise after an $SL(2,\mathbb{Z})_{t}$
transformation \eqref{transform}, which implies
\begin{equation}
\left( \begin{array}{c}
       \tilde{Y}_1 \\
       \tilde{Y}_2
       \end{array}
\right)
=
\left( \begin{array}{cc}
       a & -b \\
       -c & d
       \end{array}
\right)
\left( \begin{array}{c}
       Y_1 \\
       Y_2
       \end{array}
\right)\ ,
\end{equation}
if at the same time the summation variables in \eqref{eisen} are
transformed according to
\begin{equation}
\left( \begin{array}{c}
       \tilde{m} \\
       \tilde{n}
       \end{array}
\right)
=
\left( \begin{array}{cc}
       d & -c \\
       -b & a
       \end{array}
\right)
\left( \begin{array}{c}
       m \\
       n
       \end{array}
\right)\ .
\end{equation}
Thus, we see that the $SL(2,\mathbb{Z})_{t}$ non-invariance of the
K\"ahler potential \eqref{KP} has a similar origin as the rho-problem.
It is just an artifact, arising from taking the 1-loop corrections
only partly into account.\footnote{A recent similar discussion of a 
restoration of $SL(2,\mathbb{Z})_{u}$ in $\mathcal{N}=1$ Type IIB 
compactifications by non-perturbative effects appeared in \cite{Grimm:2007xm},
building on \cite{Green:1997tv} and \cite{RoblesLlana:2006is}.}


\subsubsection{Scalar potential}
The scalar potential is another obvious place where the
failure of the $SL(2,\mathbb{Z})_{t}$-invariance of the
theory derived from the cubic prepotential (\ref{prep})
plays an important role. More precisely, we have to
distinguish between the scalar potential due to the bulk
three-form fluxes and the non-perturbative scalar potential
from gaugino condensation on D7-branes which is supposed to
fix the $K3$-volume.

\vskip5mm
\noindent\textbf{Flux potential}\\
Although one should strictly speaking use a manifestly
$\mathcal{N}=2$ supersymmetric potential for the bulk
fluxes, we follow the standard procedure of \cite{Tripathy:2002qw}
and use $\mathcal{N}=1$ language with  the GVW-superpotential
$W \sim \int G_3 \wedge \Omega_3$.
We begin by noting that the flux itself
is independent of the coordinates chosen on the torus, but the
expansion in harmonics $d x^1$ and $d x^2$ does depend on the coordinates, of course. Following \cite{Tripathy:2002qw}, we introduce the notation
\begin{eqnarray}
G_3 & = & n_{x^1} \wedge d x^1 + n_{x^2}  \wedge d x^2 \non
&=& \frac{1}{\bar t - t} (G_x \wedge d x + G_{\bar x} \wedge d \bar x)
\end{eqnarray}
with
\begin{equation}
G_x = n_{x^1} \bar t - n_{x^2} \quad , \quad G_{\bar x} = - (n_{x^1} t - n_{x^2})\ .
\end{equation}
Using the general transformation \eqref{generaltrafo}, one can show
that
\begin{equation} \label{ntrafo}
\left( \begin{array}{c}
       \tilde n_{x^1} \\
       \tilde n_{x^2}
       \end{array}
\right)
= \left( \begin{array}{cc}
       d & c \\
       b & a
       \end{array}
\right)
\left( \begin{array}{c}
        n_{x^1}\\
        n_{x^2}
       \end{array}
\right)
\end{equation}
and
\begin{equation} \label{Gy}
\tilde G_{\tilde x} = \frac{1}{c \bar t + d} G_x
\quad , \quad \tilde G_{\bar {\tilde x} } = \frac{1}{c t + d} G_{\bar x}\ .
\end{equation}
The transformation \eqref{Gy} assures that $G_3$ is invariant (while the components
$G_x$ and $G_{\bar x}$ transform). Nevertheless, due to the transformation of the
$d x$ in the $(3,0)$ form $\Omega_3 = \Omega_2 \wedge d x$,
the flux superpotential $W \sim \int G_3 \wedge \Omega_3$ transforms according to
\begin{equation} \label{Wtrans}
\tilde W = \frac{1}{c t + d} W\ .
\end{equation}
As we just saw that the 1-loop corrected K\"ahler potential transforms
as in \eqref{kfulltrans}, the combination $e^{K}|W|^{2}$
is indeed manifestly $SL(2,\mathbb{Z})_{t}$-invariant.

Let us also check that \eqref{ntrafo} is consistent with
the $SL(2,\mathbb{Z})_t$ transformation of the value of the complex
structure modulus in the (${\cal N}=1$) supersymmetric minimum.
According to \cite{Tripathy:2002qw}, this is given by
\begin{equation}
t_0 = \frac{\bar n_{x^2} \cdot n_{x^1}}{n_{x^1} \cdot \bar n_{x^1}}\ ,
\end{equation}
cf.\ their formula (4.6), where the dot means the inner product
$n_1 \cdot n_2 = \int_{K3} n_1 \wedge n_2$. Using \eqref{ntrafo}
(and the supersymmetry condition $G_x = 0$, cf.\ formula (4.2) of
\cite{Tripathy:2002qw}), one derives for the value of
the complex structure modulus in the transformed coordinates
\begin{equation}
\tilde t_0 = \frac{\bar{\tilde{n}}_{x^2} \cdot
\tilde n_{x^1}}{\tilde n_{x^1} \cdot \bar{\tilde{n}}_{x^1}}
= \frac{a t_0 + b}{c t_0 + d}\ .
\end{equation}
Hence, the $SL(2,\mathbb{Z})_t$ transformed complex
structure $\tilde t_0$ is given by the
same formula (4.6) of \cite{Tripathy:2002qw}, using the transformed expansion
coefficients \eqref{ntrafo}. Thus, one and the same flux $G_3$ can lead to either
value of the complex structure modulus, depending on whether one chooses
the original $x$ coordinates or the transformed ones $\tilde x$, indicating
again that values of $t$ (in the supersymmetric minimum)
differing only by an $SL(2,\mathbb{Z})_t$ transformation
should not be distinguishable physically.


\vskip5mm
\noindent\textbf{Non-perturbative superpotential}\\
As reviewed in section \ref{gcsupo}, after breaking supersymmetry to ${\cal N}=1$,
gaugino condensation on a stack of D7-branes
results in a non-perturbative superpotential of the form
\begin{equation} \label{wnp}
W_{np}= A_0 \exp\Big(\frac{8 \pi^2 f_{D7}}{c}\Big)\ ,
\end{equation}
where $f_{D7}$ denotes the gauge kinetic function of the D7-brane gauge fields
\eqref{fd7} which we repeat here for convenience (for $y_7 = \mu = 0$),
\begin{equation}
f_{D7}= i s - \frac{1}{(2\pi)^2}\ln \vartheta_1(\sqrt{2 \pi} y_{3},t) + \ldots \ , \label{fullf}
\end{equation}
and $c$ was defined in \eqref{c}.
Using the cubic prepotential
(\ref{prep}), on the other hand, only reproduces (we are choosing the vevs
$u=t=-i, y_{7}=0$ for simplicity) the first term, i.e.,  $f_{D7}= i s$.
If that was really the full answer, we would have
\begin{equation}
W_{np}=A_0 \exp(-ias),
\end{equation}
with some constants $A$ and $a$, which, under an inversion
$t\rightarrow -t^{-1}$, transforms as
\begin{equation}
W_{np}\rightarrow \tilde{W}_{np}= W_{np}\exp (ia y_{3}^2/(2t)),
\end{equation}
which is not of the same form as the K\"{a}hler transformation (\ref{Wtrans})
of the flux superpotential.

It is natural to expect that using the full gauge
kinetic function (\ref{fullf}) in $W_{np}$ might improve its
$SL(2,\mathbb{Z})_{t}$-transformation properties, as it is
the case for the K\"ahler potential. A similar phenomenon also
occurs in the heterotic string
\cite{Font:1990nt}.
Thus, we consider the transformation of
the full $f_{D7}$ next. Let us start with the inversion, i.e.\
$t \rightarrow \tilde t = -t^{-1}$, which implies $y_3 \rightarrow \tilde y_3
= -y_3 t^{-1}$ and
$s \rightarrow \tilde s = s - \frac{y_3^2}{2}t^{-1}$. For the moment
we just consider one non-vanishing $y_3$, even though this is not consistent
with \eqref{centreofmass}. Of course we have a sum over
different D3-brane positions in mind but we would like to
keep the formulas simple at the beginning and then generalize to
the actual case at hand in a second step.

With our conventions for the complex structure modulus $t$ the
transformation of the theta function is as in \eqref{newsl2z}, which
amounts to
\begin{equation} \label{thetatrans}
\vartheta_1 (\sqrt{2 \pi} \tilde y_3, \tilde t) =  - i (i t)^{1/2}
e^{-2 i \pi^2 y_3^2 t^{-1}} \vartheta_1(\sqrt{2 \pi} y_3, t)\ ,
\end{equation}
where we also used that $\vartheta_1$ is an odd function in its
first argument.
Eq.\ \eqref{thetatrans}
leads to a transformation of the gauge kinetic function
according to
\begin{eqnarray} \label{ftrans}
\tilde f_{D7} & = & i \tilde s - \frac{1}{(2 \pi)^2}
\ln \vartheta_1 (\sqrt{2 \pi} \tilde y_3, \tilde t) \non
& = & i s - \frac{i y_3^2}{2} t^{-1} - \frac{1}{(2 \pi)^2} \ln[-i (i t)^{1/2}
e^{-2 i \pi^2 y_3^2 t^{-1}} \vartheta_1(\sqrt{2 \pi} y_3, t)] \non
& = & i s - \frac{1}{(2 \pi)^2} \ln \vartheta_1 (\sqrt{2 \pi} y_3, t)
- \frac{1}{(2 \pi)^2} \ln[-i (i t)^{1/2}] \non
& = & f_{D7} - \frac{1}{8 \pi^2} \ln t + \frac{i}{16 \pi}\ .
\end{eqnarray}
It is very promising that the gauge kinetic function indeed
transforms with a term proportional to $\ln t$ which is
a prerequisite for the non-perturbative superpotential
to transform like the flux superpotential (\ref{Wtrans}).

This can now be generalized to the gauge kinetic function with several
D3-branes and for concreteness we consider $N$ D7-branes at the origin
(however, the discussion for $N$ D7-branes at any of the other
fixed points would be analogous). The gauge group on this stack of D7-branes
is $SU(N)$ and by choosing $N=4$ one could ensure local
tadpole cancellation.\footnote{This gauge group arises if the $K3$
is at its $\mathbb{Z}_2$-orbifold limit. The resulting theory is T-dual
to the torus compactification of \cite{Bianchi:1990yu,GP}. It is
this theory at the orbifold point where one can actually perform
the world sheet calculation. The relation to the smooth $K3$ case was discussed in
\cite{Berkooz:1996iz,Sen:1997kw}.}
For the discussion of the $SL(2,\mathbb{Z})_{t}$-transformation of $f_{D7}$ it is
important to include also the terms independent of the D3-brane
positions, i.e.\ those proportional to $\ln \eta(t)$, as these also transform
non-trivially, cf.\ \eqref{newsl2zeta}. As we know the
explicit form including all factors only in the case without fluxes,
we will restrict our further discussion to that case. Let us first consider the
physical D7-brane gauge coupling which, including
open string 1-loop corrections,
is (in the notation of the present paper) \cite{Berg:2004ek}
\begin{eqnarray} \label{gd7}
g_{{\rm D7}}^{-2} &=&  \left[
\frac{T_3 h_{0}\tilde{V}^{K3}}{2 (2\pi)^2}  \right] e^{4U_1}
- \frac12 \sum_r N_r \frac{{\rm Im}(y_3^r)}{{\rm Im} t} \\
&& + \frac{b^{{\cal N}=2}}{8 \pi^2} \Big( K_{(0)}
+\ln \frac{M_{{\rm Pl}}}{p^2} \Big) -
\frac{1}{(2 \pi)^2} \sum_r N_r \ln \left|
\frac{\vartheta_1 (\sqrt{2 \pi} y_3^r,t)}{\eta(t)} \right|
- \frac{b^{{\cal N}=2}}{2 \pi^2} \ln |\eta(t)|\ . \nonumber
\end{eqnarray}
As before, $N_r$ is the number of D3-branes in the $r$-th stack,
$K_{(0)}$ was defined in \eqref{k0with} and $b^{{\cal N}=2} = -4$ in the case
at hand. This can be seen from formula \eqref{bn2} for $b^{{\cal N}=2}$.
If none of
the D3-branes are at the origin as well, the only massless hypermultiplets
arise from strings starting and ending on the D7-branes
(and their $\Omega$-images). There are two of these multiplets
and they transform in the
antisymmetric representation of $SU(N)$ \cite{GP}. As the index of the
antisymmetric representation is $T(a) = \frac12 (N-2)$ (cf.\
footnote \ref{indices}), we get the announced result
$b^{{\cal N}=2} = -4$.

The first line in \eqref{gd7} is $-{{\rm Im}}(s)$ and, thus, \eqref{ftrans} and
\eqref{newsl2zeta} show that the
transformation of the physical gauge coupling is
\begin{equation} \label{deltag}
\delta g_{{\rm D7}}^{-2} = \frac{b^{{\cal N}=2}}{4 \pi^2} \ln|t| - \frac{1}{8 \pi^2}
\sum_r N_r \ln |t| + \frac{3}{\pi^2} \ln |t| = 0\ ,
\end{equation}
i.e.\ it is invariant. In \eqref{deltag} we have used that $\sum_r N_r = 16$ and
$b^{{\cal N}=2} = -4$.

For the gauge kinetic function, \eqref{gd7} implies
\begin{eqnarray}
f_{{\rm D7}} &=& i s - \frac{1}{(2 \pi)^2} \sum_r N_r \ln
\vartheta_1 (\sqrt{2 \pi} y_3^r,t)
+ \frac{6}{\pi^2} \ln \eta(t) \nonumber \\
& \stackrel{t \rightarrow -t^{-1}}{\longrightarrow} &
f_{{\rm D7}} + \frac{1}{\pi^2} \ln t + \frac{5}{2 \pi} i\ . \label{ftransinv}
\end{eqnarray}
Moreover, from \eqref{newsl2z} and \eqref{newsl2zeta}, it is not difficult to see that
under $t \rightarrow t+1$, the gauge kinetic function transforms according to
\begin{equation}
f_{{\rm D7}} \stackrel{t \rightarrow t+1}{\longrightarrow} f_{{\rm D7}}
+ \frac{i}{2 \pi}\ , \label{ftransshift}
\end{equation}
which again implies that the physical gauge coupling (i.e.\ the
real part of $f_{{\rm D7}}$) is invariant.

Any discussion of the transformation of the non-perturbative
superpotential would require a more precise knowledge of the
charged field content, which determines the constant $c$
in \eqref{wnp} and also the form of the prefactor $A$ (which
in general depends on the light charged matter fields that also
transform non-trivially under $SL(2,\mathbb{Z})_{t}$, in a way
similar to $y_3$ but potentially with a different weight). As this requires
a better understanding of the massless spectrum in the actual
flux background, we refrain from discussing the transformation
properties of $W_{np}$ further. We still consider it suggestive
that the transformation of the gauge kinetic function has the right
$t$-dependence in order to be able to cancel (together with a
potential transformation of the prefactor $A$) the transformation
of the K\"ahler potential \eqref{kfulltrans}, even though we can not
conclusively verify that the factors work out
correctly.\footnote{A final comment about the imaginary shifts in
\eqref{ftransinv} and \eqref{ftransshift}: These shifts do not mean that
$W_{np}$ obtains a phase under an $SL(2,\mathbb{Z})_{t}$ transformation.
It should rather not, in order to transform exactly as the flux
superpotential \eqref{Wtrans}. To verify this, one again needs to know
the actual spectrum in the flux background. This could modify the factors
in \eqref{ftransinv} and \eqref{ftransshift} and would determine the
value of $c$. Only then could one see whether the phase in
$e^{8 \pi^2 f/c}$ becomes a multiple of $2 \pi$ and, thus, drops out from
the transformed superpotential. We expect this to happen.}


\section{Uplifting potential and inflaton mass}
\label{appmarco}

In section \ref{cosmology}, we studied a potential of the form
\begin{equation}
V=V_{inf}+V_{F} \ ,
\end{equation}
where $V_{inf}$ denotes the D-term inflaton potential
induced by the   FI D7-brane with world volume fluxes, and
$V_{F}$ is the F-term potential that stabilizes the $K3$ volume modulus $s$.
When $V_{inf}$ vanishes at the end of inflation (i.e. for $y_3=0$),
the cosmological constant due to $V_{F}$ alone would be negative, and one thus also has to add an uplifting potential, $V_{up}$,
\begin{equation}
V=V_{inf}+V_{F}+V_{up} \ ,
\end{equation}
that could (nearly) cancel the cosmological constant after inflation.\footnote{We are ignoring any contributions to the cosmological constant by other sectors of the theory, such as, e.g., a Standard Model sector.} As already mentioned in section \ref{cosmology}, this uplifting potential would generically modify the inflaton mass because of the following two effects:
\begin{enumerate}
\item $V_{up}$ may, in general, depend explicitly on $y_3$. As we work on an orientifold of the torus which identifies $y_3$ and $-y_3$, this dependence on $y_3$ must be even, and generically leads to non-trivial additional quadratic terms in $y_3$ (plus higher even powers).
 \item $V_{up}$ generically also depends on $s$. If we use
 \begin{equation}
 \partial_s V(s,y_3)=0
 \end{equation}
to integrate out $s$, this leads to a $y_3$-dependent solution (we still assume ${\rm Im} (y_{3})=0$),
\begin{equation}
s(y_3,\ldots)=\tilde{s}+i \tilde{h}(y_3)\ ,
\label{solves}
\end{equation}
where the ellipsis stands for various parameters in the potentials,
such as, e.g., the flux quantum numbers, and $\tilde{s}$ denotes a complex
constant and $\tilde{h}$ a function of $y_3$ (with $\tilde{h}(0)=0$)
that both depend
on these parameters (see, for instance,
eq.\ (\ref{W0}) and the discussion following it).
Re-inserting this solution for $s$
into the uplifting potential introduces an additional $y_3$ dependence beyond the explicit $y_3$-dependence mentioned in the previous item.
\end{enumerate}
The combined effect of items 1. and 2. on the inflaton mass is in general model dependent. Here we will focus on the case when the uplifting potential is due to a D-term potential from another (stack of) D7-brane(s) (i.e., we are now talking about altogether three different types of D7-branes: The FI-D7-brane for the        original inflaton potential, the stack of volume stabilizing D7-branes with gaugino condensation and the uplifting (``UP'') D7-branes just introduced).

In order to generate a D-term potential, the UP-D7-branes also have to carry a world volume flux.
This world volume flux, $\mathcal{F}$, can be expanded in the harmonic two-forms, $\omega_{\alpha}$, of $K3\times T^{2}/\mathbb{Z}_{2}$,
\begin{equation}
\mathcal{F}=f^{\alpha}\omega_{\alpha},
\end{equation}
with constant coefficients $f^{\alpha}$.

Denoting the four-cycles of $K3\times T^{2}/\mathbb{Z}_{2}$ by $\Sigma^{\alpha}$, there are two types of four-cycles: First, there is the four-cycle,
$\Sigma^{s}$, that is given by $K3$ itself. Second, there are the four-cycles, $\Sigma^{i}$, that
are of the form $\Gamma^{i} \times T^{2}/\mathbb{Z}_{2}$ with $\Gamma^{i}$ denoting the two-cycles in $K3$. The four-cycle that is wrapped by the UP-D7-brane is denoted by $\Sigma^{\ast}$ and can, a priori, be either
$\Sigma^{s}$ or one of the $\Sigma^{i}$.

The worldvolume flux on the UP-D7-brane wrapped on $\Sigma^{\ast}$ induces the gauging of axionic shift symmetries of the form $T_{\alpha} \rightarrow T_{\alpha} + Q_{\alpha \ast}$,
where $T_{\alpha}$ denote the K\"{a}hler moduli of $K3\times T^{2}/\mathbb{Z}_{2}$. As indicated, the constants $Q_{\alpha \ast}$ depend on the wrapped four-cycle $\Sigma^{\ast}$, but also on the world volume fluxes via
\begin{equation}
Q_{\alpha \ast}=f^{\beta}K_{\alpha\beta\ast} \ ,
\end{equation}
with $K_{\alpha \beta \gamma}$ being the triple intersection numbers. The charges $Q_{\alpha \ast}$ can (and in general will) be zero for some indices $\alpha$.
This axionic gauging induces a D-term potential of the form
\begin{equation}
V_{up}=V_{D}\sim g_{UP-D7}^{2} \Big[ (\partial_{T_{\alpha}} K) Q_{\alpha \ast} +\ldots \Big]^2 \ , \label{Vwf}
\end{equation}
where $g_{UP-D7}$  is the gauge coupling of the UP-D7-brane, and the ellipsis denotes possible charged matter fields which we will ignore.  $g_{UP-D7}^{2}$ is inversely proportional to the volume of the four-cycle $\Sigma^{\ast}$ wrapped by the UP-D7-brane.

We can now distinguish the following cases:
\begin{enumerate}
\item The UP-D7-brane wraps $K3$. This means $\Sigma^{\ast}=\Sigma^{s}$, and hence, at one-loop,
\begin{equation}
g_{UP-D7}^{2}=(-s_{2} - \textrm{Re}(h(y_3)))^{-1},
\end{equation}
where the holomorphic function $h(y_3)$ is even in $y_3$ and can be read off from eq.\ (\ref{fd7}). Using a solution of the form (\ref{solves}) and assuming large
$(-\tilde{s}_{2})$, we obtain the expansion
\begin{equation}
g_{UP-D7}^{2}=\frac{\Big[1-\frac{{\rm Re}(\tilde{h}(y_{3}))+{\rm Re}(h(y_3))}{\tilde{s}_{2}} + \mathcal{O}(\tilde{s}_{2}^{-2})\Big]}{(-\tilde{s}_{2})} \ . \label{gform}
\end{equation}
Let us now consider the derivative of the K\"{a}hler potential in eq.\ (\ref{Vwf}). The K\"{a}hler potential $K$ decomposes into a sum of a K\"{a}hler potential, $K_{s}$, for the $K3$ volume modulus $s$ (as well as $y_3$) and a K\"{a}hler potential, $\hat{K}$, for the remaining K\"{a}hler moduli. This can, e.g., be understood by recalling that the latter descend from $\mathcal{N}=2$ hypermultiplets, whereas
$s$
descends from an $\mathcal{N}=2$ vector multiplet. Alternatively, one can use
that the non-vanishing triple intersection numbers are of the form $K_{sij}$ to directly compute the K\"{a}hler potential from a dimensional reduction.
This implies that $\partial_{T_{i}}K=\partial_{T_{i}}\hat{K}$ is independent of $s$. Since we are choosing $\Sigma^{\ast}=\Sigma^{s}$, we have, again remembering the particular form of the triple intersection numbers,
$Q_{s\ast}=K_{s\alpha\ast}f^{\alpha}=K_{s\alpha s}f^{\alpha}=0$, and hence
$(\partial_{T_{\alpha}}K)  Q_{\alpha \ast}=\partial_{T_{i}}K Q_{i\ast}$, which is independent of $s$ (and $y_3$). Thus, to summarize, if the UP-D7-brane wraps
the $K3$, the only $s$ and $y_3$ dependence of $V_{up}$ comes from the D7-brane gauge coupling $g_{UP-D7}$ in the form (\ref{gform}).

\item Let us now assume the UP-D7-brane wraps a four-cycle of the form
$\Gamma^{i}\times T^{2}/\mathbb{Z}_{2}$, i.e., $\Sigma^{\ast}=\Sigma^{i}$.
In this case, the UP-D7-brane gauge coupling does not depend on the volume of
$K3$, but rather on the K\"{a}hler moduli that descend from hypermultiplets.
As we explain in footnote 3, one does not expect the threshold corrections to $g_{UP-D7}$ to introduce a $y_3$ dependence either. Thus to the approximation we are calculating, $g_{UP-D7}^{2}$ is independent of $s$ and $y_3$ for this way of wrapping the branes. Let us now consider the term in (\ref{Vwf}) that involves the first derivative of the K\"{a}hler potential. If the only nonvanishing flux is in $T^{2}/\mathbb{Z}_{2}$, only the component $f^{\alpha}=f^{s}$ would be non-zero.
We would thus have $Q_{s}=K_{s\alpha\ast}f^{\alpha}= K_{ss\ast}f^{s}=0$ due to our particular intersection numbers. Thus, $(\partial_{T_{\alpha}}K)Q_{\alpha\ast}=(\partial_{T_{i}}K)               Q_{i\ast}$, which is independent of $s$ and $y_3$. Thus, putting everything together, if we choose to
wrap the UP-D7-brane on a four-cycle of the form $\Sigma^{i}=\Gamma^{i} \times T^{2}/\mathbb{Z}_{2}$ and turn on world volume flux only along $T^{2}/\mathbb{Z}_{2}$, the entire uplifting potential would be independent of
$s$ and $y_3$, and would therefore not induce any new inflaton mass terms.
Unfortunately, tadpole cancellation cannot be fulfilled whenever we wrap D7-branes on four-cycles other than $K3$, because we only have O7-planes
that are wrapped on $K3$ as well. Putting a D7-brane in the way described in this item would require a different orientifold projection. It would be very interesting if a suitable orientifold could be constructed where the above decoupling between the uplifting potential and the inflaton mass term could be realized.
\end{enumerate}
We finally note that a brane setup of the type 1. does not lead to the conflict
with gaugino condensation that was studied in \cite{Choi:2005ge}, because the two
types of D7-branes, i.e.\ the D7-branes responsible for uplifting and the
volume-stabilizing D7-branes with the gaugino condensate, do not intersect.


\section{Quartic term in the inflaton potential}
\label{appquartic}

At several places in the main text we approximated the F-term contribution to the inflaton potential by its quadratic term.
In this appendix we would like to check that the quartic term can indeed be fine-tuned to be much smaller than the quadratic one, as is required 
for the self-consistency of this approximation.\footnote{This appendix grew out of discussions with Cliff Burgess, Jim Cline and Marieke Postma.}
To this end, we use the expansion
\begin{equation} \label{WnpF}
W_{np}= A\Big[1-\Delta(t_{0})(y_{3})^{2}-\Sigma(t_{0})(y_3)^{4}+\ldots\Big]e^{-ias},
\end{equation}
with some functions $\Delta(t)$ and $\Sigma(t)$ that can be obtained by expanding the theta function (see below). As in the main text, we will assume $\textrm{Im}(y_{3})=0$.

We first integrate out $s$ by imposing
\begin{equation}
D_{s}W|_{min}=0, \label{DsF}
\end{equation}
which is equivalent to
\begin{equation}
W_{0}=W_{np}(2s_{2}a -1). \label{HalloF}
\end{equation}
The solution to this equation yields a value for  $s$ that depends on $W_{0}$ as well as the momentary value of
$\textrm{Re}(y_{3})$. We expand this solution in powers of $\textrm{Re}(y_{3})$:
\begin{equation}
s=\tilde{s} +i\lambda [\textrm{Re}(y_{3})]^2 +i\mu[\textrm{Re}(y_{3})]^{4} + \ldots. \label{lllF}
\end{equation}
Inserting this in (\ref{HalloF}) implicitly  determines $\tilde{s}$ via
\begin{equation}
W_{0}=Ae^{-ia\tilde{s}}(2\tilde{s}_{2}a -1)
\end{equation}
and yields
\begin{eqnarray}
\textrm{Re}(\lambda)&=&\frac{(x-1)}{a(x+1)}\textrm{Re}(\Delta)\ , \label{RelF}\\
\textrm{Im}(\lambda)&=&\frac{\textrm{Im}(\Delta)}{a}\label{ImlF}\ , \\
\textrm{Re}(\mu)&=&\left[\frac{(x-1)}{2a(x+1)^{3}}\left[(x+1)^2+4\right)\right][\textrm{Re}(\Delta)]^{2}-\frac{(x-1)}{2a(x+1)}[\textrm{Im}(\Delta)]^{2} \nonumber\\
&& + \frac{(x-1)}{a(x+1)}\textrm{Re}(\Sigma)\ , \label{RemF}\\
\textrm{Im}(\mu)&=&\frac{1}{a}\Big[\textrm{Re}(\Delta)\textrm{Im}(\Delta)+\textrm{Im}(\Sigma)\Big] \label{ImmF}  ,
\end{eqnarray}
where we have introduced
\begin{equation}
x\equiv 2a\tilde{s}_{2}.
\end{equation}
After one has integrated out $s$, one obtains an effective field theory for the only remaining dynamical field $\textrm{Re}(y_{3})$. Setting $M_P=1$, its 
kinetic term in this effective field theory is given by 
\begin{equation}
 \frac{1}{2} \frac{(\partial_{\mu} \textrm{Re}(y_{3}))^{2}}{2s_{2}t_2}, \label{kkkF}
\end{equation}
where $s_{2}$ now depends on $\textrm{Re}(y_{3})$ via (\ref{lllF}):
\begin{equation}
s_{2}=\tilde{s}_{2} + \textrm{Re}(\lambda) [\textrm{Re}(y_{3})]^2
+\textrm{Re}(\mu) [\textrm{Re}(y_{3})]^4 + \ldots.
\end{equation}
Expanding the kinetic term (\ref{kkkF}), the canonically normalized field $\phi$ is then  given by
\begin{eqnarray}
\partial_{\mu}\phi &=& \frac{\partial_{\mu} [\textrm{Re}(y_{3})]}{\sqrt{2t_2 \tilde{s}_{2}}}\Big[ 1- \frac{\textrm{Re}(\lambda)}{2\tilde{s}_{2}} [\textrm{Re}(y_{3})]^{2} + \mathcal{O}([\textrm{Re}(y_{3})]^{4}) \Big] \nonumber \\  
\Rightarrow \phi &=& \frac{\textrm{Re}(y_{3})}{\sqrt{2t_2 \tilde{s}_{2}}}\Big[ 1- \frac{\textrm{Re}(\lambda)}{6\tilde{s}_{2}} [\textrm{Re}(y_{3})]^{2} + \mathcal{O}([\textrm{Re}(y_{3})]^{4}) \Big] . 
\end{eqnarray}
Note that this is not the same as if one had just expanded the naive relation
$\phi=\textrm{Re}(y_{3})/\sqrt{2 s_{2}t_{2}}$ according to (\ref{lllF}).
In the F-term potential, on the other hand, this difference only affects the quartic or higher order terms, leaving  
 the discussion of the quadratic terms in the main text unchanged. 

Using all this in the F-term potential, we obtain
\begin{equation} 
V_{F}=\frac{|A|^{2}e^{2a\tilde{s}_{2}}\tilde{s}_{2}}{u_{2}}\left[\frac{3a^2}{2t_{2}}-\tilde{m}^{2}\phi^{2}+\tilde{\lambda}\phi^{4}+\ldots \right] \label{potentialF}
\end{equation}
with
\begin{eqnarray} \label{mlambdaF}
\tilde{m}^{2}&=& 4t_{2}|\Delta|^{2} + 3a \textrm{Re}(\Delta)\\
\tilde{\lambda}&=&2 t_{2}\tilde{s}_{2}\left[ -16t_{2}\, \textrm{Re}(\Delta \bar{\Sigma}) -\frac{a}{2x(x+1)}\left[3x^2 -5x -4 \right][\textrm{Re}(\Delta)]^2 \right. \nonumber \\
&& \left. +\frac{3a}{2}[\textrm{Im}(\Delta)]^2  -3a \textrm{Re}(\Sigma) -8 t_{2}\frac{(x-1)(3x+1)}{3x(x+1)}    \, |\Delta|^{2}\textrm{Re}(\Delta) \right] . \nonumber
\end{eqnarray}

Now we would like to compare the actual values for $\tilde{m}^{2}$ and $\tilde{\lambda}$ which appear in the case at hand. For that, we need explicit formulas for $\Delta$ and $\Sigma$. 
Starting point is, as in section \ref{braneconfigs}, 
\begin{equation} \label{WcorrectedF}
W_{np} = \tilde{A} \Big( \vartheta_1( \sqrt{2\pi}y_3-1/2, t ) \vartheta_1( \sqrt{2\pi}y_3+1/2, t ) \Big)^{-1/c}  e^{-ias}\ .
\end{equation}
Expanding this for small $y_3$ using (cf.\ appendix B.1 of \cite{Kiritsis:2000zi})
\begin{eqnarray}  \label{thetaexpandF}
&& \vartheta_1(1/2-\nu,t)
= \vartheta_2(\nu,t) = \vartheta_2(0,t) \Big(1 -
\frac{\pi^2}{6} \Big[E_2(t) + \vartheta_3^4(0,t) + \vartheta_4^4(0,t)\Big] \nu^2 \nonumber \\ 
&& \hspace{0.6cm} + \frac{\pi^4}{72} \Big[-2 E_4(t) + E_2^2(t) + 2E_2(t) (\vartheta_3^4(0,t) + \vartheta_4^4(0,t)) + 3 \vartheta_2^8(0,t)\Big] \nu^4 \Big)\ , \nonumber \\
\end{eqnarray}
we obtain 
\begin{eqnarray} \label{DeltaSigmaF}
\Delta & = & -\frac{2 \pi^3}{3 c} \Big[E_2(t) + \vartheta_3^4(0,t) + \vartheta_4^4(0,t)\Big] \\
\Sigma & = &-\frac{2 \pi^6}{9c^2} \Big[ c (\vartheta_3^4 \vartheta_4^4 - \vartheta_2^8 +  \vartheta_3^8 +  \vartheta_4^8) + E_2^2 + 2 E_2 (\vartheta_3^4 + \vartheta_4^4) + (\vartheta_3^4 +  \vartheta_4^4)^2 \Big]\ , \nonumber
\end{eqnarray}
where we also used 
\begin{equation} 
E_4(t) = \frac12 [\vartheta_2^8(0,t)+\vartheta_3^8(0,t)+\vartheta_4^8(0,t)]
\end{equation}
in order to get rid of the Eisenstein series $E_4$. Note that there is an overall factor 
of $\vartheta_2(0,t)$ in \eqref{thetaexpandF} (which can thus be absorbed in the overall factor $\tilde{A}$, leading to the factor $A$ of \eqref{WnpF}) and that all $\vartheta$-functions in the second line of \eqref{DeltaSigmaF} have vanishing first argument. 

\begin{figure}[tb]
\centering{\includegraphics[height=7cm]{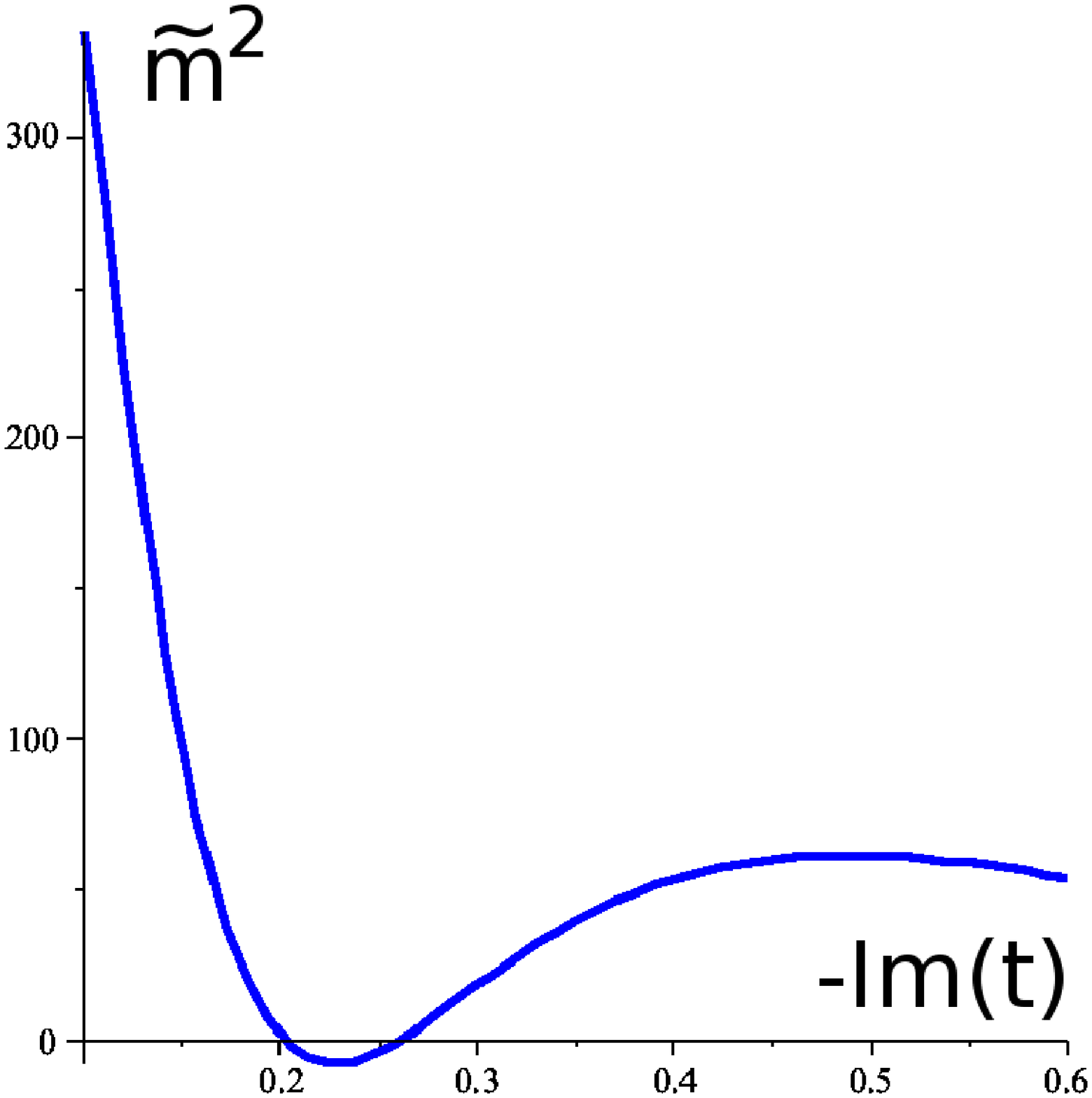}   \hskip 0.3cm   \includegraphics[height=7cm]{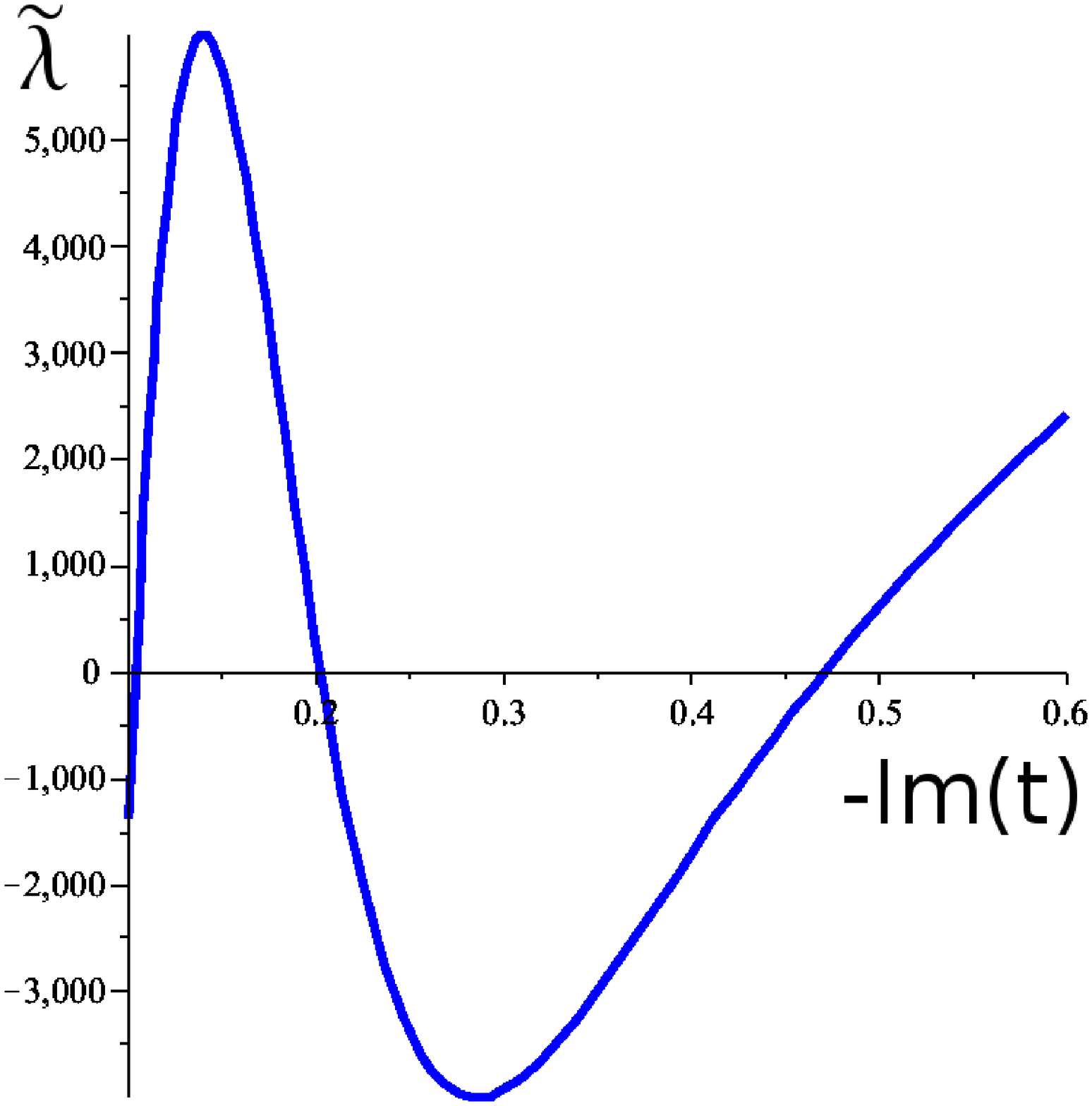} }
 \caption{$\tilde{m}^2$ and $\tilde{\lambda}$ as functions of $-t_2$ for $t_{1}=0.3155$ and $a=8\pi^{2}/10$ (note that the value for $t_1$ differs from the one used in sections \ref{braneconfigs} and \ref{cosmology}, which explains the difference to fig.\ \ref{mfig}).}
  \label{quadraticquartic}
\end{figure}

Using the formulas \eqref{DeltaSigmaF} in \eqref{mlambdaF} and, for concreteness, 
$c=-10, a=8 \pi^2/10, t_1=0.3155$ and $\tilde{s}_2=-1$, we obtain the results
for $\tilde{m}^2$ and $\tilde{\lambda}$ shown in figure \ref{quadraticquartic}. Obviously, the ratio $\tilde{\lambda}/\tilde{m}^2$ is not generically small, but can be fine-tuned to be so (for example at the first and third zero of $\tilde{\lambda}$). Note also that we only plotted the coefficients $\tilde{m}^2$ and $\tilde{\lambda}$. The quartic term is of course even more suppressed compared to the quadratic one for small values of the canonically normalized field $\phi$. This is important in order to ensure that the higher powers in \eqref{potentialF} become less and less important and one has to fine-tune their coefficients to less and less accuracy in order to be able to neglect them. 

\begin{figure}[tb]
\centering{\includegraphics[height=7cm]{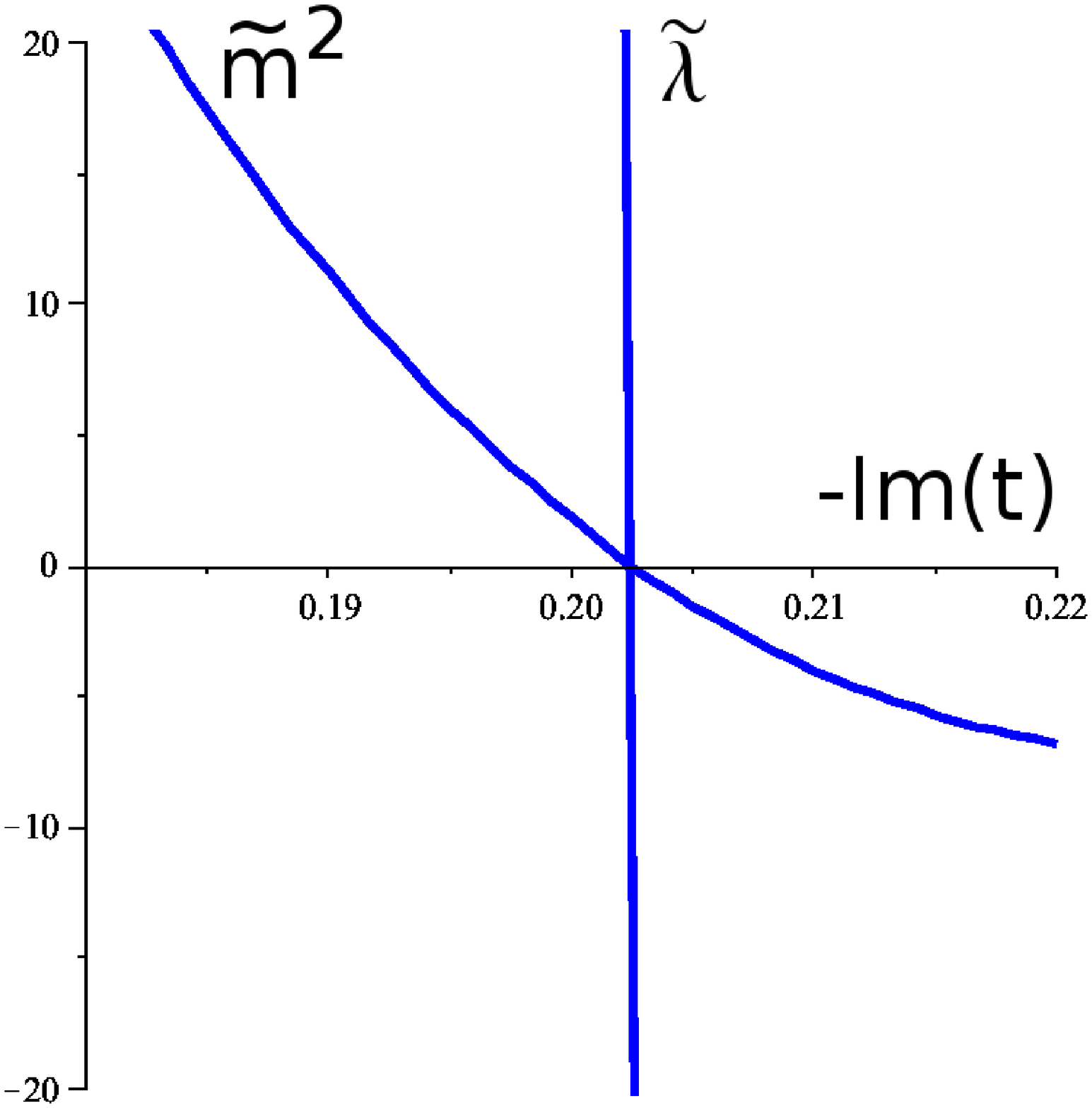}   \hskip 0.3cm   \includegraphics[height=8.2cm]{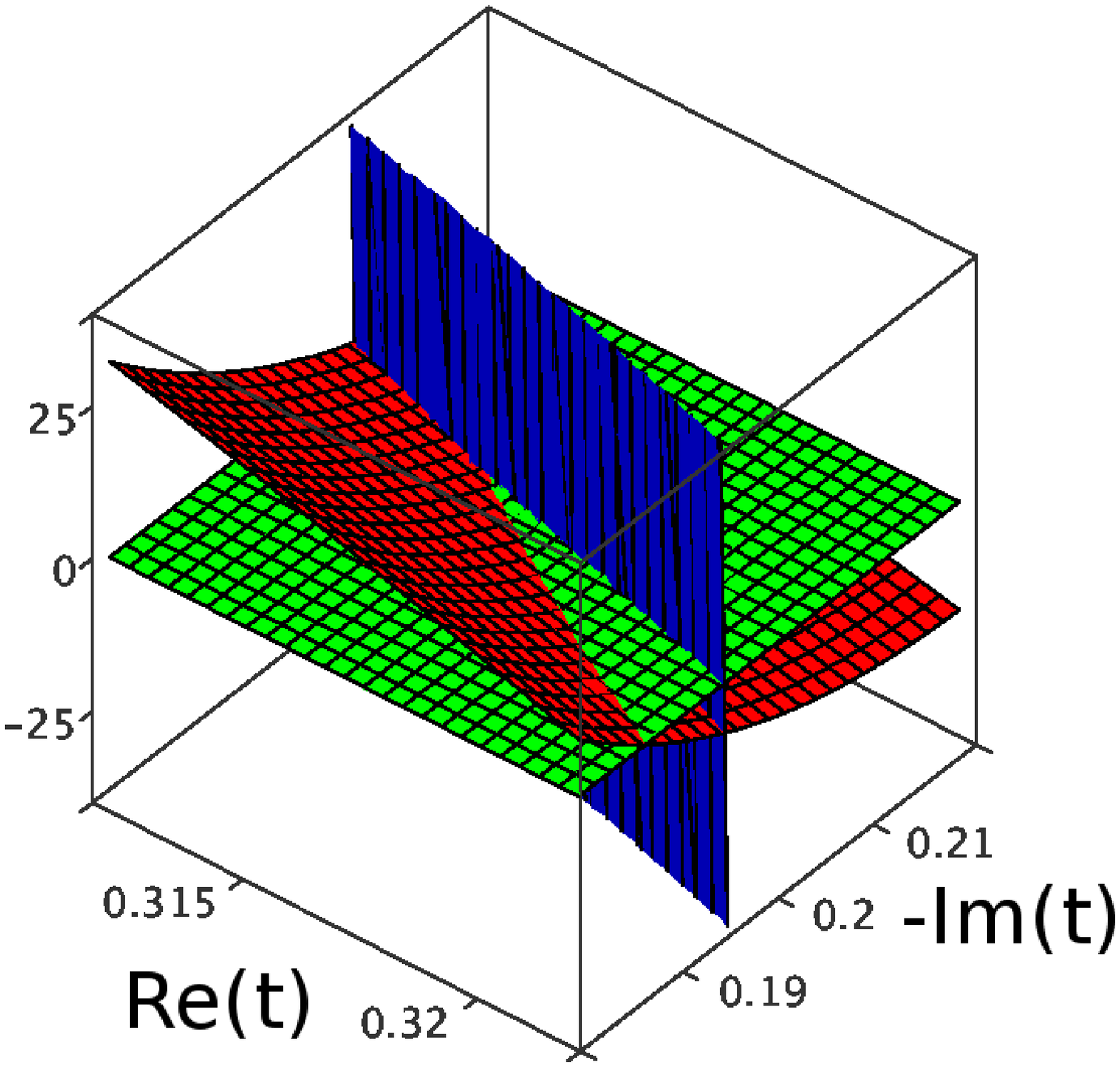} }
 \caption{The left graph shows a close-up of the region around $-t_2=0.2$ of fig.\ \ref{quadraticquartic}, i.e.\ for $t_{1}=0.3155$ and $a=8\pi^{2}/10$. The value for $t_1$ has to be fine-tuned in order to simultaneously allow for a small value of $\tilde{m}^2$ and $\tilde{\lambda}$, as can be seen from the right graph. This shows the dependence of $\tilde{m}^2$ and $\tilde{\lambda}$ on $t_1$ and $t_2$ in the vicinity of $t_{1}=0.3155$ and $-t_2=0.2$. The steep (blue) surface shows $\tilde{\lambda}$, the mildly curved (red) surface shows $\tilde{m}^2$ and the flat (green) surface is the zero section.}
  \label{quadraticquartic2}
\end{figure}

Some more comments are in order here. The value $t_1=0.3155$ differs from the one used in sections \ref{braneconfigs} and \ref{cosmology} and was chosen as it allows to fine-tune both parameters, $\tilde{m}^2$ and $\tilde{\lambda}$, to small values at the same time, cf.\ fig.\  
\ref{quadraticquartic2}. Doing so dispenses us from the need to use the exponential suppression factor in order to obtain a small mass parameter and allows us to use the value $\tilde{s}_2=-1$. For this value (and the values for $g$ and $\xi$ used in section \ref{near}, for instance) the volume does not get destabilized when the D-term potential is added to the F-term potential (cf.\ the discussion in \cite{Kallosh:2004yh}). A simple analysis shows that the D-term potential $V_{D} \sim 2\times 10^{-20}$ is generically much smaller than the 
constant piece of the F-term potential, given that $e^{2a\tilde{s}_{2}}$ is of order $10^{-7}$. 


\end{appendix}

\pagebreak

\end{document}